\documentclass{aa}  

\usepackage{float}
\usepackage{natbib}
\usepackage{amsmath}
\usepackage{amssymb}
\usepackage{subfig}
\usepackage{placeins}
\usepackage{commath}
\usepackage{graphicx}
\usepackage{txfonts}

\usepackage[hyperindex,breaklinks]{hyperref}
\hypersetup{
    draft=false,%
    breaklinks=true,%
    colorlinks=true,%
    linkcolor=blue,%
    citecolor=blue,%
    linktocpage=true,%
    pdftitle={},%
    pdfauthor={Marius Lehmann},%
    pdfsubject={},%
    }

\makeatletter
\newcommand*{\hyperlinkcite}[1]{\hyper@link{cite}{cite.#1}}
\makeatother

\begin{document}

   \title{Impact of Local Pressure Enhancements on Dust Concentration in Turbulent Protoplanetary Discs}


   \author{M. Lehmann
          \inst{1}
          \and
          M.-K.~ Lin\inst{1,2}
          }

   \institute{Institute of Astronomy and Astrophysics, Academia Sinica, Taipei 10617, Taiwan
    \and Physics Division, National Center for Theoretical Sciences, Taipei 10617, Taiwan\\
              \email{mlehmann@asiaa.sinica.edu.tw}
             }


 
  \abstract{
The standard core accretion model for planetesimal formation in protoplanetary discs (PPDs) is known to bear a number of challenges. One is the vertical settling of dust to the disc mid-plane against turbulent stirring. This is particularly relevant in presence of the vertical shear instability (VSI), a purely hydrodynamic instability applicable to the outer parts of PPDs, which drives moderate turbulence characterized by large-scale vertical motions. 
%

We investigate the evolution of dust and gas in the vicinity of local pressure enhancements (pressure bumps) in a PPD with turbulence sustained by the VSI. Our goal is to determine the morphology of dust concentrations and if dust can concentrate sufficiently to reach conditions that can trigger the Streaming Instability (SI). We perform a suite of global 2D axisymmetric and 3D simulations of dust and gas for a range of values for  $\Sigma_{d}/\Sigma_{g}$ (ratio of dust-to-gas surface mass densities or metallicity), particle Stokes numbers $\tau$, and pressure bump amplitude $A$. Dust feedback onto the gas is included. 

For the first time we demonstrate in global 3D simulations the collection of dust in long-lived vortices induced by the VSI. These vortices, which undergo slow radial inward drift, are the dusty analogs of large long-lived vortices found in previous dust-free simulations of the VSI. 
Without a pressure bump and for solar metallicity $Z\approx 0.01$ and Stokes numbers $\tau \sim 10^{-2}$ we find that such vortices can reach dust-to-gas density ratios slightly below unity in the discs' mid-plane, while for $Z \gtrsim 0.05$ long-lived vortices are largely absent. 
In the presence of a pressure bump, for $Z\approx 0.01$ and $\tau \sim 10^{-2}$ a dusty vortex forms that reaches dust-to-gas ratios of a few times unity, such that the SI is expected to develop, before it eventually shears out into a turbulent dust ring. At intermediate metallicities $Z \sim 0.03$ this occurs for $\tau \sim 5 \cdot 10^{-3}$, but with a weaker and more short-lived vortex, while for larger $\tau$ only a turbulent dust ring forms. For $Z\gtrsim 0.03$ we find that the dust ring  becomes increasingly axisymmetric for increasing $\tau$ and dust-to-gas ratios reach order unity for $\tau \gtrsim 5 \cdot 10^{-3}$.  Furthermore, the vertical mass flow profile of the disc is strongly affected by dust for $Z \gtrsim 0.03$, such that gas is transported inward near the mid-plane and outward at larger heights, which is the reversed situation compared to simulations with zero or small amounts of dust. We find Reynolds stresses or $\alpha$-values to drop moderately as $\sim  10^{-3}- 10^{-4}$ for metallicities increasing as $Z=0-0.05$.
Our results suggest that the VSI can play an active role in the formation of planetesimals through the formation of vortices for plausible values of metallicity and particle size. Also it may provide a natural explanation for the presence or absence of asymmetries of observed dust rings in PPDs, depending on the background metallicity.
}

   \maketitle
%

\section{Introduction}\label{sec:intro}

The standard core accretion scenario for planet formation \citep{safronov1972} proposes the growth of micron-sized dust particles to become planetesimals of size 1km-1000km, which can subsequently accrete solids \citep{wetherill1990,kenyon1999} and, if sufficiently massive, also gas \citep{mizuno1980}, to become terrestrial or gas giant planets. This process faces several difficulties. For one, it has been shown that micron size dust particles hit a so called bouncing barrier when reaching sizes of mm's to cm's \citep{blum2018} upon which further growth is hindered by means of collisions. On the other hand, gas drag causes rapid loss of pebbles due to radial drift \citep{weidenschilling1993,johansen2014}. 
One possible way to avoid these problems is by a direct self-gravitational collapse of dust grains following their settling to the disc mid-plane \citep{goldreich1973}. However, for this to happen self-gravity must overcome stellar tidal forces as well as collective pressure forces of the dense dust-layer. This in turn requires dust-to-gas ratios much larger than those typically expected in newly formed protoplanetary discs \citep{shi2013}.

Therefore, processes that can efficiently concentrate dust are under active research. The most popular of these is the SI \citep{youdin2005,johansen2007,youdin2007}, a linear dust-gas drag instability.
In its original form, formulated for an unstratified, monodisperse, and unmagnetized dusty disc, the SI draws energy from the radial dust-gas relative motion. In its nonlinear state it can lead to strong dust clumping and hence planetesimal formation \citep{johansen2009a,bai2010,simon2016}. 
However, strong clumping of pebbles typically requires super-solar values of the vertically-integrated dust-to-gas ratio or metallicity \citep{johansen2009a,bai2010,carrera15,yang17,li21}. One therefore assumes that additional processes are required to trigger the SI. Among these processes are particle concentration in zonal flows \citep{johansen2009b}, pressure bumps \citep{haghighipour2003a,haghighipour2003b,taki2016,onishi2017,huang2020}, vortices \citep{barge1995,johansen2004,klahr2006,fu2014,rubsamen2015,raettig2015,miranda2017,surville2019}, or other instabilities such as the Dust Settling Instability (DSI: \citet{squire2018,krapp2020}).

In this work, we are particularly interested in dust trapping by pressure bumps and vortices. We are motivated by recent ALMA observations that indicate dust rings -- presumably reflective of an underlying pressure bump -- are common in bright discs \citep{andrews18,long18}; while asymmetric dust distributions -- possibly reflective of vortices -- constitute a smaller, but non-negligible fraction of observed disc morphologies \citep{vdmarel2021}. \citet{carrera2021a} showed the SI can indeed be triggered in the vicinity of a moderate pressure bump embedded in a disc with solar metallicity and cm-sized dust particles, which then leads to planetesimal formation. On the other hand, while dust trapping by vortices has been shown to be efficient, in razor-thin disc simulations vortices eventually get disrupted due to dust-gas instabilities once the dust-to-gas ratio reaches order unity \citep{fu2014,rubsamen2015,raettig2015,surville2019}. However, \citet{lyra2018} and \citet{raettig2021} found that in 3D, vortices do not suffer from destruction because dust feedback is only important in the disc mid-plane, while their vortices are vertically extended. It is also possible to have a combination of pressure bumps and vortices, for example through the RWI \citep{lovelace1999,li01}, in which case long-lived, dust trapping vortices do form \citep{meheut2012}. Understanding the evolution of dust rings and asymmetries has direct application to interpreting observations.

One important element that may influence the aforementioned dust trapping processes is external turbulence. Ever since the discovery of the magneto-rotational-instability [MRI: \citep{balbus1991}], accretion in protoplanetary discs was thought to be mediated by MRI turbulent stresses. However, due to low ionization rates in protoplanetary discs, the MRI is most likely extinguished within a region of about $\sim$1-10 AU, which has been named ``dead zone'' \citep{gammie1996,turner2009,armitage2011,turner2014}. In fact, recent state-of-the-art magneto-hydrodynamical models show that the inner parts of protoplanetary discs between roughly $\sim$1-20 AU are indeed largely laminar and exhibit angular momentum transport induced by magneto-thermal winds and laminar Maxwell stresses  \citep{gressel2015,bai2015,bai2017}. 

Nonetheless, turbulence in this planet-forming region may still be present owing to the possible occurrence of purely hydrodynamic instabilities. Among these are the VSI \citep{urpin1998,urpin2003,nelson2013,barker2015,lin2015}], requiring a vertically sheared angular velocity profile coupled with rapid cooling of the gas, the Sub-critical Baroclinic Instability [SBI: \citep{klahr2003, lesur2010,lyra2011}], the Convective Overstability [COS: \citep{klahr2014,lyra2014,latter2016}], which requires an unstable radial entropy gradient coupled with a local gas cooling time on the order of the orbital time scale; and the zombie vortex instability [ZVI: \citep{marcus2015,lesur2016}], which requires slow cooling. A common outcome of these hydrodynamic instabilities is vortex formation, which could seed the SI by accumulating dust, as described above. However, vortex dust trapping has not been simulated explicitly in the case of the VSI and the ZVI, nor has the effect of global disc structures (such as a pressure bump) on any of the above hydrodynamic instabilities.

The aim of this paper is to study the efficiency of dust concentration at pressure bumps and vortices in VSI-turbulent discs. We focus on the VSI because it is a generic phenomenon in the sense that the only structural requirement for it to occur is a radial gradient in temperature or entropy of in principle arbitrary sign, which produces a vertical gradient in the disc's rotation. Based on models that account for a finite disc cooling time, the VSI is expected to be active at tens of AU \citep{lin2015}. 
A few studies considered the nonlinear evolution of gas and dust in VSI-turbulent protoplanetary discs by means of global hydrodynamic simulations. \citet{stoll2014,stoll2016} ran 3D simulations with a thermal disc structure governed by stellar irradiation and radiation transport. They found that the VSI is able to generate particle clumps that can in principle trigger the SI. \citet{flock2017} and \citet{flock2020}, employing a similar method, but covering a larger radial domain, (\citet{flock2020} in addition covering the full 360 degree azimuth), as well as higher resolution, concluded that the VSI is rather an impediment for the early and late phases of
planet formation. That is, they found that on the one hand the strong vertical gas motions generated by the VSI effectively lift 0.1-1 mm-sized dust particles such that these are prevented from settling to larger densities at the disc mid-plane. On the other hand, they concluded that accretion of mm-sized pebbles onto planetary embryos in the terrestial mass range is rendered inefficient by the VSI as the dust-layer thickness likely exceeds the planetary Hill sphere.  \citet{picogna2018} studied the accretion of pebbles with a wide variety of sizes onto planetary cores in a VSI turbulent disc. They found that at $\sim 5$AU the fastest growth of protoplanets is achieved for pebbles with $\tau \sim 1$ on account of their fast drift rates. Moreover, they found that the effect of the VSI on this process can be well described through an $\alpha$-viscosity accompanied by stochastic 'kicks' on particles.

However, these studies did not include the dust's back reaction force onto the gas that arises from their mutual frictional drag. 
\citet{lin2015} showed that while the VSI is driven by a vertical shear in the gas velocity, it is mitigated by buoyant forces, which are usually associated with an adiabatic gas. On the other hand, \citet{lin2017} showed that an isothermal, dusty gas can effectively be described by an imperfectly adiabatic gas for which the finite coupling time between gas and dust, as well as global gas temperature gradients act as sources/sinks of the effective entropy of the dust-gas mixture. One important aspect resulting from this model is that the presence of dust leads to an effective buoyancy frequency of the dusty gas that under normal conditions is larger than that of the gas in isolation. Moreover, linear stability calculations of \citet{lin2017} suggest that this dust-induced buoyancy indeed leads to a weakening of the VSI which can promote dust settling. This was confirmed by 
\citet{lin2019} with nonlinear axisymmetric hydrodynamic simulations adopting the single fluid model of \citet{lin2017}.   
Furthermore, \citet{lin2019} showed that dust-to-gas density ratios of a few times the solar value are sufficient to enable efficient settling of particles with Stokes numbers in the range $\tau \sim 10^{-3}-10^{-2}$. However, \citet{lin2019} did not consider the possible role of pressure bumps and their axisymmetric model precludes vortex formation. 

In this work we employ global, 2D axisymmetric and non-axisymmetric 3D simulations of a PPD to investigate under which conditions a gas pressure bump in a VSI-turbulent disc can raise the local dust-to-gas ratio to such levels that the SI can be triggered to facilitate planetesimal formation. Another question that we aim to answer is whether dust particles will tend to concentrate in vortices or in rings, and whether rings are axisymmetric or not, depending on the disc and dust parameters. Regarding the first point, we will show in this work that a moderate pressure bump ($A\gtrsim 0.2$) can collect sufficient amounts of dust to reach order unity dust-to-gas ratios even for solar metallicities $Z=0.01$, provided dust particles have sizes with  $\tau\gtrsim 10^{-2}$.  Moreover, we will show that the shapes of dust concentrations at a pressure bump become more axisymmetric with increasing metallicity and Stokes number. Generally, a non-axisymmetric appearance of dust rings or the occurrence of vortices requires metallicities $Z \lesssim 0.03$ in our model.

The paper is structured as follows. In Section \ref{sec:model} we describe the basic hydrodynamic equations and the model disc that will be the initial state of our hydrodynamical simulations. We will review basic aspects of dust-gas interaction and the effect of a pressure bump on dust drift. Also a brief description of the single fluid model of dust and gas, its resulting stability criteria, and the VSI are provided. These will be useful in interpreting results from our full two-fluid simulations. In Section \ref{sec:twodim} we will present and discuss the results of 2D simulations and in Sections \ref{sec:threedim} and \ref{sec:vortices}  those of 3D simulations. In Section \ref{sec:summary} we provide a summary and some conclusions following from our results, as well as prospects for future research.

\section{Hydrodynamic Disc Model}\label{sec:model}

\subsection{Basic Equations}\label{sec:eqn}
We consider a global hydrodynamic model of a PPD consisting of gas and a single species of dust, governed by the set of dynamical equations
\begin{equation}
\left(\frac{\partial}{\partial t} + \vec{v}_{g}\cdot\vec{\nabla}\right) \, \rho_{g} =  - \rho_{g} \left( \vec{\nabla} \cdot \vec{v}_{g} \right), \label{eq:contrhog}
\end{equation}
\vskip -0.5cm
\begin{equation}
 \left(\frac{\partial}{\partial t} + \vec{v}_{g}\cdot\vec{\nabla}\right) \, \vec{v}_{g}    =  - \frac{1}{\rho_{g}} \vec{\nabla} P +  \frac{1}{\rho_{g}} \vec{\nabla} \cdot \hat{T} - \vec{\nabla} \Phi_* - \frac{\epsilon}{t_{s}} \left(\vec{v}_{g} -\vec{v}_d \right), \label{eq:contvg}  
\end{equation}
\vskip -0.5cm
\begin{equation} 
\left(\frac{\partial}{\partial t} + \vec{v}_{d}\cdot\vec{\nabla}\right) \, \rho_{d} =  - \rho_{d} \left( \vec{\nabla} \cdot \vec{v_{d}} \right) +\vec{\nabla} \cdot \left( \nu \rho \vec{\nabla} f_{d}  \right), \label{eq:contrhod}
\end{equation}
\vskip -0.5cm
\begin{equation} 
\left(\frac{\partial}{\partial t} + \vec{v}_{d}\cdot\vec{\nabla}\right) \, \vec{v}_{d}  =   -\vec{\nabla}\Phi_*   - \frac{1}{t_{s}} \left(\vec{v}_{d} - \vec{v}_{g} \right), \label{eq:contvd} 
\end{equation}
where $\rho_{g}$, $\rho_{d}$, $\vec{v}_{g}$ and $\vec{v}_{d}$ are the gas and dust volume mass densities and three-dimensional gas and dust velocities, respectively, in a non-rotating frame with cylindrical coordinates $(r,\varphi,z)$ and with origin on a central star of mass $M_{*}$ with gravitational potential $\Phi_* = - GM_*/\sqrt{r^2+z^2}$, where $G$ is the gravitational constant. The indirect gravitational term, stemming from the fact that the center of mass of the system does not coincide with the center of the star, is omitted. Furthermore, we neglect self-gravity and magnetic fields. The remaining symbols in the above equations are explained below. 

We adopt a locally isothermal equation of state for the gas such that the pressure 
\begin{equation}\label{eq:eos}
P= c_{s}^2 \rho_{g},
\end{equation}
where the squared sound-speed follows a power-law with constant index $-q$: 
\begin{equation}
c_{s}^2(R)= c_{s0}^2 \left(\frac{r}{r_{0}}\right)^{-q}.
\end{equation}
The reference sound-speed $c_{s0} = c_s(r_0)$, where $r_0$ is a reference radius. Unless otherwise stated, we take $q=1$. This value is a bit larger than what is typically expected at large radii in PPDs where the temperature profile is set by stellar irradiation [e.g. \citet{andrews2009}]. However, our choice facilitates a comparison with the isothermal simulations by \citet{nelson2013}, \citet{stoll2016}, \citet{richard2016} and \citet{manger2018,manger2020} who used the same value. Moreover, it also has the advantage of a radially constant disc aspect ratio such that the required vertical resolution in our simulations is independent of radius. The characteristic gas pressure scale height $H_g = c_s/\Omega_K$, where $\Omega_K \equiv \sqrt{GM_*/r^3}$ is the Keplerian frequency. 

The dust component is treated as a pressureless fluid that interacts with the gas through a friction force \citep{johansen2014}
quantified by the stopping time\footnote{Strictly $c_{s}$ should be replaced by the mean thermal gas velocity \citep{weidenschilling1977} which differs from $c_{s}$ by a factor $\sqrt{8/\pi}$. For convenience we absorb this factor in the particle size $r_{d}$}.
\begin{equation}
t_{s}= \frac{r_{d} \rho_{d}}{\rho_{g} c_{s}},
\end{equation}
where $r_{d}$ and $\rho_{d}$ stand for the particle radius and bulk density, respectively. The stopping time is the characteristic timescale for a grain to reach velocity equilibrium with its surrounding gas. The fluid approximation for dust is valid for sufficiently small $t_s$ \citep{jacquet2011}. 
Usually we will work with the dimensionless Stokes number
\begin{equation}
\tau= \Omega_{K} t_{s}.
\end{equation}
 Grains with $\tau \ll 1$ are tightly (but not necessarily perfectly) coupled to the gas, which either corresponds to small grain sizes or to small distances to the star where the gas density is appreciably higher. The former case is the one that applies to this paper. The Stokes number at $r=r_{0}$, $z=0$, and time $t=0$, denoted by $\tau_{0}$, will be a freely specifiable parameter in our model. This means that, for a given particle radius and bulk density, its stopping time at a certain location and at a certain time is given by 
\begin{equation}\label{eq:tstop}
t_{s}= \frac{\tau_{0}}{\Omega(r_{0})} \frac{ \rho_{g}(r_{0},z=0,t=0)\, c_{s}(r_{0}) }{\rho_{g} c_{s} }.
\end{equation}
We also define
\begin{align}
\epsilon & = \frac{\rho_{d}}{\rho_{g}},\label{eq:eps}\\
f_{d} & = \frac{\rho_{d}}{\rho},\label{eq:fd}
\end{align}
as the dust-to-gas density ratio and the dust fraction, respectively, with the total density $\rho=\rho_{g}+\rho_{d}$. Note that $f_d = \epsilon/(1+\epsilon)$. 

Finally,
\begin{equation*}
\hat{T}= \rho_{g} \nu \left[ \vec{\nabla} \vec{v}_{g} + \left( \vec{\nabla} \vec{v}_{g} \right) ^{\dagger} - \frac{2}{3} \hat{U} \vec{\nabla} \cdot \vec{v}_{g} \right]
\end{equation*}
is the viscous stress tensor with a (constant) kinematic viscosity $\nu$, which also describes dust diffusion via the diffusion term in Equation (\ref{eq:contrhod}). 
The symbol $\dagger$ denotes the conjugate transpose and $\hat{U}$ stands for the unit tensor. 
Viscous terms are only included to ensure numerical stability. 
Hence, $\nu$ is chosen to be very small and in derivations which follow below viscous terms are neglected. In particular, $\nu$ is much smaller than the typical value of the $\alpha$-viscosity (defined in Section \ref{sec:sim}) measured in simulations.

\subsection{One-Fluid Model for Dust and Gas}\label{sec:onef}
Our numerical simulations evolve Equations \ref{eq:contrhog}--\ref{eq:contvd} directly (Section \ref{sec:sim}). However, as shown by \citet{lin2017} many important aspects of dust-gas dynamics can be described within a reduced one-fluid model, governed by the set of equations
\begin{equation}
\left(\frac{\partial}{\partial t} + \vec{v}\cdot\vec{\nabla}\right) \, \rho =  - \rho \left( \vec{\nabla} \cdot \vec{v} \right), \label{eq:contrho1f}
\end{equation}
\vskip -0.35cm
\begin{equation}
 \left(\frac{\partial}{\partial t} + \vec{v}\cdot\vec{\nabla}\right)\, \vec{v}     =  - \frac{1}{\rho} \vec{\nabla} P  - \vec{\nabla} \Phi_* , \label{eq:contv1f}  
\end{equation}
\vskip -0.34cm
\begin{equation} 
\left(\frac{\partial}{\partial t} + \vec{v}\cdot\vec{\nabla}\right) \, \mathrm{S_{\mathrm{eff}}} =  - \vec{v} \cdot \vec{\nabla} \ln \, c_{s}^2 +\frac{c_{s}^2}{P} \vec{\nabla} \cdot \left( t_{\mathrm{eff}} f_{d} \vec{\nabla} P   \right), \label{eq:contS1f}
\end{equation}
with density
\begin{equation*}
\rho = \rho_{g} + \rho_{d},
\end{equation*}
the center of mass velocity
\begin{equation}\label{eq:cms}
\vec{v} = \frac{\rho_{g} \vec{v}_{g} + \rho_{d} \vec{v}_{d} }{\rho}
\end{equation}
and where the \emph{effective} (dimensionless) entropy of the dust-gas mixture is defined by
\begin{equation}\label{eq:seff}
\mathrm{S_{\mathrm{eff}}} = \ln \frac{P}{\rho}
\end{equation}
and the effective particle stopping time by
\begin{equation}
 t_{\mathrm{eff}} = t_{s} \frac{\rho_{g}}{\rho}.
\end{equation}
These equations can be derived from Equations (\ref{eq:contrhog})-(\ref{eq:contvd}) and (\ref{eq:eos}) if dust and gas relative velocities are fixed by the terminal velocity approximation
\begin{equation}\label{eq:vterm}
\vec{v}_{g} - \vec{v}_{d} = -\frac{t_{\mathrm{eff}}}{\rho_{g}}  \vec{\nabla} P,
\end{equation}
valid for small particles which are tightly coupled to the gas \citep{youdin2005}. For a more general set of one-fluid equations, see \cite{laibe2014}. We use (\ref{eq:contrho1f})-(\ref{eq:contS1f}) to motivate the ground state of our disc. 

Within this formalism the dust fraction is related to the gas pressure (and temperature) via (\ref{eq:eos}) and (\ref{eq:fd}) such that
\begin{equation}
f_{d} = 1- \frac{P}{c_{s}^2 \rho}.
\end{equation}
Similarly, one can define a reduced temperature of the dust-gas mixture
\begin{equation}
T_{\mathrm{red}} \equiv \frac{\mu}{\mathcal{R}}\frac{P}{\rho} = T \left(1 - f_{d} \right),
\end{equation}
with the gas constant $\mathcal{R}$ and the mean molecular weight $\mu$,
such that an increased dust fraction corresponds to a reduced temperature. Furthermore, the flux term (second term on the right hand side of Equation (\ref{eq:contS1f})) stems from the fact that dust drifts into the direction of increasing pressure, as reflected by (\ref{eq:vterm}). Hence, the mixture of a locally isothermal gas with tightly coupled dust behaves as a single-component adiabatic fluid with a non-vanishing energy flux due to the exchange of dust between adjacent fluid parcels and an entropy source term resulting from the imposed global gas temperature profile.  

A detailed derivation of this model and applications to various dusty analogs of pure gas instabilities can be found in \citet{lin2017}. Here we merely summarize important quantities that govern the ground state of our disc and its stability, which can be directly derived from Equations (\ref{eq:contrho1f})-(\ref{eq:contS1f}).
Following \citet{lin2019} we assume a Gaussian dust-to-gas ratio distribution
 \begin{equation}\label{eq:epsmid}
\epsilon(r,z) = \epsilon_{mid}(r) \exp\left( -\frac{z^2}{2 H_{\epsilon}^2} \right), 
\end{equation}
where the gas and dust scale heights are implicitly defined via
\begin{equation}
\frac{1}{H_{\epsilon}^2} = \frac{1}{H_{d}^2}-\frac{1}{H_{g}^2}.
\end{equation}
This ansatz suggests that $\rho_{d}$ follows a Gaussian with scale height $H_{d}$ and $\rho_{g}$ follows a Gaussian with scale height $H_{g}$, which is indeed nearly the case as shown below.
The initial mid-plane dust-to-gas ratio $\epsilon_{mid}(r)$ is chosen such that the radial contribution to the effective entropy source vanishes, i.e.
\begin{equation}
\frac{\partial}{\partial r} \left( t_{\mathrm{eff}} f_{d} \frac{\partial P}{\partial r}   \right) = 0
\end{equation}
to reduce radial evolution of the initial state \citep[see also][]{chen2018}. Of course, in a stratified disc the vertical contribution cannot be zero without diffusion, which results in dust settling (so strictly speaking $v_z\neq 0$). In practice we will set $H_{d}\lesssim H_{g}$ such that $H_{\epsilon}\gg H_g$. That is, the dust is initially well-mixed with the gas. This enables us to define the ground state metallicity
\begin{equation}
Z\equiv \left(\frac{\Sigma_{d}}{\Sigma_{g}}\right)_{0} = \epsilon_{mid} \left(\frac{H_{d}}{H_{g}}\right)_{0},
\end{equation}
with the dust and gas surface mass densities $\Sigma_{d}$ and $\Sigma_{g}$, respectively. Since $H_{d}\lesssim H_{g}$ this also means that $\epsilon_{mid}(r=r_{0})\approx Z$. In addition to $\tau_{0}$ defined above, $Z$ will also be a used as an adjustable parameter in our simulations.

The radial and vertical components of Equation (\ref{eq:contv1f}) at equilibrium and under the assumption of axisymmetry read
\begin{equation}\label{eq:eqrad}
\Omega^2 r = \Omega_{K}^2 \left(1-\frac{3}{2} \frac{z^2}{r^2}\right) + \frac{1}{\rho} \frac{\partial P}{\partial r}
\end{equation}
and
\begin{equation}\label{eq:eqz}
0= -\Omega_{K}^2 z - \frac{1}{\rho} \frac{\partial P}{\partial z},
\end{equation}
where we defined the orbital frequency $\Omega=v_{\phi}/R$ and applied the thin disc approximation, i.e. a Taylor expansion of the gravitational force term in (\ref{eq:contv1f}) with respect to the quantity $z^2/r^2$ to next leading order and used the condition that the radial component of $\vec{v}$ identically vanishes. Furthermore, we neglected the small contribution $\sim (f_{d}\, \tau)^2 \Omega_{K}^2 |z| \ll \Omega_{K}^2 |z|$ due to dust settling\footnote{This term can be estimated from (\ref{eq:contvd}) and (\ref{eq:cms}) in the limit of small $\tau$ and assuming $v_{gz}=0$.}.
From Equation (\ref{eq:eqrad}) we obtain the orbital frequency
\begin{equation}\label{eq:omega}
\Omega(r,z)= \Omega_{K} \left[1-\frac{3}{2}\frac{z^2}{r^2} -\frac{2 \rho_{g}}{\rho}\eta   \right]^{1/2},
\end{equation}
where we defined
\begin{equation}\label{eq:eta}
\eta = -\frac{1}{2 \rho_{g} (\Omega_{K} r )^2} \frac{\partial P}{\partial \ln r},
\end{equation}
which was also defined in \citet{youdin2005}. Near the disc mid-plane we have $\eta>0$ such that the dusty gas rotates with sub-Keplerian frequency.
Integration of Equation (\ref{eq:eqz}) using (\ref{eq:epsmid}) yields the equilibrium gas density profile
\begin{equation}\label{eq:rhogas}
\rho_{g} = \rho_{g,mid} \exp\left\{-\frac{z^2}{2 H_{g}^2} -\epsilon_{mid} \frac{H_{\epsilon}^2}{H_{g}^2} \left[1- \exp\left(-\frac{z^2}{2 H_{\epsilon}^2} \right) \right]\right\}.
\end{equation}
As expected, for $\epsilon_{mid} \to 0$ we recover a Gaussian gas density profile with scale height $H_{g}$. In practice, deviations of the gas density from the latter are small, since we consider $\epsilon_{mid}\ll 1$, $H_{\epsilon}\gg H_g$, and $|z|$ is $O(H_g)$. 

We define
\begin{equation}\label{eq:gasmid}
\rho_{g,mid}(r)=\frac{\Sigma_{g}(r)}{\sqrt{2 \pi} H_{g}(r)},
\end{equation}
such that the vertically integrated gas density yields the surface density $\Sigma_{g}(r)$,
which we set to be
\begin{equation}\label{eq:sigmagas}
\Sigma_{g}(r) = \Sigma_{g,0} \left(\frac{r}{r_{0}} \right)^{-s},
\end{equation}
with $s=3/2$. The total surface density is $\Sigma = \Sigma_g + \Sigma_d$. 

\subsection{Dust Drift and Pressure Bumps}\label{sec:pbump}
Since our aim is to study the effect of a pressure bump on the dust evolution in the disc we modify the mid-plane density as
\begin{equation}\label{eq:gasmidpb}
\begin{split}
\rho_{g,mid}(r) & \to \rho_{g,mid}(r) \times  \left[1+ A \exp\left(-\frac{\left(r-r_{0}\right)^2}{2 H_{g0}^2} \right) \right]\\
\quad & = \frac{\Sigma_{0}  }{\sqrt{2 \pi} H_{g0}} \left(\frac{r}{r_{0}} \right)^{-p} \left[1+ A \exp\left(-\frac{\left(r-r_{0}\right)^2}{2 H_{g0}^2} \right) \right]
\end{split}
\end{equation}
with $p\equiv s+(3-q)/2$. That is, we add a Gaussian density bump of amplitude $A$ and width $H_{g}(r_{0})=H_{g0}$ to the gas mid-plane density profile. While the width of the bump will be kept fixed throughout all simulations, its amplitude $A$ will be varried. Similar density bumps have been employed in numerous previous studies (e.g. \citet{meheut2012,taki2016,carrera2021a}). With (\ref{eq:gasmidpb}) the equilibrium azimuthal velocity (\ref{eq:omega}) now reads 
\begin{equation}\label{eq:omega2}
\begin{split}
\Omega(r,z)  =  \Omega_{K} & \left\{ 1-\frac{3}{2} \frac{z^2}{r^2} -\frac{\rho_{g} H^2}{\rho r^2} \left[p+q+ r\frac{r-r_{0}}{H_{g0}^2} \right. \right. \\
 \quad & \left. \left. \times \frac{A \exp\left(-\frac{\left(r-r_{0}\right)^2}{2 H_{g0}^2} \right)}{1+A \exp\left(-\frac{\left(r-r_{0}\right)^2}{2 H_{g0}^2} \right)} \right] \right\}^{1/2}.
\end{split}
\end{equation}
If we restrict our considerations to a narrow region about the mid-plane for the time being, we can neglect vertical gravity and the results of unstratified discs apply.
In particular, since our disc is effectively inviscid the ground state planar velocities of dust and gas are those derived by \citet{nakagawa1986}, i.e.
\begin{equation}
\vec{v}_{d}  =  \left[ r \Omega   -\frac{1}{2}\frac{\rho_{g}}{\rho} \tau \, v_{\mathrm{drift}} \right]  \vec{e_{\varphi}}    +v_{\mathrm{drift}} \, \vec{e}_{r} ,
\end{equation}
\begin{equation}
\vec{v}_{g}  =  \left[ r \Omega  +\frac{1}{2} \frac{\rho_{d}}{\rho} \tau \, v_{\mathrm{drift}}  \right]  \vec{e_{\varphi}}    -\frac{\rho_{d}}{\rho_{g}} v_{\mathrm{drift}} \,  \vec{e}_{r} ,
\end{equation}
where we defined
\begin{equation}\label{eq:vdrift}
 v_{\mathrm{drift}}=-2 \left(\frac{\rho_{g}}{\rho}\right)^2 \frac{\eta \Omega_K r \tau }{1 + \left(\tau \rho_{g} /\rho \right)^2}.
\end{equation}
These ground state velocities can be derived from (\ref{eq:contvg}) and (\ref{eq:contvd}) under the neglect of vertical stratification.
Solutions for $\alpha$-viscous discs that take into account the vertical disc structure can be found for instance in \citet{takeuchi2002} and \citet{kanagawa2017}.

\begin{figure}[h!]
\centering
\includegraphics[width = 0.4 \textwidth]{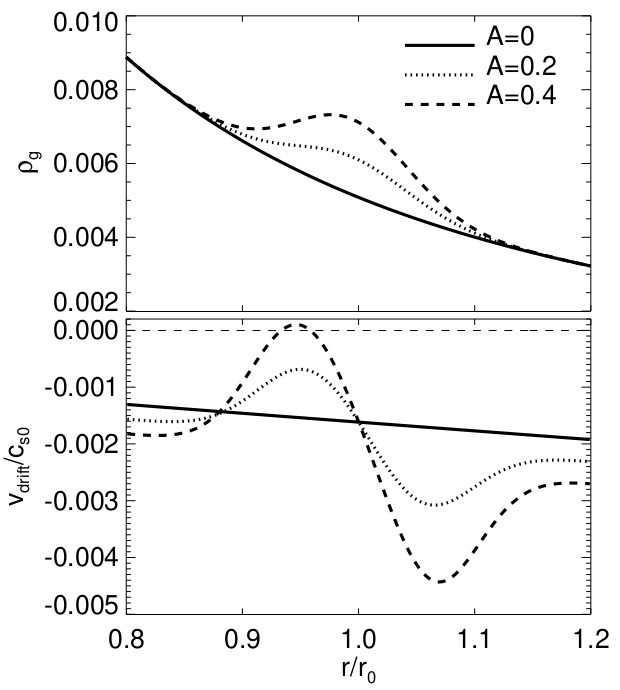}
\caption{Illustration of the initial mid-plane gas density profile (top panel) according to (\ref{eq:gasmidpb}) and the dust drift velocities (bottom panel) computed from (\ref{eq:vdrift}) with $\tau_{0}=10^{-2}$ for different pressure bump amplitudes $A$.}
\label{fig:gasdens}
\end{figure}

The effect of a pressure bump ($A>0$) in Eq. (\ref{eq:omega2}) has a profound effect on dust drift. This is illustrated in Figure \ref{fig:gasdens} for different amplitudes $A$ and $\tau=10^{-2}$.
This figure shows how $v_{\mathrm{drift}}$ is essentially controlled by the magnitude of $\eta$. Since $\eta \sim  h^2 \equiv (H_{g}/r)^2 $ in PPDs ($h=H_{g}/r$ being the disc aspect ratio), typical values are $\eta \sim 10^{-2}-10^{-3}$. For the most part $\eta>0$ so dust drifts inwards, but $\eta$ is non-uniform. The bump thus leads to a 'traffic jam'-like accumulation of dust within a region $\Delta r \sim H_{g}$ surrounding the bump center. For the case $A=0.4$ the particle drift comes to a complete halt at a certain radius. This is one reason why pressure bumps in PPDs are expected to be preferred locations for planetesimal formation. 

\subsection{Disc Stability and the VSI}\label{sec:vsi}
By combining Equations (\ref{eq:eqrad}) and (\ref{eq:eqz}) we obtain
\begin{equation}\label{eq:vshear}
\begin{split}
r \frac{\partial \Omega^2}{\partial z}  = & \frac{c_{s}^2(r)}{\left(1+\epsilon \right)^2} \\
 \quad & \times \left[ \frac{\partial \epsilon }{\partial r} \frac{\partial \ln \rho_{g}}{\partial z} -\frac{\partial \epsilon}{\partial z} \frac{\partial \ln P}{\partial r} +\frac{q}{r} \left(1+\epsilon\right) \frac{\partial \ln \rho_{g}}{\partial z} \right].
 \end{split}
\end{equation}
Hence, the disc's equilibrium velocity profile is subjected to vertical shear, which constitutes the driving force of the VSI. As pointed out in \citet{lin2017} and \citet{lin2019},
in typical situations the main source of vertical shear is the disc's global temperature gradient, quantified through $q$. However, in regions where sufficiently large gradients of $\epsilon$ occur, also the first two terms in the bracket may play a role. More on this follows below.

Another important quantity to characterize our disc is the vertical Buoyancy frequency
\begin{equation}\label{eq:buoyan}
N_{z}^2 \equiv -\frac{1}{\rho} \frac{\partial P }{\partial z } \frac{\partial S_{\mathrm{eff}}}{\partial z} =c_{s}^2  \frac{\partial \ln \rho_{g}}{\partial z} \frac{\partial f_{d}}{\partial z},
\end{equation}
which quantifies the stabilizing effect of vertical entropy stratification with respect to adiabatic perturbations. In our vertically isothermal disc this quantity is entirely due to dust-layering.
As shown by \citet{lin2017}, `dusty' buoyancy mitigates the VSI by stabilizing vertical motions, similar to classical buoyancy if $N_{z}^2  \gtrsim r \abs{ \frac{\partial \Omega^2}{\partial z} } $, which is satisfied within sufficiently settled dust-layers. However, in the presence of strong VSI turbulence corrugation of the dust-layer can result in situations where $\partial f_{d} / \partial_{z} >0$ for $z>0$ and hence $N_{z}^2 <0$, such that vertical convection may occur.

 The occurrence of the VSI in a locally isothermal (dust-free) disc as the one considered here has been established analytically by \citet{nelson2013}, \citet{barker2015} and \citet{lin2015}. In addition, \citet{nelson2013} conducted 3D hydrodynamical simulations.
Linear stability analyses presented in these papers show that the VSI excites inertial waves that are destabilized by free energy extracted from the disc's vertical shear.
The dominant modes are so called ``body modes'', which constitute vertically global ($l_{z} \sim h r$), radially local ($l_{x}\sim h^2 r$) low frequency ($ \sim h  \Omega_{K}$) traveling inertial waves that result in either corrugation or 'breathing' motion of the disc with respect to the mid-plane.
The maximum linear growth rates $\sigma$ of the body modes in an isothermal disc are governed by the maximum shear in the domain under consideration, i.e.
\begin{equation*}\label{eq:grat}
\abs{\sigma} < \mathrm{max} \abs{ r \frac{\mathrm{d} \Omega}{\mathrm{d} z}} \sim \Omega_{K} q h,
\end{equation*}
where the latter similarity assumes a domain $\abs{ z} \lesssim H_{g}$. This approximation remains applicable even in the presence of dust \citep{lin2017}. However, dust limits the amplitude of turbulent velocity perturbations wherever dust-induced buoyancy becomes significant.

Furthermore, \citet{lin2017} provided the corresponding Solberg-H\o iland stability criteria for a dusty gas in the limit of perfect coupling $t_{\mathrm{eff}}=0$ and vanishing radial temperature gradient $q=0$, such that the source terms in Equation  (\ref{eq:contS1f}) vanish. These criteria read 
\begin{equation}\label{eq:sh1}
\kappa^2 + \frac{1}{\rho_{g}} \vec{\nabla} P \cdot \vec{\nabla} f_{d} >0,
\end{equation}
\begin{equation}\label{eq:sh2}
-\frac{1}{\rho_{g}} \frac{\partial P}{\partial z} \left( -\kappa^2 \frac{\partial f_{d}}{\partial z}  + r \frac{\partial \Omega^2}{\partial z} \frac{\partial f_{d}}{\partial r} \right) >0,
\end{equation}
where $\kappa^2=r^{-3}\partial_r\left(r^4\Omega^2\right)$ is the epicycle frequency squared, and determine the conditions for stability with respect to adiabatic perturbations \citep{tassoul1978}. These are appropriate since the dust-gas mixture under the aforementioned approximations behaves like an adiabatic gas that conserves the entropy (\ref{eq:seff}). 
As discussed by \citet{lin2017}, the first criterion (\ref{eq:sh1}) is expected to be fulfilled in typical situations due to alignment of the two gradients in the second term and since the disc is rotationally supported. 
The second criterion (\ref{eq:sh2}) can in principle be violated at locations where dust is well mixed vertically but $\epsilon$ undergoes sufficiently strong radial variations. Furthermore, if $\partial f_{d} / \partial_{z} > 0$ occurs for $z>0$ the first term in the brackets in (\ref{eq:sh2}) is expected to be destabilizing. We will indeed see that both situations can be realized under certain circumstances in our simulations which involve strong corrugations of the dust-layer due to the VSI. 
However, this also means that the VSI is required in first place to violate any of the two criteria in our simulations. 
We will come back to this issue in Section \ref{sec:pbumpres}.

Finally, an important instability that is directly involved in the nonlinear saturation of the VSI is the Rossby Wave Instability (RWI, \citet{lovelace1999}) which can result in the formation of vortices. This instability can be expected when the generalised vortensity  
\begin{equation}\label{eq:vorten}
\mathcal{L} = \frac{\Sigma}{2 \omega_{z}}  \left(\frac{\Pi}{\Sigma^{\gamma}}\right)^{2/\gamma} 
\end{equation}
possesses a local extremum, where $\Pi = c_{s}^2 \Sigma_{g}$ denotes the vertically integrated pressure and where the $z$-component of vorticity defined in the inertial frame is 
\begin{equation}\label{eq:vort}
\omega_z =\vec{e}_{z}  \cdot \vec{\nabla} \times (\vec{v} + \Omega_{0} r \, \vec{e}_{\varphi}),
\end{equation}
and $\gamma$ is the adiabatic index. It should be noted that although (\ref{eq:vort}) and other quantities above are defined using the center of mass velocity (\ref{eq:cms}), the latter is nearly equal to both the dust and gas velocities since dust is tightly coupled to the gas in our model.
Although the above condition was originally derived for gaseous discs, our locally isothermal gas with tightly coupled dust is equivalent to an adiabatic gas disc with $\gamma = 1$ \citep{lin2017}. We therefore expect $\mathcal{L}$ to also play a role in our simulations. Note that here, $\Pi$ arises from the gas only, but $\Sigma$ accounts for both gas and dust. For brevity we will simply refer to $\mathcal{L}$ as the vortensity. 

As outlined in \citet{richard2016} the VSI generates axisymmetric vortensity rings. These become unstable to the RWI, which consequently leads to the formation of vortices while the vortensity extrema are destroyed \citep[see also][]{manger2018,manger2020}. \citet{latter2018} argue that in an isothermal disc where the vertical shear is strictly forced by the imposed global temperature gradient,
the amplitude of the VSI-related velocity perturbations is limited by the emergence of Kelvin-Helmholtz parasitic modes that drain energy from the VSI modes and transfer it to smaller scales until viscous dissipation sets in. Based on theoretical arguments \citet{latter2018} estimate a maximal turbulent velocity amplitude of saturated VSI modes of a few percent to roughly ten percent of $c_{s}$. This is indeed in agreement with results from previous isothermal simulations of the VSI and also with results presented below. The parasitic modes may also result in the formation of small-scale vortices. 
Moreover,  the mid-plane dustlayer can in principle undergo Kelvin Helmholtz Instability [KHI; \citet{johansen2006,lee2010}] but this is not expected to be resolved in our simulations. We also note that the pressure bump that is initially placed in our simulations is not expected to be Rossby Wave unstable when comparing the values of the width and amplitude used here with those applied in linear stability analyses of \citet{ono2016} for a Gaussian density bump. That is, the largest amplitude bump considered here ($A=0.4$) might be marginally unstable to the RWI. However, in our
\begin{table*}
	\centering
	\caption{List of all 2D simulations.}
\label{tab:2d}
	\begin{tabular}{lccc} 
		\hline \hline
	   & Sect. \ref{sec:twodim} & Sect. \ref{sec:turbulence} & App. \ref{sec:qvar}  \\
		\hline
   $Z$ [$10^{-2}$]   &    1, 2, 3, 4, 5, 10  & 0, 1, 3, 5, 10 & 1, 3   \\
 $\tau_{0}$ [$10^{-3}$] & 1, 2, 3, 4, 5, 6, 10  & 1, 5, 10 & 6    \\
       $A$ &     0, 0.1, 0.2, 0.3, 0.4  & 0  & 0.4 \\
           $p$  &  2.5  & 2.5  & 2.5 \\ 
           $q$  &  1 & 1 & 0.1, 0.5, 0.75, 1 \\ 
             $N_{R}\times N_{\theta}$    &   $2160 \times 480$  & $1088 \times 256$  &   $2160 \times 480$  \\
		\hline
	\end{tabular}
	        	\tablefoot{Only one simulation presented in Section \ref{sec:turbulence} adopted $Z=0$ and  corresponds to the dust-free simulation.}
\end{table*}
\begin{table*}
	\centering
	\caption{List of all 3D simulations.} 
\label{tab:3D}
	\begin{tabular}{lcccccc} 
		\hline \hline
	   & Sect. \ref{sec:turbulence} & Sect. \ref{sec:dust rings} & Sect. \ref{sec:vort_dustfree} & Sect. \ref{sec:nopbdust} & Sect. \ref{sec:pbdust} & App. \ref{sec:pvar} \\
		\hline
   $Z$ [$10^{-2}$]  &  0, 1, 3, 5, 10  &  1, 3, 5, 10  &  0 & 1, 3, 5, 10  & 1, 3, 5, 10  & 0    \\
 $\tau_{0}$ [$10^{-3}$] & 1, 5, 10 & 1, 4, 5, 10 & - & 1, 5, 10  & 1, 4, 5, 10  & -    \\
       $A$ &     0  & 0.2, 0.4  & 0, 0.2, 0.3, 0.4 & 0 & 0.2, 0.4 & 0 \\
           $p$  &  2.5  &  2.5 & 2.5  & 2.5 & 2.5 & -1, 0, 1, 2.5, 3.5  \\ 
           $q$  &  1 &  1 &  1 & 1 & 1 & 1  \\ 
             $N_{R}\times N_{\theta}\times N_{\varphi}$  &   $1088 \times 256 \times 512$  &   $\cdots$ & $\cdots$  & $\cdots$ & $\cdots$& $\cdots$\\
		\hline
	\end{tabular}
	 \tablefoot{Only one 3D simulation presented in Section \ref{sec:dust rings} adopted $\tau_{0}=4\cdot 10^{-3}$ together with $Z=0.03$ and $A=0.4$ (Figure \ref{fig:dustring_us}). Only one 3D simulation presented in Section \ref{sec:pbdust} adopted $\tau_{0}=4\cdot 10^{-3}$ together with $Z=0.03$ and $A=0.2$ (Figure \ref{fig:pres_evol_dust}). Simulations with $Z=0$ correspond to dust-free simulations.}
\end{table*}
\noindent
simulations we do not find a qualitative difference between the cases $A=0.2$ and $A=0.4$.

\subsection{Hydrodynamic Simulations}\label{sec:sim}

We solve Equations (\ref{eq:contrhog})-(\ref{eq:contvd}) with the multi-fluid code FARGO3D\footnote{\url{http://fargo.in2p3.fr}} \citep{fargo3d,llambay2019}.  
All quantities are subjected to periodic boundary conditions in azimuth.
As for the radial and vertical
boundary conditions equilibrium values of gas density and azimuthal velocity are extrapolated into the ghost zones, while gas radial and vertical velocities are set to zero at the boundaries. The only difference for the dust is that densities are subjected to symmetric boundary conditions. Simulations are carried out in spherical coordinates $(R,\varphi,\theta)$. Units are such that $R_{0}=\Omega_{0}=G=1$ ($R_{0}$ equals the radius $r_{0}$ defined in Section \ref{sec:model}).
We adopt $h_{0}=H_{g0}/R_{0} =0.05$ in all runs. 
The numerical grid covers $0.5\leq R\leq 1.5$, $-3H_{g}\leq z \leq +3 H_{g}$, where $z=R \tan(\pi/2+\theta)$, and in 3D simulations $0\leq \varphi \leq \pi$.
The grid resolution of most of our 2D simulations is $N_{R}\times N_{\theta} = 2160 \times 480$. Our 3D simulations are conducted on a grid with $N_{R} \times N_\theta \times N_\varphi = 1088 \times 256 \times 512$. This corresponds to a resolution of $\sim (54  \times 42 \times 8) \, H_{g0}^{-1}$. While the radial and vertical resolutions are high compared to previous studies, we adopt a rather moderate azimuthal resolution for reasons of computational resources and since we expect structures to form in our simulations to be predominantly axisymmetric or elongated in azimuthal direction. 
Selected 2D simulations that are used for a direct comparison with 3D simulations have the same radial and vertical grid cells. The latter are carried out on a GPU cluster which is necessary to cover a substantial domain in parameter space while adopting a reasonable spatial resolution. All simulations are run for 1000 reference orbits. A list of all conducted 2D and 3D simulations with references to the corresponding sections is provided in Tables \ref{tab:2d} and \ref{tab:3D}, respectively.

For analysis of our simulations we will often consider the diagnostic quantities defined as follows.
Turbulent vertical momentum transport will be quantified through the Reynolds stress component
\begin{equation}\label{eq:reynolds}
\langle \mathcal{R}_{\theta\varphi}(\theta,t) \rangle = \langle \mathrm{sgn}(z) \, \rho_{g} v_{g\theta} \Delta v_{g\varphi} \rangle_{R\varphi}
\end{equation}
where $\Delta v_{g\varphi}$ denotes the azimuthal velocity deviation from its ground state value. Averages $\langle \, \rangle$ are in all cases taken over $R$, $\varphi$ (in 3D simulations) and in addition either $\theta$ or $t$. The additional factor $\mathrm{sgn}(z)$ is included here in order to facilitate averaging of values above and below the mid-plane $z=0$ when presenting its time evolution, which enables a better comparison with the radial stress component defined below. Positive values of $\langle \mathcal{R}_{\theta\varphi} \rangle$ correspond to angular momentum transport away from the mid-plane. Unless otherwise stated, the diagnostic domain of our simulation region is $0.8 \leq R \leq 1.2$. 

In 3D simulations the radial turbulent angular momentum transport is similarly described by 
\begin{equation}
\langle \mathcal{R}_{R\varphi}(\theta,t) \rangle = \langle \rho_{g} v_{gR} \Delta v_{g\varphi} \rangle_{R\varphi}.
\end{equation}
The turbulent $\alpha$-parameter \citep{shakura1973} is then defined through
\begin{equation}\label{eq:alpha}
\alpha(\theta,t) = \frac{\langle \mathcal{R}_{R\varphi} \rangle_{R\varphi}}{\langle P \rangle_{R \varphi}}.
\end{equation}
Furthermore, the dust scale height $H_{d}$ is computed by fitting a Gaussian along the vertical grid $z$ to the distribution $\rho_{d}(z)$.

\section{Results of Axisymmetric 2D Simulations}\label{sec:twodim}

\subsection{Dust Settling in the Absence of a Pressure Bump}

The settling of dust toward the mid-plane of a VSI-turbulent PPD in the absence of a pressure bump was studied in some detail
\begin{figure*}[h!]
\centering
\includegraphics[width = 0.9 \textwidth]{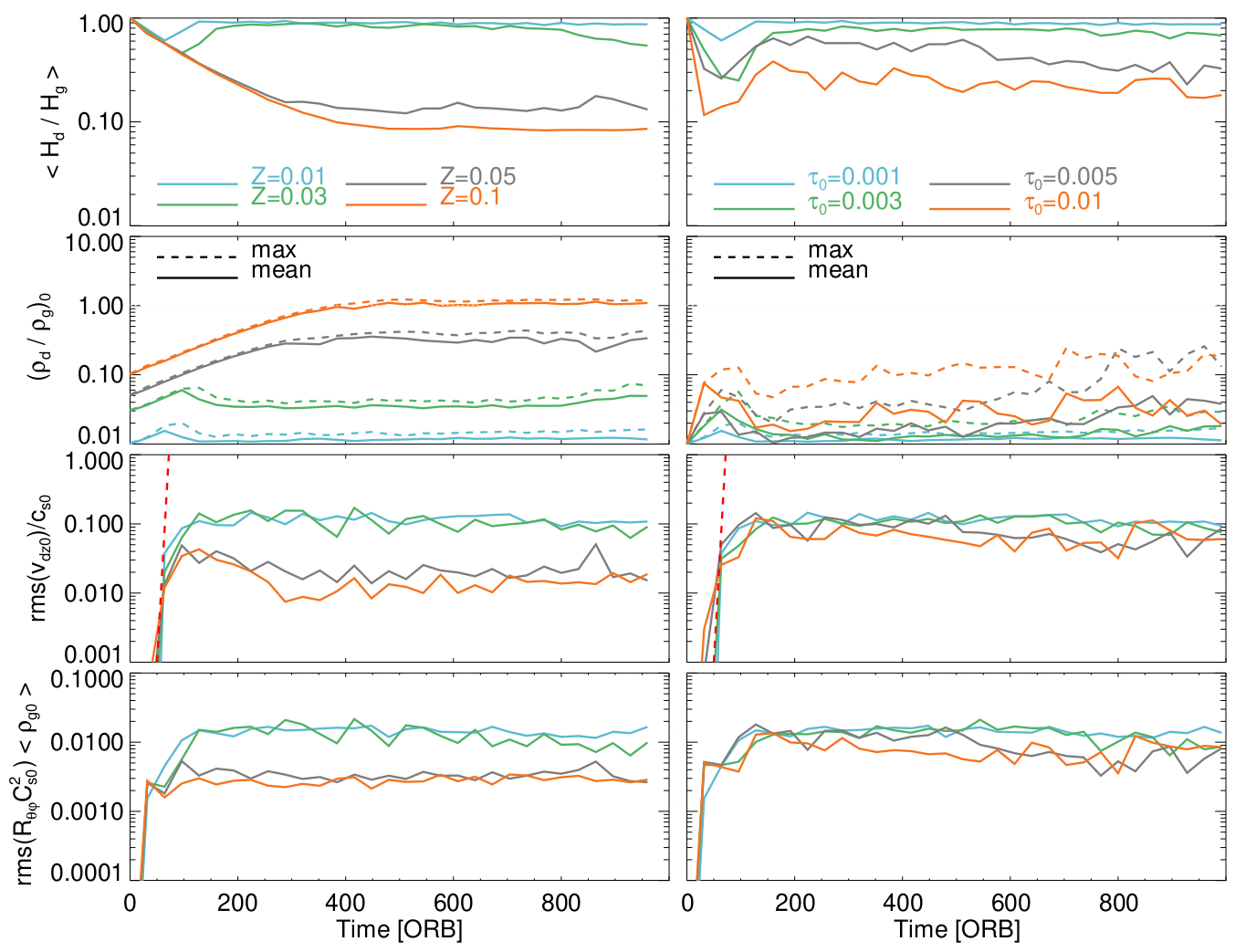}
\caption{Time evolution of quantities that describe the dust settling process against the emergence of VSI turbulence. From top to bottom these are the ratio of dust-to-gas scaleheight, the mid-plane dust-to-gas density ratio, the root mean squared mid-plane vertical gas velocity and the root mean squared vertical Reynolds's stress. The left panels compare different metallicities with fixed $\tau_{0} = 10^{-3}$ while the right panels compare different Stokes numbers $\tau_{0}$ with $Z=0.01$. The dashed red lines in the panels of rms($v_{gz0}$) are the exponential $\exp \left(q \Omega_{k} h \right)$ which describes the predicted linear growth of VSI-related velocity perturbations for an isothermal dust-free gas \citep{nelson2013}.  }
\label{fig:z_comp_lin}
\end{figure*}
by \citet{lin2019} who employed the one-fluid model to describe a dusty gas devised by \citet{lin2017} (Section \ref{sec:onef}).
Figure \ref{fig:z_comp_lin}  illustrates the effect of the particle size, i.e. the Stokes number $\tau_{0}$, as well as the background metallicity $Z$ on the dust's ability to settle in the mid-plane of the disc.
Shown is the time evolution of (from top to bottom) dust-to-gas scale height ratio, dust-to-gas mid-plane mass-density ratio, rms mid-plane vertical dust velocity, rms vertically averaged Reynolds stress as defined in section \ref{sec:sim}, averaged here over the radial domain $R_{0}-H_{0}<R<R_{0}+H_{0}$.
The results in the left panels of Figure \ref{fig:z_comp_lin}, which correspond to a Stokes number $\tau_{0}=10^{-3}$, show a clear systematic trend with increasing metallicity $Z$.
That is, turbulence generated by the VSI is increasingly mitigated with increasing amount of dust in the system. As a result, the latter is able to settle to a layer of decreasing thickness, which is also reflected in decreased values
of the mid-plane vertical dust velocity $\mathrm{rms}(v_{dz})$ and the vertical Reynolds stress parameter $\mathcal{R}_{\theta \phi}$.
It can be seen from these curves that the VSI reaches nonlinear saturation after some 100 orbits, in agreement with previous studies \citep{nelson2013,richard2016,manger2018}.
Furthermore, these results agree well with those of \citet{lin2019}, although the impact of dust settling appears to be slightly weaker in our simulations. 
A substantially weaker trend is seen in the right panel, which shows the effect of an increasing particle size ($\tau_{0}$) for a fixed metallicity $Z=0.01$.

In agreement with \citet{lin2019} the VSI growth rates are largely unaffected by dust.
However, there appears to be a weak increase in the early values of $\mathrm{rms}(v_{dz})$ with increasing $\tau_{0}$ and also with increasing $Z$, the former increase being stronger. Since initially dust settles such that the center of mass velocity $v_{settle} \sim \tau_{0} f_{d} c_{s}$ (at $z \sim H_{g0}$) it can be expected that the collective vertical movement of dust and gas toward the mid-plane serves as a seed for VSI modes, which are hence stronger initially for larger $\tau_{0}$ and to a lesser extent for larger $\epsilon_{0}$ (recall that $f_{d}=\epsilon/(1+\epsilon)$), explaining the observed trends.
Furthermore, since $v_{settle}$ roughly increases linearly with $\tau_{0}$, dust can consequently settle to a thinner layer before the VSI turbulence develops. In the thinner layer buoyancy (\ref{eq:buoyan}) is increased so as to stabilize the VSI which reduces stirring of the dust-layer. Outside of the dust layer away from the mid-plane the VSI still drives turbulence, albeit weaker than in the dust-free case.
The effect of particle size observed here is weaker than in the simulations of \citet{lin2019}. 
Thus, it appears that the overall impact of dust in the system is stronger in the latter study. 
One reason might be that the single fluid description implicitly assumes the terminal velocity approximation \citep{youdin2005,lovascio2019} for the dust. 
Furthermore, the different numerical setup used in \citet{lin2019} is possibly more dissipative such that the overall level of turbulence is expected to be slightly weaker.
Nevertheless, the plots in Figure \ref{fig:z_comp_lin} demonstrate that the back-reaction force from the dust onto the gas is essential for the weakening of the VSI. 
The level of turbulent velocity perturbations of a few percent to about 10 percent in the nonlinear saturation of the VSI is consistent with the theoretical estimates by \citet{latter2018}.

\subsection{Dust Settling in the Presence of a Pressure Bump}\label{sec:pbumpres}

\begin{figure}[h!]
\centering
\includegraphics[width = 0.47 \textwidth]{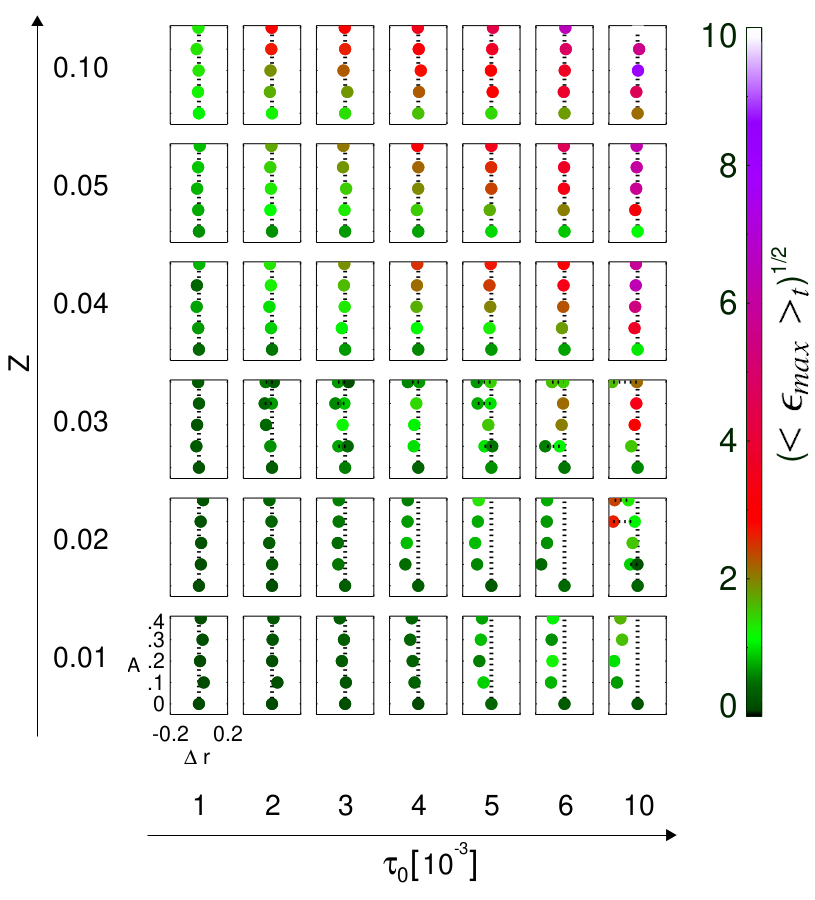}
\caption{2D simulation survey of the effect of a pressure bump on dust evolution. Shown is the final location of $\epsilon_{max}=(\rho_{d}/\rho_{g})_{max}$ within the domain $0.8 \leq R \leq 1.2$ and its value is indicated through the coloring. The different panels along the vertical direction compare different metallicities, while those along the horizontal direction compare different Stokes numbers. Within each panel different amplitudes $A$ are compared. In cases where two dots are drawn the original dust ring produced a secondary ring through a dusty gas instability as described in the text.}
\label{fig:pgrid}
\end{figure}

In the following we investigate the impact of a pressure bump on the accumulation of dust in 2D simulations which will be compared to results of 3D simulations in Section \ref{sec:threedim}.
Figure \ref{fig:pgrid} summarizes a 2D simulation survey for the evolution of dust near a pressure bump over 1000 orbital periods. Results are presented for varying initial amplitude $A$ and for different Stokes numbers $\tau_{0}$ and ground state metallicities $Z$.
Each solid circle represents a single simulation in terms of the radial location of the maximal 
mid-plane dust-to gas ratio $\langle\epsilon_{max}\rangle_{t} \equiv \langle\mathrm{max}(\rho_{d}/\rho_{g})_{z=0}\rangle_{t}$ averaged over the last 100 orbits, which in all simulations with $A>0$ is the result of dust accumulation in the vicinity of the initially imposed pressure bump. The color of each circle represents the value of $\langle\epsilon_{max}\rangle_{t}$. 
This figure reveals several noteworthy features. 
First of all, in absence of a pressure bump ($A=0$) we find that in order to achieve mid-plane dust-to-gas ratios on the order unity or larger (and hence triggering of the SI and possibly planetesimal formation), a metallicity of at least $Z\approx 0.04$ and a Stokes number $\tau \gtrsim 0.01$ are necessary. 
On the other hand, for non-vanishing bump amplitude this can be achieved with solar metallicity ($Z\sim 0.01$) for the same Stokes number. 

For lower metallicity ($Z=0.01-0.02$) we observe an increasing (radial) inward shift of the radial location of $\langle\epsilon_{max}\rangle_{t}$ with increasing $\tau_{0}$,
while for higher metallicity ($Z \gtrsim 0.04$) this trend does not exist and $\langle\epsilon_{max}\rangle_{t}$ is in all cases located at $R\sim R_{0}$. 
The inward shift at lower metallicities implies that the gas pressure bump has drifted inward while dust accumulated within its vicinity and migrated along with the gas bump, in contrary to the cases with higher metallicity where the pressure bump remains at its origin. 

In some cases with lower metallicity and larger bump amplitude ($A \gtrsim 0.3$) the resulting dust ring becomes unstable such that it gets disrupted (either partially or entirely) and a new dust ring forms just inside the original one. Thus, in the case of partial disruption two dust rings remain at the end of the simulation, both of which have resulted from the pressure bump. Such simulations are accordingly represented by two filled circles connected by a dotted line in Figure \ref{fig:pgrid}. For large parts of the considered parameter space though, the simulation outcomes show only a subtle dependence on the bump amplitude (as long as it not too small). A similar observation was also reported by \citet{carrera2021a}.

\begin{figure*}[h!]
\centering
\includegraphics[width = 0.75 \textwidth]{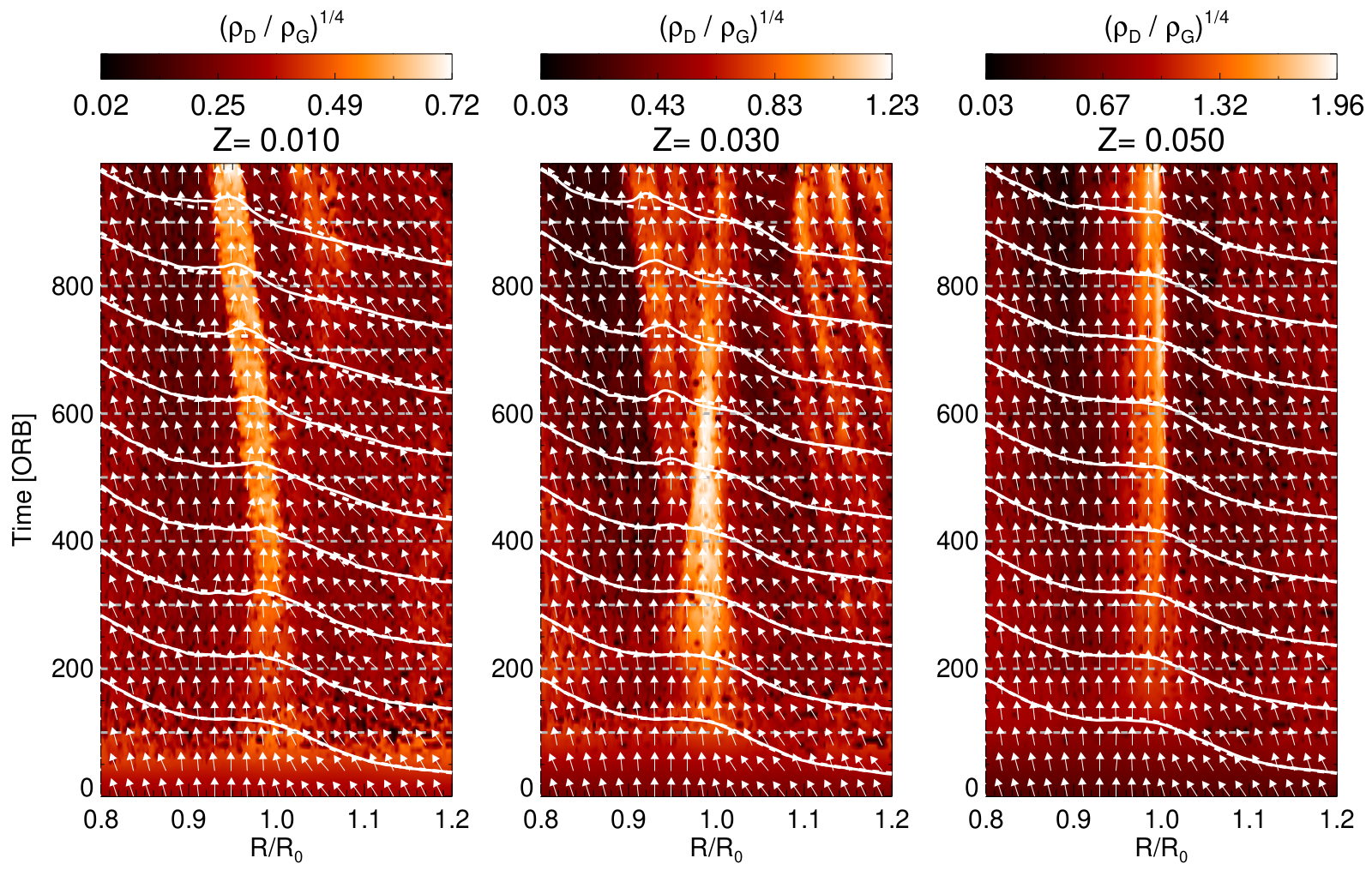}
\caption{Time evolution of the dust-to-gas ratio in the vicinity of a pressure bump with amplitude $A=0.4$. Compared from left to right are different metallicities $Z=0.01,0.03,0.05$ with the same $\tau_{0}=4 \cdot 10^{-3}$. The over-plotted solid curves are the mid-plane gas pressure at times indicated by horizontal dashed lines. The dashed curves are the initial gas pressure for comparison. The arrows indicate the direction of dust movement as predicted by (\ref{eq:vdrift}) and are all normalized to have the same length.}
\label{fig:sptd_zcomp}
\end{figure*}

The different outcomes of our simulations are illustrated in Figure \ref{fig:sptd_zcomp} which displays the time evolution of $\epsilon$ (we plot its fourth root for improved visibility) for three simulations with $Z=0.01$, $Z=0.03$ and $Z=0.05$ (in all cases $A=0.4$), respectively.
 The over-plotted profiles illustrate the gas pressure at different times which are indicated by horizontal dashed lines. Since our simulations are locally isothermal, these curves effectively correspond to the evolution of the gas density. 
 The arrows indicate dust drift velocities as computed from (\ref{eq:vdrift}), which, as expected describe the movement of dusty `clumps' quite well. Interestingly, particle feedback seems to enhance the gas pressure bump for lower background metallicities, rendering the pressure bump unstable such that it eventually disrupts. In principle such an enhanced pressure bump could potentially be subjected to the RWI \citep{ono2016} which, however, is suppressed due to the imposed axisymmetry in these simulations. Disruption of the dust ring occurs much earlier for the case with intermediate metallicity $Z=0.03$. Whether the dust ring in the case $Z=0.01$ would disrupt in our simulation ultimately depends on the radial size of the domain since sufficient dust needs to be accumulated before the region outside the ring gets depleted of dust. On the other hand, no instability appears to occur for the case with $Z=0.05$. These findings are chiefly confirmed in full 3D simulations which will be presented in Section \ref{sec:threedim}.

\subsection{Instability of Dusty Rings}\label{sec:instab}
It turns out that the drifting, as well as the splitting of dust rings both are the result of a violation of at least one of the Solberg-H\o iland criteria for a dusty gas 
[Eqs. (\ref{eq:sh1}), (\ref{eq:sh2})]. 
In regions of low metallicity the VSI is sufficiently vigorous to strongly puff up the dust-layer such that dust adjacent to a high density ring possesses a substantially larger scale height than the former, such that a relatively strong radial gradient $\abs{\partial\epsilon /\partial R} \gg  0$ can occur away from the mid-plane and at the same time a small vertical gradient $\partial\epsilon /\partial z \approx 0$.
If $\partial\epsilon /\partial R \ll / \gg 0$, accommodated by a vertical shear around the mid-plane $\partial\Omega^2 / \partial z >/<0$, this results in a violation of (\ref{eq:sh2}).

VSI turbulence can also result in situations\footnote{in what follows we restrict to values $z>0$, keeping in mind that a similar argument applies to the case $z<0$.} where $\partial\epsilon /\partial z >0$ for $z>0$, which, along with $\kappa^2 >0$ leads to a violation of (\ref{eq:sh2}) provided $\partial\epsilon/\partial R \approx 0$. Such situations are realized for instance within the dust "bubble" directly inside the dust ring in the simulation with $Z=0.03$. Moreover, a positive $\partial \epsilon / \partial z$ above the mid-plane translates to vertical convection since then $N_{z}^2<0$ [Equation (\ref{eq:buoyan})].
\begin{figure*}[h!]
\centering
\includegraphics[width = 0.9 \textwidth]{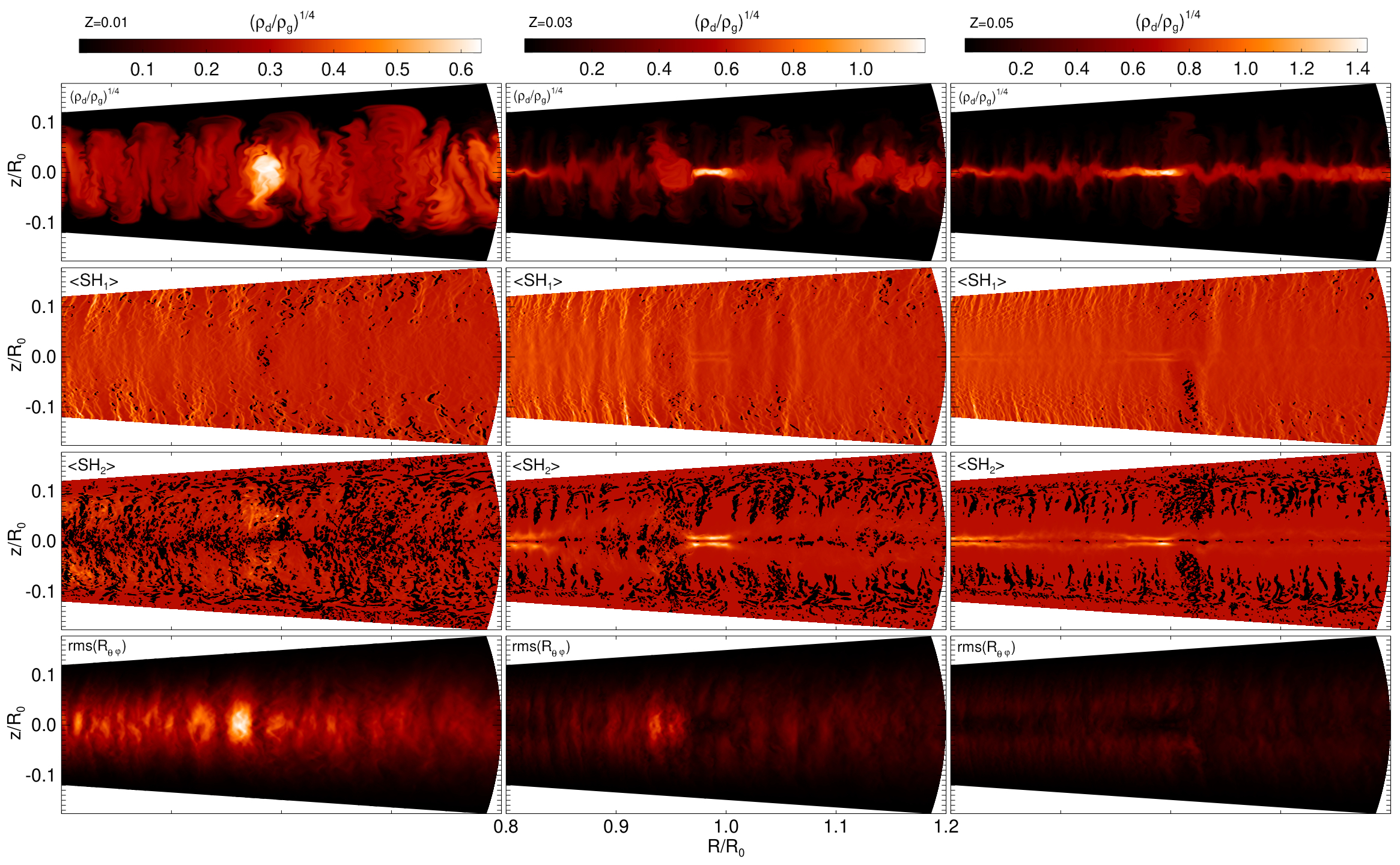}
\caption{Meridional cuts of (from top to bottom) the dust-to-gas ratio at 450 orbits, the averaged Solberg-H\o iland expressions (\ref{eq:sh1}), (\ref{eq:sh2}) (the left hand side terms) and root mean squared vertical Reynolds stress. Averages are taken over the time 350-450 orbits. In the panels $\langle \textrm{SH}_{1}\rangle$ and $\langle \textrm{SH}_{2}\rangle$ negative values are indicated by a black color.
These plots compare the same simulations as in Figure \ref{fig:sptd_zcomp} and illustrate the emergence of a dusty gas instability through a violation of the Solberg-H\o iland criteria.}
\label{fig:instab}
\end{figure*}

Furthermore, in low metallicity regions the VSI modes, when attaining sufficiently large amplitude, can directly violate the Rayleigh criterion such that (\ref{eq:sh1}) is violated since the first term on the right hand side of (\ref{eq:sh1}) dominates the second term practically everywhere in our simulated discs. In parts of such a region we also find (\ref{eq:sh2}) to 
be violated where $\partial\epsilon /\partial z < 0 $ for $z>0$, which is expected near the mid-plane.
The dominance of the first term over the second in (\ref{eq:sh1}) is due to the system still being rotationally supported, so that the buoyancy term $\vec{\nabla} P \cdot \vec{\nabla} f_d$ is small in comparison. 
Radial profiles of $\kappa^2$ appear noisy on the grid level and they vary in time. It is not unexpected that $\kappa^2$ can drop to negative values. The VSI operates on small radial length scales and the linear modes are also nonlinear solutions, at least in the incompressible limit \citep{latter2018}. Therefore one can assume that linear VSI modes are able to grow to large amplitudes. Specifically the vertical component of vorticity, which amounts to $\kappa^2/2 \Omega$ in axisymmetric models. Therefore, this quantity can grow until the Rayleigh criterion is violated. In 3D simulations, however, linear modes will undergo KHI / RWI and the flow develops non-axisymmetry before the Rayleigh criterion is actually violated.

Figure \ref{fig:instab} illustrates the unstable behavior of dust rings in the same simulations as those shown in Figure \ref{fig:sptd_zcomp}. 
The upper panels display the dust-to-gas ratios for a time near 450 orbits, which is just prior to the dust ring in the simulation with $Z=0.03$ being disrupted at the expense of a new dust ring, which subsequently drifts inward. This is also when the dust ring in the simulation with $Z=0.01$ starts to drift inward. The remaining panels from top to bottom are the two time-averaged (over the range 350-450 orbits) Solberg-H\o iland expressions $\langle \textrm{SH}_{1}\rangle$, $\langle \textrm{SH}_{2}\rangle$, corresponding to (\ref{eq:sh1}), (\ref{eq:sh2}), and the rms averaged Reynolds stress $\mathrm{rms}(\mathcal{R}_{\theta \varphi})$, corresponding to (\ref{eq:reynolds}). What these plots mainly show is that the Solberg-H\o iland criteria are violated in various places, mostly away from the mid-plane. 
However, both stability criteria are most severely violated in the puffed up dust-layers adjacent to the dust ring of each simulation. In the unstable regions just inside of the dust rings for the cases $Z=0.01$ and $Z=0.03$ we find a substantially increased Reynolds stress $R_{\theta\varphi}$ and hence an increased vertical angular momentum transport (away from the mid-plane). We expect this to be responsible for the inward drift / disruption of the dust rings in these simulations.

\subsection{Discussion}

Previous works that investigated the evolution of dust in the vicinity of a pressure bump ( \citet{taki2016,onishi2017,huang2020,carrera2021a}) did not report unstable behavior of dust rings that formed at a pressure bump, as described in the previous sections.  Although \citet{taki2016} found that the pressure bump gets destroyed by the particle feedback within hundreds of orbital periods without sufficient reforcing, this turned out to be caused by the neglect of vertical stellar gravity and hence dust sedimentation in the mid-plane \citep{onishi2017}. Also, for the parameters used by \citet{taki2016} ($Z=0.1$, $\tau_{0}=1$) or \citet{huang2020} ($Z=0.05$, $\tau_{0}\sim 0.25$) we do not expect drifting or disruption in our simulations to occur as well, based on Figure \ref{fig:pgrid}.

\citet{huang2020} found an instability of dust rings with dust-to-gas ratios of order unity, but their simulations were 2D (vertically integrated) and thus as well omitted the vertical disc 
\begin{figure*}[h!]
\centering
\includegraphics[width = 0.99 \textwidth]{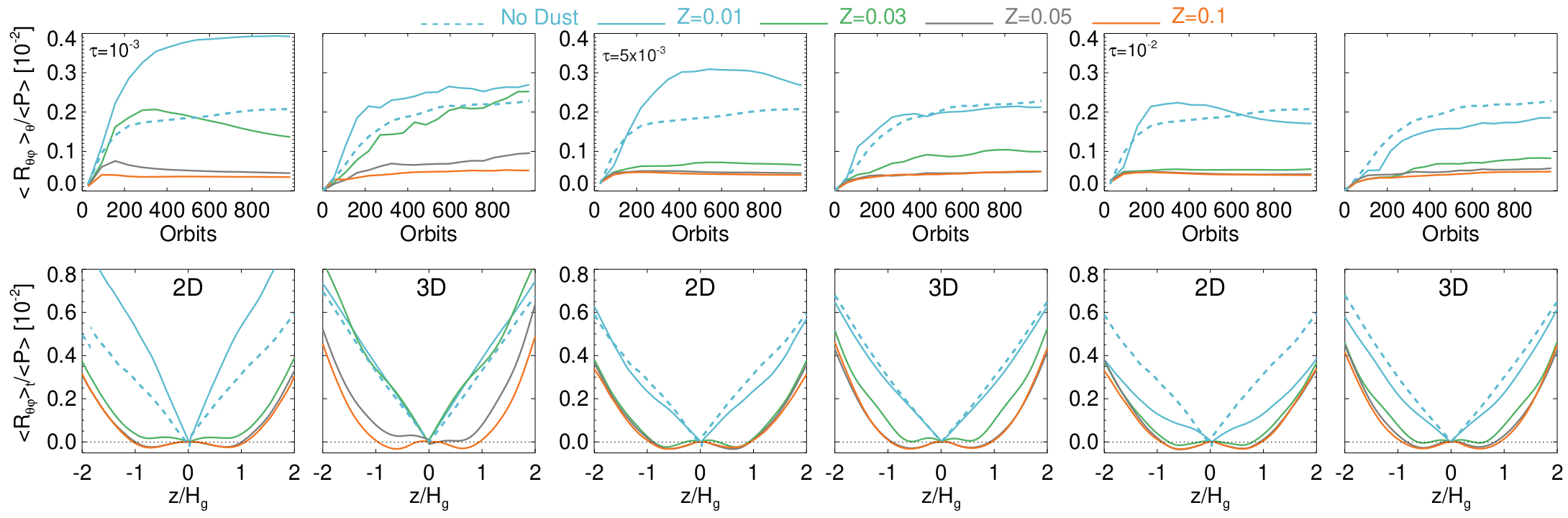}
\caption{Comparison of the vertical Reynolds stress (\ref{eq:reynolds}) in 2D and 3D simulations for different metallicities $Z$ and Stokes numbers $\tau_{0}$. The upper panels show the running time-averages (spatially averaged over the entire diagnostic domain $0.8 \leq R \leq 1.2$) while the lower panels display the time-averaged (over the last 400 orbits) vertical profiles. The dashed curve is in all panels for 2D and 3D the same, respectively. The horizontal dotted line indicates ``zero'' such that positive and negative values correspond to angular momentum transport away from the mid-plane and towards the mid-pane, respectively.}
\label{fig:reynolds_comp}
\end{figure*}
%
%
%
stratification. Furthermore, the instability found by \citet{huang2020}
does in principle only require a dust ring with sufficiently sharp edges and order unity dust-to-gas ratio such that gas within the dust ring is forced to attain Keplerian velocity, and they speculate that the edges of the ring could be unstable to the RWI due to a sharp drop of the vorticity $\omega_{z}$, which is not captured in our 2D axisymmetric simulations. Also the finding that our simulation with $Z=0.05$ does not show unstable behaviour despite having developed a dense sharp dust ring would then be hard to explain.

We have verified that neither drifting nor disruption of dust rings occurs in 1D radial simulations or 2D vertically unstratified, axisymmetric simulations. Furthermore, we ran simulations with reduced temperature gradient (i.e. smaller values of $|q|$) and found that the unstable behavior diminishes, the dust rings behaving essentially laminar, with decreasing $|q|$. These findings, which are presented in Appendix \ref{sec:qvar}, support our picture of a dusty gas instability that is triggered by the VSI turbulent motions and which lead to the phenomena described above. Indeed, the VSI does not occur in an unstratified disc as there is no vertical shear. Also a reduction of $|q|$ weakens the vertical shear and hence the VSI.


\section{Results of 3D Simulations: Turbulent Properties and Dust Rings}\label{sec:threedim}

In this section we present the results of 3D simulations where the main focus lies on the collection of dust into rings or vortices 
%
%
%
%
\noindent
which will be discussed in Sections \ref{sec:dust rings} and \ref{sec:vortices}, respectively. Before turning to these topics we first present a brief investigation of the turbulent properties of the dusty gas disc. Where possible we will draw comparisons with our 2D simulations.

\subsection{Turbulent Flow Properties}\label{sec:turbulence}

\begin{figure*}[h!]
\centering
\includegraphics[width = 0.7 \textwidth]{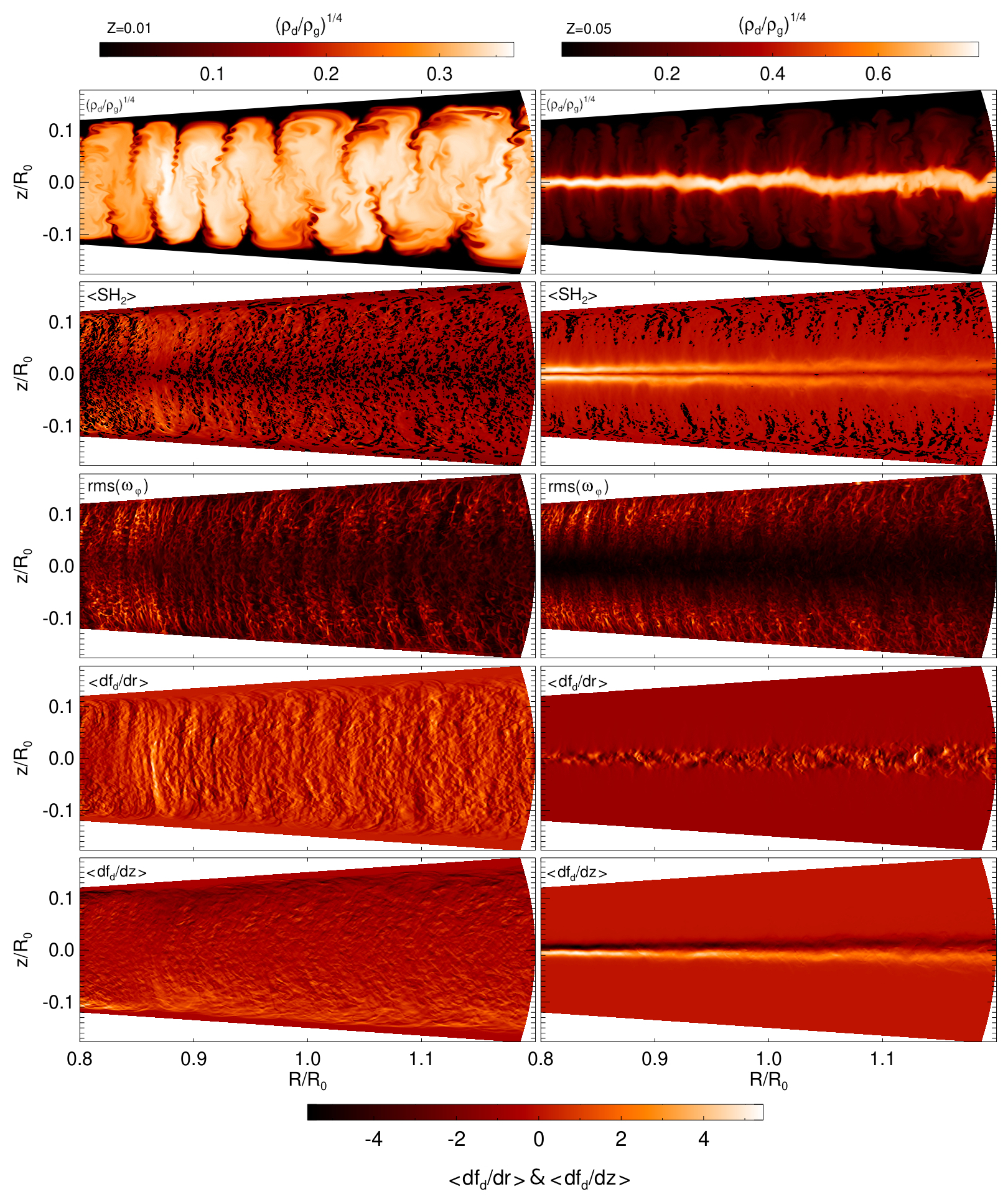}
\caption{Illustration of the origin of the increased vertical Reynolds stress $\mathcal{R}_{\theta\varphi}$ in simulations with small $Z$ and $\tau_{0}= 10^{-3}$ compared to the dust-free simulation as seen in Figure \ref{fig:reynolds_comp}. In the left panels, corresponding to $Z=0.01$, the VSI strongly corrugates the dust-layer which results in non-vanishing radial and vertical gradients of $f_{d}$ which result in a violation of (\ref{eq:sh2}), which is indicated by black colors in the plots labeled $\langle \textrm{SH}_{2}\rangle$. The largest contribution to the violation is by radial gradients. The right panels correspond to $Z=0.05$ where the VSI is too much weakened to cause significant corrugation.}
\label{fig:activ}
\end{figure*}

We start by comparing the strength of VSI turbulence as it occurs in 3D and 2D simulations without an initial pressure bump.
Figure \ref{fig:reynolds_comp} shows the time evolution of the spatially averaged vertical Reynolds stress (\ref{eq:reynolds}), which represents a running time-average to smooth out strong fluctuations (upper panels),
as well as its time-averaged vertical profile (lower panels). Time-averages have in the lower panel been taken over the last 400 orbits. 
Overall, turbulence takes comparable magnitudes in 2D and 3D simulations.
The influence of dust on the strength of turbulence appears to be complicated.
At low metallicity and small Stokes number, the presence of dust appears to result in additional stirring of the gas such that the stress $\mathcal{R}_{\theta\varphi}$ is enhanced as compared to the dust free case. This applies to both 2D and 3D simulations, although the effect is stronger in 2D.
We conjecture that this increased turbulent activity in the low metallicity and Stokes number cases shares its origin with the dusty gas instability discussed in Section \ref{sec:instab}. This is illustrated in Figure \ref{fig:activ}, where the plots correspond to $~900$ orbits. In the plots of the dust-to-gas ratios small-scale vortices are visible along the edges of the strongly corrugated dust-layers in the lower metallicity case. The second row displays, similarly as in Figure \ref{fig:instab} the time-averaged (900-1000 orbits) expression $\langle \textrm{SH}_{2}\rangle$, corresponding to (\ref{eq:sh2}). The third row represents the time-average of the  gas-vorticity component $\omega_{\varphi} \equiv \vec{e}_{\varphi} \cdot \left(\vec{\nabla} \times \vec{v}_{g} \right)$. The last two rows show the radial and vertical gradients of the dust fraction $f_{d}$. This figure underlines that increased vorticity production stems mainly from radial structuring and to a smaller extent from vertical structuring of the dust-layer, which we verified by inspection of the two terms within the brackets on the right hand side of (\ref{eq:sh2}). This structuring of the dust-layer is caused by corrugation (an idea first put forward by \citet{aguilar2015}). In principle, since there are locations where $\partial \epsilon / \partial z >0 $ for $z >0$ and hence $N_{z}^2<0$, the condition for vertical convection is locally fulfilled at these locations during some time. It is unclear if the latter has a notable influence on the disc turbulence. The figure also shows how the vertical gradient $\partial f_{d}/\partial z$ has a strongly stabilizing effect in the case $Z=0.05$ that overcompensates potentially destabilizing effects of the radial gradient $\partial f_{d}/\partial R$.
Thus, at larger metallicity dust sufficiently weakens the VSI such that dust can settle to a thinner and denser layer which is less prone to corrugation motions. This leads to a strongly reduced production of vorticity $\omega_{\varphi}$ around the mid-plane. Subsequently, the Reynolds stresses fall below the dust-free values. In principle an increased vorticity could also be due to KHI as adjacent dusty gas layers shear past each other in the vertical direction due to corrugation. However, if KHI were the reason for the observed increased vorticity and Reynolds stress we would expect this effect to be strongest in the dust-free case since the KHI does not rely on the presence of dust and the VSI is causing the corrugation which would be the origin of the KHI. The observed strong increase in $R_{\theta\varphi}$ when adding small amounts of dust would not be expected in this scenario. Also the finding that the Solberg-H\o iland criteria are actually violated in the corrugated dusty layer points to the dusty gas instability for being the origin of the increased hydrodynamic activity. Moreover, our resolution is likely not sufficiently high to resolve such parasitic KHI modes (C. Cui, private communication).

Interestingly, \citet{aguilar2015}, who conducted smoothed particle hydrodynamics simulations of a PPD with very similar setup (3D, locally isothermal) found a
similar corrugation of the mid-plane dust-layer.
However, they concluded this to be the result of a baroclinic instability which arises in response to a violation of (\ref{eq:sh2}) which they defined for a pure gas with an entropy $S \sim   \log\left(P/ \rho_{g}^{\gamma}\right)$ in contrast to our definition (\ref{eq:seff}), which is also that of \citet{lin2017} and where $\rho_{g}$ is replaced by the total density $\rho$. This subtle difference in the definition of $S$ leads to a change of sign in the criterion for the onset of vertical convection, i.e. $\partial \epsilon / \partial z > 0$ for $z>0$ in contrast to Equation (7) of \citet{aguilar2015}. From the results presented in Figure \ref{fig:activ} we conclude that the criterion adopted here is adequate since the settled dust-layer itself is stable, whereas regions above and below it are marginally unstable. Corrugation of the dust-layer in our simulations is a direct result of velocity perturbations of VSI modes and it is the corrugation that facilitates the conditions for a violation the second Solberg-H\o iland criterion or a locally unstable vertical entropy stratification. Furthermore, it remains questionable that the VSI did not develop in the simulations of \citet{aguilar2015} despite the required physical conditions being fulfilled.

\begin{figure*}[h!]
\centering
\includegraphics[width = 0.9 \textwidth]{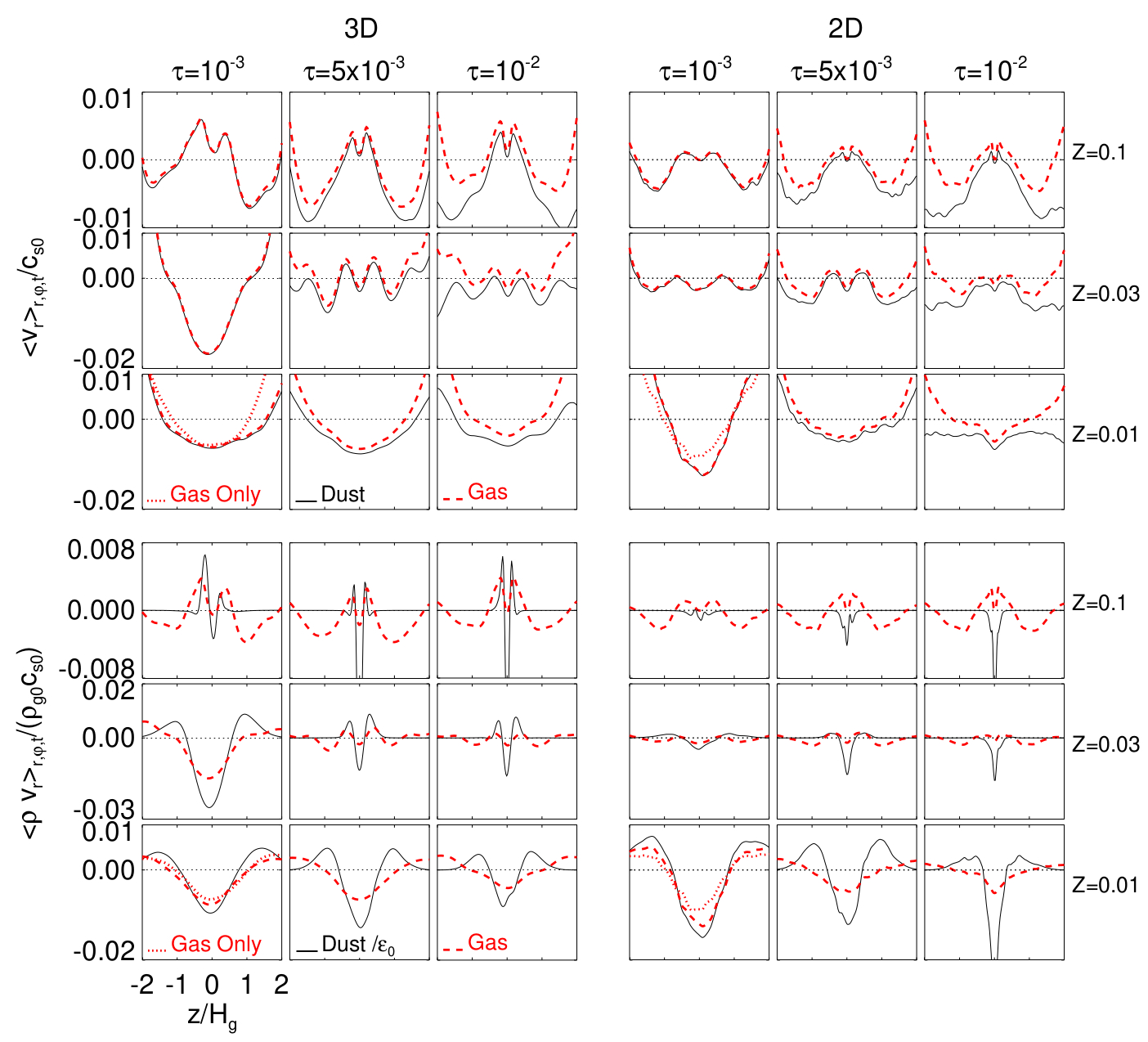}
\caption{Vertical profiles of the averaged radial dust and gas velocities (upper 3 rows) and the dust and gas mass flows (lower 3 rows) in 3D (left panels) and 2D (right panels) simulations. The Stokes number $\tau_{0}$ increases from left to right while $Z$ increases from the lower to upper panels.
For comparison, in the panels corresponding to $Z=0.01$ and $\tau_{0}=10^{-3}$ we additionally plot the result of a dust-free simulation. Averages have been taken over the last 400 orbits and over $0.8 \leq R \leq 1.2$. The dust mass flows are additionally scaled with a factor $1/\epsilon_{0}$ as compared to the gas mass flows. The horizontal axis displays $-2 H_{g} \leq z \leq 2 H_{g}$ in all panels.}
\label{fig:flow}
\end{figure*}

To further illustrate the impact of dust on the turbulent properties of the disc we compare in Figure \ref{fig:flow} the vertical dependence of the time and radially/azimuthaly averaged radial mass flow and radial velocity of gas and dust in 2D and 3D simulations, respectively.
At low metallicity (lower panels in all cases) we recover the results of \citet{stoll2016} and \citet{manger2018}. That is, gas flows inward near the disc's mid-plane, whereas at larger heights it flows outward. 
A similar flow pattern is also found in our 2D simulations, albeit of slightly different magnitude, which, in some cases exceed the 3D values.
The observation that inflow velocities near the mid-plane for low metallicities  in 2D and 3D simulations take comparable values most likely stems from the  anisotropic (Reynolds) stress generated by the VSI and that the most prominent perturbations excited by the VSI are vertically global modes of relatively short radial wavelength $\lesssim H_{g}$.
This anisotropy was studied by \citet{stoll2017} who attempted to model the turbulent accretion flow induced
 by the VSI in 3D simulations by a laminar viscous accretion flow in 2D axisymmetric simulations and found that 
the vertical viscosity coefficient (characterizing $\mathcal{R}_{\theta\varphi}$) in their models needed to be 650 times larger than
 the radial component (characterizing $\mathcal{R}_{r\varphi}$) in order to make the vertical structure of the accretion flow pattern in the two simulations agree. In passing, we note that we obtained a total gas accretion rate of $\sim -1.2 \cdot 10^{-5}\, \Sigma_{g0} \Omega_{K}$ for the dust-free 3D simulation by vertically integrating the mass flow 
%
%
 profile shown in the bottom left panel of Figure \ref{fig:flow}. This value seems consistent with the value $\sim -2.4 \cdot 10^{-5}\, \Sigma_{g0} \Omega_{K}$ reported by \citet{manger2018} who adopted a larger disc aspect ratio of $h=0.1$. However, a quantitative study of accretion rates requires carefully defined boundary conditions and will not be considered here.

\begin{figure*}[h!]
\centering
\includegraphics[width = 0.99 \textwidth]{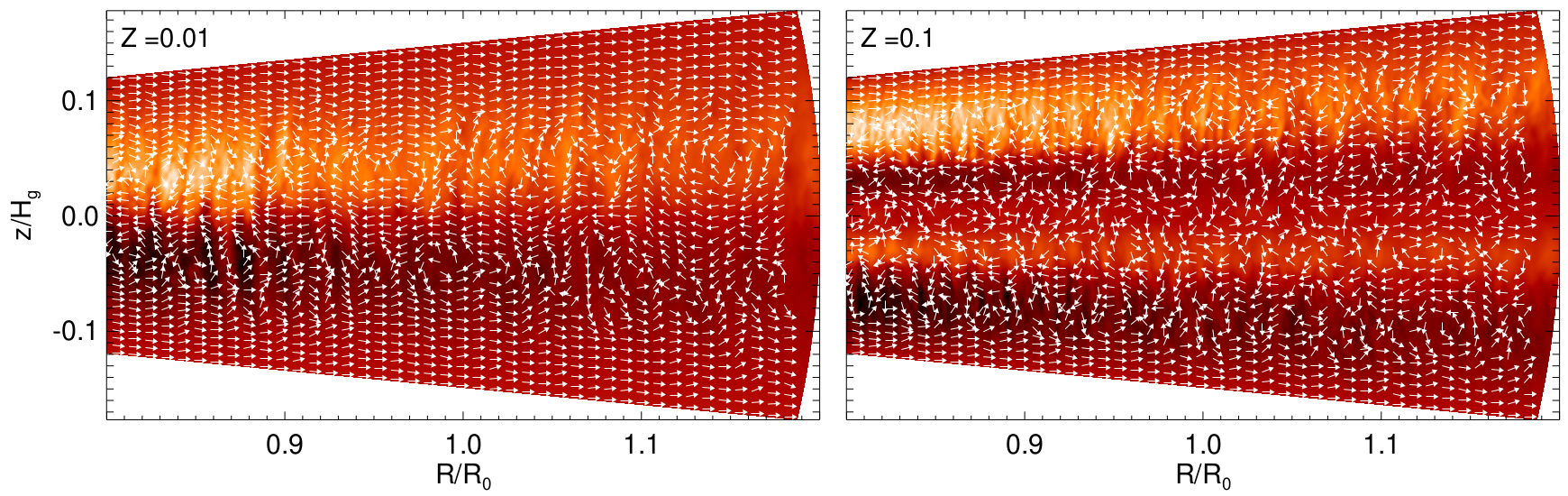}
\caption{Meridional plots of the time-averaged Reynolds stress $\langle\mathcal{R}_{\theta\varphi}\rangle_{t}$ over the last 400 orbits corresponding to the simulations with $Z=0.01$ and $Z=0.1$ with $\tau_{0}=10^{-3}$ in Figure \ref{fig:flow}. Bright/Dark colors represent positive/negative stress-values. The arrows indicate the direction of the corresponding time-averaged dust velocity. All arrows are normalized to have the same length. The different vertical profiles of $\langle\mathcal{R}_{\theta\varphi}\rangle_{t}$ are the reason for the different vertical velocity profiles seen in Figure \ref{fig:flow}.}
\label{fig:ffield}
\end{figure*}

Many aspects of the profiles plotted in Figure \ref{fig:flow} can be understood by considering basic aspects of dust-gas drag (Section \ref{sec:pbump}). Considering the 3D results for the time being, for small Stokes numbers $\tau_{0}=10^{-3}$ dust and gas averaged velocities are essentially equivalent, regardless of $Z$, which is also in agreement with results in Figure 17 of \citet{stoll2017}. For larger $\tau_{0}$, dust velocities become increasingly negative and gas velocities increasingly positive, which is also expected. Moreover, In the dense dusty mid-plane layer that forms in simulations with higher values of $\tau_{0}$ and $Z$, radial drift velocities become small as $\rho_{d} > \rho_{g}$. However, with increasing metallicity and Stokes numbers the gas and dust radial velocity profiles change substantially. For instance, for $Z=0.1$ and $\tau_{0}=10^{-3}$ (actually this is already observed for $Z=0.05$ which is not shown here) or for $Z=0.03$ and $\tau_{0}=5 \cdot 10^{-3}$ the profiles have changed such that material is flowing inward away from the mid-plane while it flows outward in narrow regions close to the mid-plane. Interestingly this flow profile resembles to some extent the standard laminar isotropic $\alpha$-viscosity case (e.g. \citet{takeuchi2002}). The reason for this change is due to different VSI modes being dominant in the different cases. This is illustrated for the examples $Z=0.01,0.1$ with $\tau_{0}=10^{-3}$ in Figure \ref{fig:ffield} and can also be seen in the vertical profiles of the Reynolds stress for these parameters in Figure \ref{fig:reynolds_comp}.
At low metallicity and for increasing Stokes number, we find that the radial mass fluxes qualitatively agree with those of \citet{stoll2016} (their Figure 18). Departures occur for
\begin{figure*}[h!]
\centering
\includegraphics[width = 0.99 \textwidth]{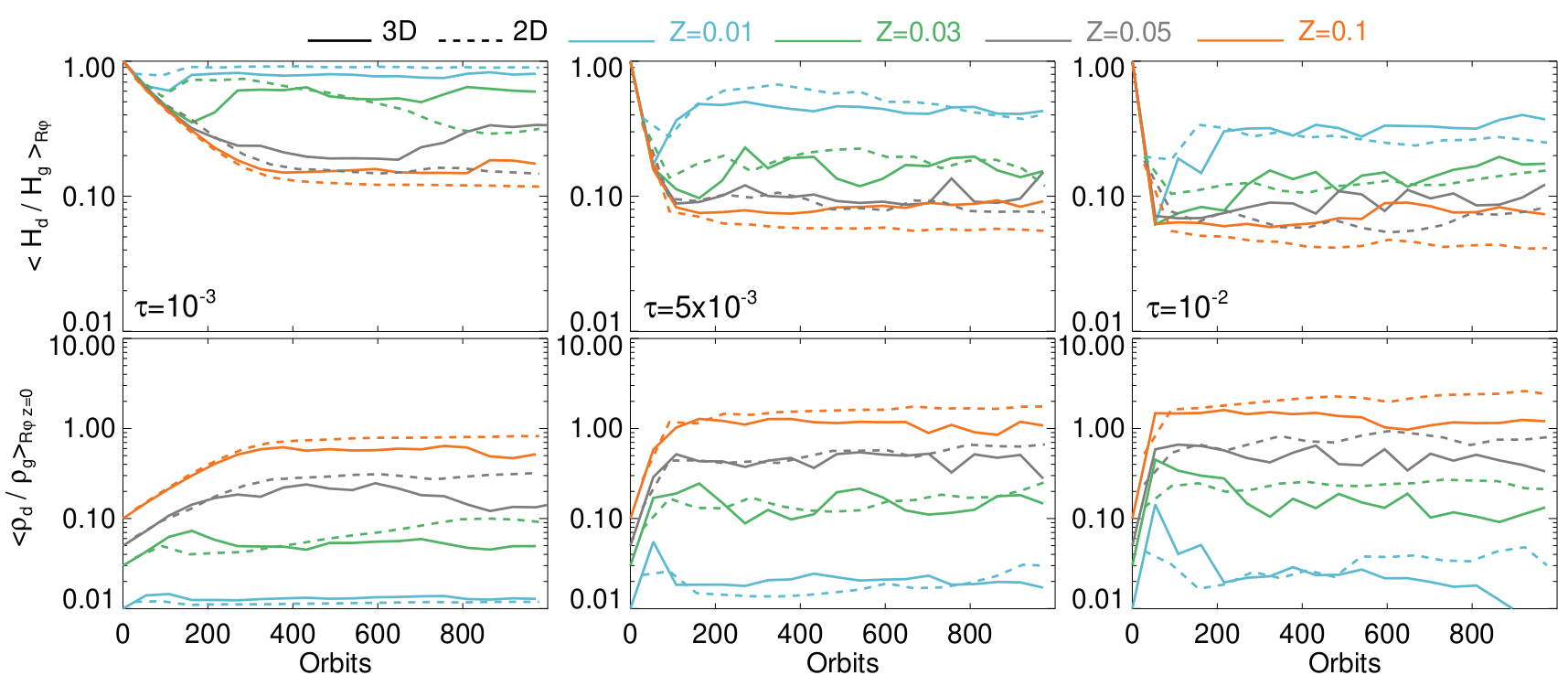}
\caption{Comparison of the dust settling process in 2D and 3D simulations for different metallicities $Z$ and Stokes numbers $\tau_{0}$. Shown are the ratios of the dust and gas scale heights (upper panels) and the mid-plane dust and gas densities. All quantities are averaged in radius and in 3D simulations also in $\varphi$.}
\label{fig:settling_comp}
\end{figure*}
%
%
the radial dust velocities at larger Stokes numbers where they found that the radial dust velocity away from the mid-plane is always directed outward and larger in magnitude than the radial gas velocities. In contrast, we find that the dust velocities become negative for larger Stokes numbers, as mentioned above. These discrepancies with \citet{stoll2016} are most likely due to the neglect of particle feedback in their simulations.  Moreover, they added particles at a time at which the VSI had reached a quasi steady state whereas we simulate both dust and gas coevally. However, the overall qualitative agreement between 2D and 3D simulations lends additional credibility to our results.
Comparing averaged velocities with averaged mass fluxes for the gas is generally straight forward. The gas is nearly incompressible such that its averaged mass flow is essentially the averaged gas velocity multiplied with the initial Gaussian density profile. This does not, however, apply to the dust whose mid-plane density can exhibit complex time variations owing to corrugation, and a detailed analysis of the dust mass flow profiles is beyond the scope of this paper.
Figure \ref{fig:settling_comp} compares the dust settling process in 2D and 3D simulations with metallicities $Z=0.01-0.1$ and Stokes numbers $\tau=10^{-3}-10^{-2}$. Shown are the dust-to-gas scale height ratio and the mid-plane dust-to-gas density ratio. Overall the results compare quite well. For both the 2D and 3D simulations averages are taken within a radial domain of $0.8 \leq R \leq 1.2$. For the 3D simulations an additional averaging over azimuth has been performed. Apart from the smallest metallicity and Stokes number cases, dust settling is somewhat stronger in 2D simulations, which is expected from the results in Figure \ref{fig:reynolds_comp}. All simulations shown were run with the same radial and vertical resolution and domain sizes. For the 3D simulation with $Z=0.01$ and $\tau=10^{-2}$ the disc region under consideration already becomes significantly depleted of dust by the end of the simulation. This is in part due to vortices that form and collect inward drifting dust and eventually leave the considered disc region. This will be discussed in more detail in Section \ref{sec:vortices}.
Furthermore, Figure \ref{fig:alpha_comp} illustrates the effect of dust on the VSI's ability to transport angular momentum in radial direction
\begin{figure*}[h!]
\centering
\includegraphics[width = 0.99 \textwidth]{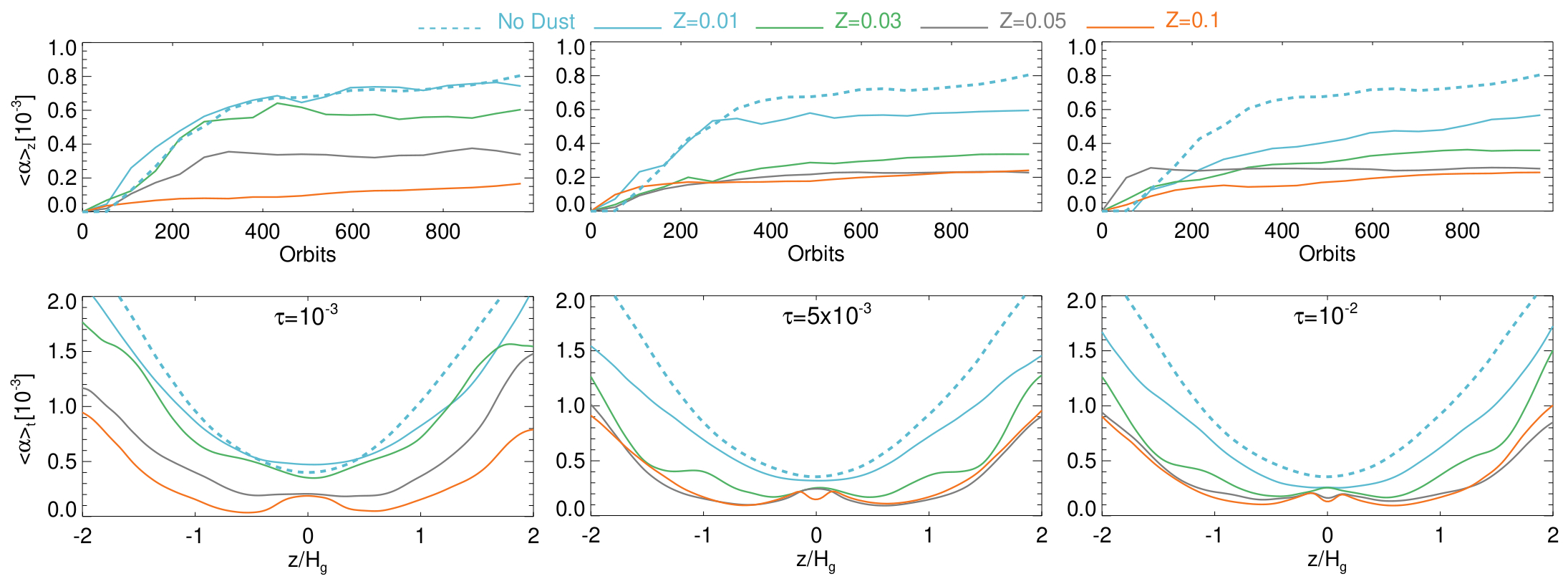}
\caption{Plots of the $\alpha$-viscosity parameter (\ref{eq:alpha}) as measured in 3D simulations for different metallicities and Stokes numbers. The dotted curves correspond to a dust-free simulation. The upper panels display the running time-average of the spatially averaged value while the lower panels show vertical profiles, averaged in time.}
\label{fig:alpha_comp}
\end{figure*}
%
%
in 3D simulations.
Shown is the time evolution, which again are a running time-average, as well as time-averaged vertical profiles of the $\alpha$-viscosity parameter (\ref{eq:alpha}). The averages are taken in the same fashion as in Figure \ref{fig:reynolds_comp}.
 In contrast to the behavior of $\mathcal{R}_{\theta\varphi}$ (Figure \ref{fig:reynolds_comp}) which describes vertical angular momentum transport, $\alpha$ decreases steadily with increasing metallicity, even at low metallicity. The reason is that the main driver of an increased $\mathcal{R}_{\theta\varphi}$ is corrugation of the dust-layer which can only occur in vertical direction. 
The values $\alpha \lesssim 10^{-3}$ extracted from our dust-free reference simulation are in good agreement with those of \citet{nelson2013}, but almost a factor 10 larger than those reported by \citet{manger2020} for the same disc aspect ratio $h=0.05$. Several other studies report smaller values \citep{stoll2016,pfeil2020,flock2020}, mainly due to adopting a more realistic equation of state, improved by at least a finite cooling time (compared to the isothermal approximation) which is known to mitigate the VSI \citep{lin2015}. The most likely reason for the difference to the values of \citet{manger2020} is the higher radial and vertical grid resolution adopted in our simulations. Furthermore, \citet{manger2020} found that the resolution in azimuthal direction has a subdominant effect on the VSI activity, in contrary to the strong dependence on the azimuthal domain size \citep{manger2018}.

\subsection{Dust Rings}\label{sec:dust rings}

One class of prominent substructures in PPDs that ALMA has revealed in recent years are rings and gaps, visible in the dust's 
mm continuum or spectral line emission (e.g. \citet{vdmarel2013,alma2015,andrews2020,vdmarel2021}). 
A fraction of the well-observed dust rings exhibit more or less strong (typically crescent shaped) non-axisymmetries whose origin is still topic of debate.
One generally assumes that there are two main mechanisms which can generate these structures. These are so called `horseshoes', which constitute non-axisymmetric gas enhancements in discs just outside the inner cavity which has been cleared by a stellar binary \citep{ragusa2017}, or vortices generated at a pressure bump adjacent to a gap which is assumed to be created by a planet of sufficiently high mass (e.g. \citet{zhu2014}). 
The latter mechanism is the one relevant to this work and here we wish to investigate if and under which conditions dust rings that form at a pressure bump in an VSI turbulent disc can attain dust to 
gas ratios that could result in planetesimal formation and if dust rings can be subjected to any persistent non-axisymmetries.

\begin{figure*}[h!]
\centering
\includegraphics[width = 0.75 \textwidth]{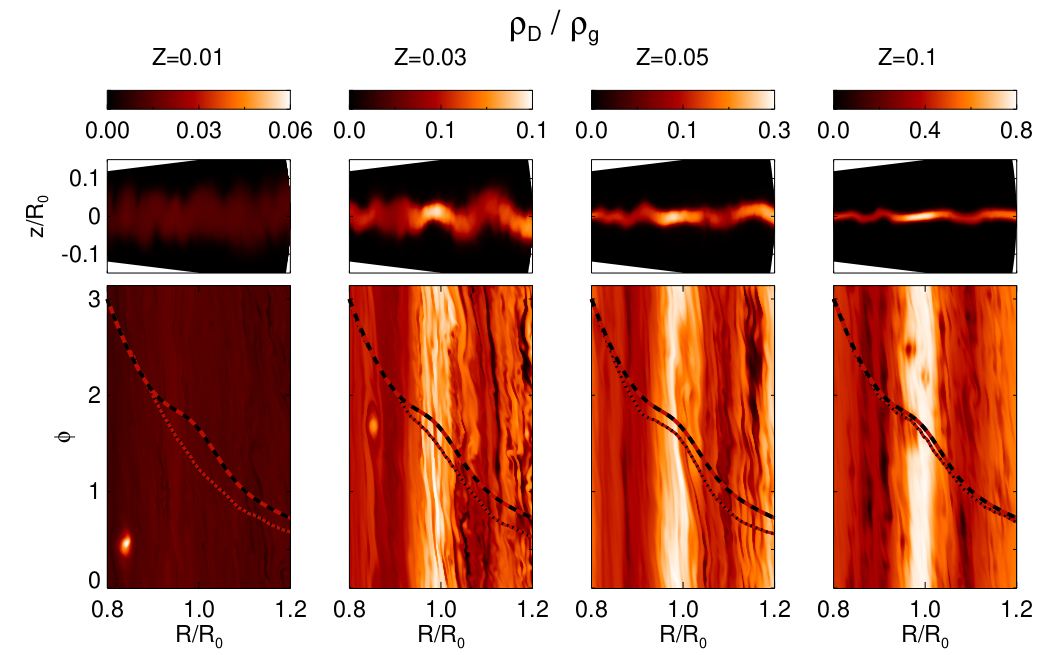}
\caption{Snap shots of dust-to-gas density ratio at late times $800-1000$ orbits in simulations including a pressure bump with amplitude $A=0.2$ and with Stokes number $\tau_{0}= 10^{-3}$. The dashed / dotted curves represent the initial / final mid-plane gas pressure profile.}
\label{fig:dust ring_1dm3}
\end{figure*}

\begin{figure*}[h!]
\centering
\includegraphics[width = 0.75 \textwidth]{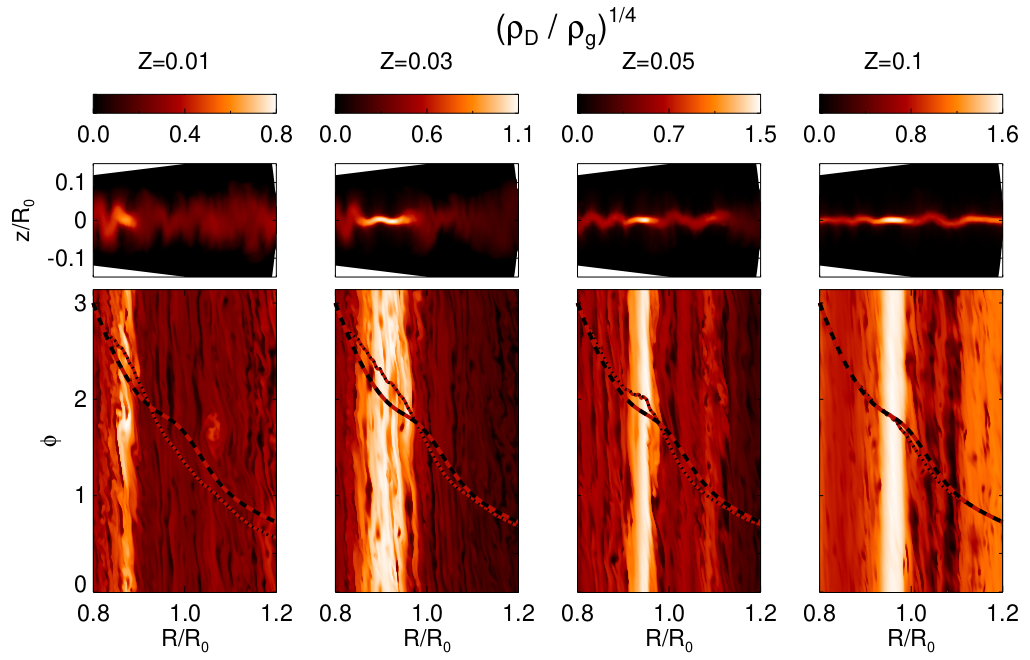}
\caption{Same as Figure \ref{fig:dust ring_1dm3} except that $\tau_{0}= 10^{-2}$.}
\label{fig:dust ring_1dm2}
\end{figure*}

\begin{figure*}[h!]
\centering
\includegraphics[width = 0.75 \textwidth]{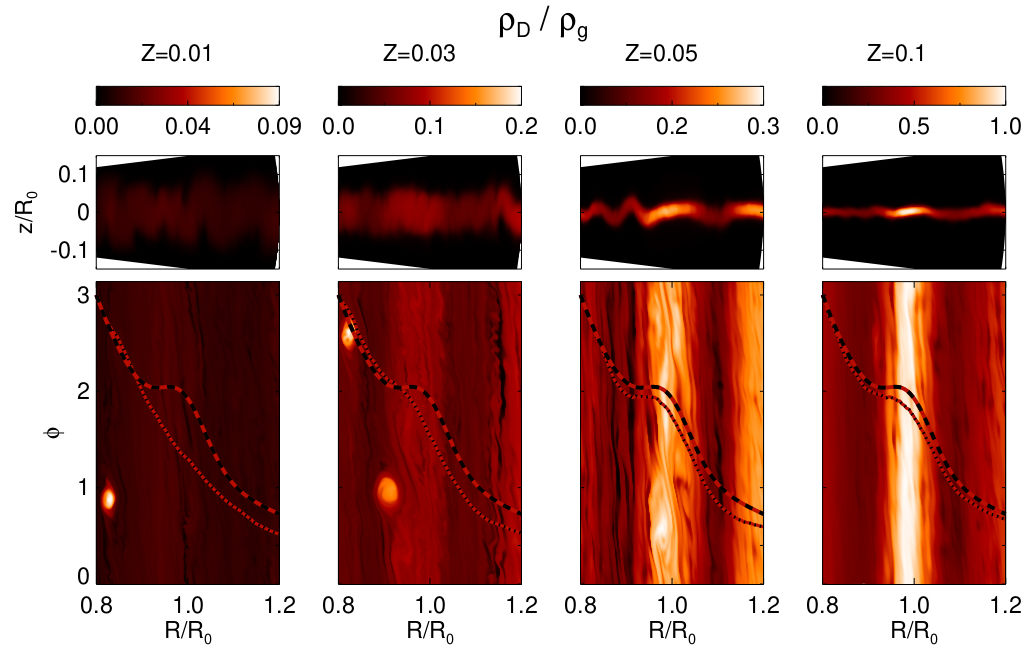}
\caption{Same as Figure \ref{fig:dust ring_1dm3} except that $A=0.4$.}
\label{fig:dust ring_1dm3_amp04}
\end{figure*}

\begin{figure*}[h!]
\centering
\includegraphics[width = 0.75 \textwidth]{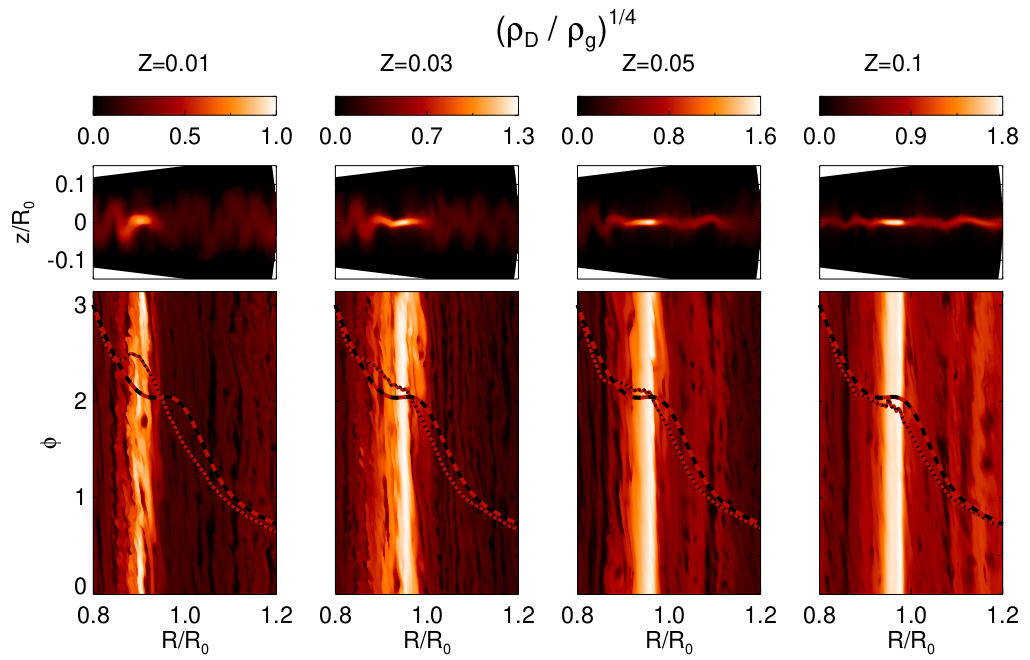}
\caption{Same as Figure \ref{fig:dust ring_1dm3} except that $\tau_{0}= 10^{-2}$ and $A=0.4$.}
\label{fig:dust ring_1dm2_amp04}
\end{figure*}

\begin{figure}[h!]
\centering
\includegraphics[width = 0.49 \textwidth]{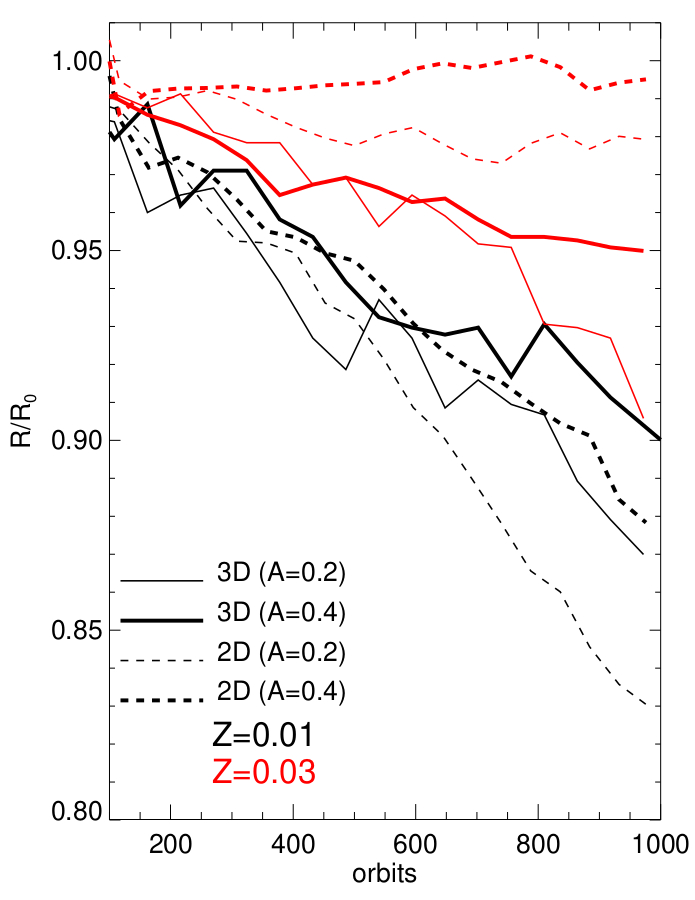}
\caption{Radial locations of dust rings in 3D (solid curves) and 2D (dashed curves) simulations for metallicities $Z=0.01$ (black curves) and $Z=0.03$ (red curves) and bump amplitudes $A=0.2$ (thin curves) and $A=0.4$ (thick curves). The locations correspond to the maximum value of $\epsilon$ at each time. The sporadic reversals of the drift direction are due to turbulent fluctuations of the dust density within the rings. The 3D simulations are the same as in Figures \ref{fig:dust ring_1dm2} and \ref{fig:dust ring_1dm2_amp04}. The 2D simulations are those shown in Figure \ref{fig:pgrid}.}
\label{fig:drift_rings}
\end{figure}

Figures \ref{fig:dust ring_1dm3}-\ref{fig:dust ring_1dm2_amp04} show dust rings that formed in 3D simulations in response to an initially seeded pressure bump, for different background metallicities $Z$, Stokes numbers ($\tau_{0}=0.001$, $\tau_{0}=0.005$ and $\tau_{0}=0.01$) and bump amplitudes ($A=0.2$ and $A=0.4$), respectively. The dotted and dashed curves illustrate the azimuthally averaged final and initial gas pressure profile, respectively.
We verified by means of test simulations that the pressure bumps in these simulations are not subjected to the RWI\footnote{In these test simulations we adopted a very small temperature gradient $q \to 0$ such that the VSI was extinguished.}. It should be kept in mind that our resolution is not sufficient to resolve small-scale instabilities such as the SI, which may disrupt the dust rings, although the SI might be less efficient in a turbulent disc.
That being said, what Figures \ref{fig:dust ring_1dm3}-\ref{fig:dust ring_1dm2_amp04} reveal is that an increase of the parameters $Z$ and $\tau_{0}$ (and to some extent also $A$) results in a transition from a preference to form dusty vortices toward a preference to form dusty rings. 
  Generally speaking, we find an increasing axisymmetry of formed dust enhancements when increasing these parameters. We hypothesize that the reason for this transition is the same as that for the increased stability of dust rings found in our 2D simulations when increasing the same parameters, since essentially this weakens the VSI.
When comparing the results presented in this section with those of our 2D simulations which are summarized in Figure \ref{fig:pgrid}, we find many similarities. For instance, the cases with the smallest Stokes number $\tau_{0}=10^{-3}$ and lower metallicity $Z\lesssim 0.03$  are lacking any notable dust enhancements at the pressure bump site, although we observe the formation of weak dusty vortices in 3D.
Furthermore, in Section \ref{sec:pbumpres} we discussed the occurrence of a dusty gas instability and how it results in disruption and/or inward drift of dust rings. The parameter regime within which this happens can be directly determined from Figure \ref{fig:pgrid}. Indeed, Figures \ref{fig:dust ring_1dm3}-\ref{fig:dust ring_1dm2_amp04} show that also in 3D simulations dust rings undergo inward drift and the magnitude of this drift is larger for 
%
%
larger/smaller Stokes number/metallicity. For illustration, Figure \ref{fig:drift_rings} 
compares the radial locations of dust rings as observed in 2D and 3D simulations for two metallicities ($Z=0.01,0.03$) and bump amplitudes ($A=0.2,0.4$) and Stokes number $\tau_{0}=10^{-2}$. For the case $Z=0.01$ dust rings in 2D simulations (dashed curves) drift slightly faster than in 3D simulations (solid curves), whereas the opposite is the case for $Z=0.03$. 
The behavior for $Z=0.01$ can be explained considering that the dusty gas instability which causes the inward drift is somewhat stronger in 2D than in 3D (cf. Figure \ref{fig:reynolds_comp}). On the other hand, the negligible inward drift in the 2D simulations with $Z=0.03$ is a consequence of the much larger value of $\epsilon$ attained in the corresponding dust rings (cf. Figure \ref{fig:pgrid}). For the same reason (i.e. increased $\epsilon$) drift speeds for the cases with $A=0.4$ are generally smaller than for the corresponding cases with  $A=0.2$. That being said, drift speeds of dust rings according to Figure \ref{fig:drift_rings} can get as fast as $\sim 0.2 H_{g0}$/(100 orbits) in 3D simulations and $\sim 0.3 H_{g0}$/(100 orbits) in 2D simulations.

Moreover, the dust rings formed in simulations with $Z\leq 0.03$ are subject to strong turbulent distortions, whereas those 
%
%
corresponding to larger $Z$ are nearly axisymmetric. In the former
cases the gas pressure bump is dragged along with the dust ring and in some cases even amplified, similar as in corresponding 2D simulations (Figure \ref{fig:sptd_zcomp}).
Also one can clearly see strong corrugation of the dust-layer adjacent to the ring for these cases in the upper panels which show meridional contour plots. This corrugation facilitates the dusty gas instability discussed in Section \ref{sec:instab}.

\begin{figure}[h!]
\centering
\includegraphics[width = 0.49 \textwidth]{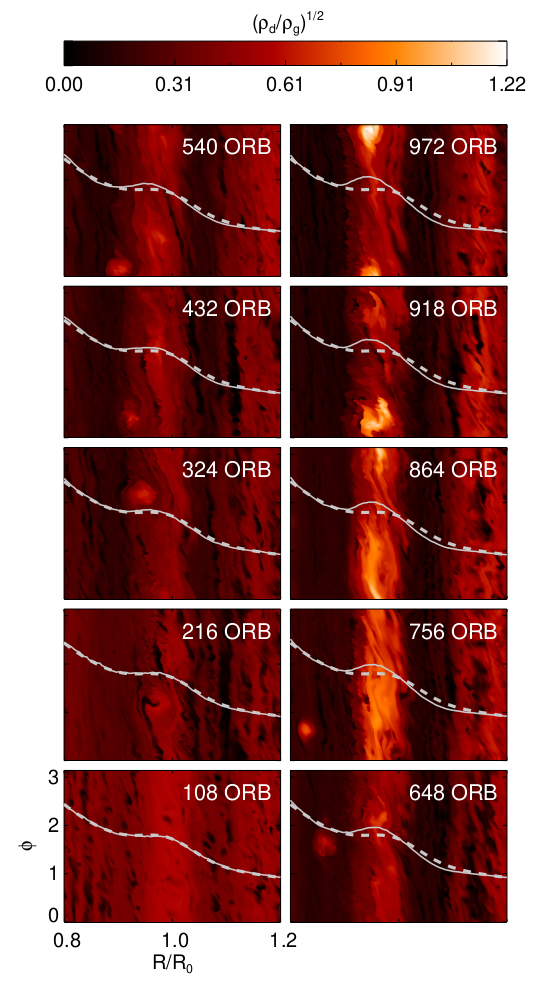}
\caption{Time evolution of an unstable dust ring in a 3D simulation with parameters $\tau_{0}= 4 \cdot 10^{-3}, Z=0.03$ and $A=0.4$. The dashed and solid curves are the initial and current mid-plane gas pressure profile, respectively. One can clearly see an amplification of the gas pressure bump and unstable behavior of the dust ring, as it develops strong non-axisymmetries and splits of a vortex which subsequently migrates inward.}
\label{fig:dustring_us}
\end{figure}

One difference is that in 3D simulations instability leads to vortex formation, which is suppressed in 2D axisymmetric simulations. An example for the evolution of an unstable dust ring in a 3D simulation is presented in Figure \ref{fig:dustring_us}. In this case the dust ring continuously produces new vortices that split off and drift inward. The driving force of this unstable behaviour is, similarly to the 2D simulations, the VSI.
The attained dust-to-gas density
ratios in 3D dust rings are not nearly as high as in 2D, which is due to planar turbulent diffusion generated by the VSI and quantified through $\alpha$ (as discussed in Section \ref{sec:turbulence}) in 3D simulations.

\section{Results of 3D Simulations: Vortices}\label{sec:vortices}

In this section we discuss the appearance of vortices in 3D simulations and their capability to concentrate dust and hence assist in planetesimal formation.

\subsection{Dust-free Simulations}\label{sec:vort_dustfree}
\begin{figure}[h!]
\centering
\includegraphics[width = 0.5 \textwidth]{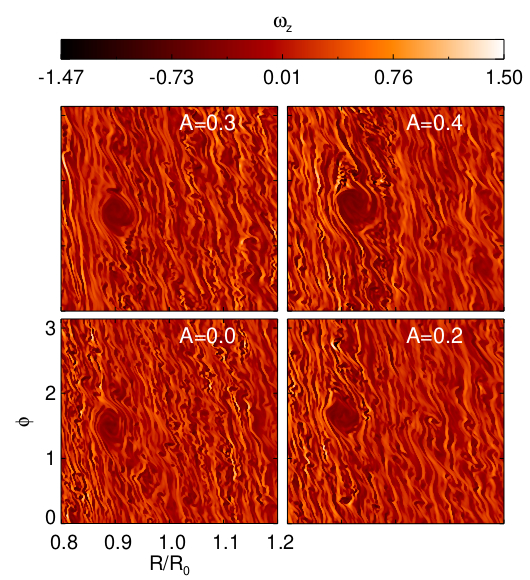}
\caption{mid-plane plots of the $z$-component of the gas vorticity (\ref{eq:vort}) after 540 orbits in dust-free simulations with an initial pressure bump of varying amplitude $A$.}
\label{fig:vort_dustfree}
\end{figure}

Figure \ref{fig:vort_dustfree} shows the mid-plane vorticity (\ref{eq:vort}) after around 600 reference orbits in 3D dust-free simulations with an initially seeded pressure bump of varying amplitude $A$.
The lower left panel corresponds to the run presented in the lower left frame of Figure 8 in \citet{manger2020} and we find good agreement between the sizes $\Delta R \sim H_{g0}$ and aspect ratios $R\Delta \varphi /\Delta R\sim 8-10$ of the largest vortices appearing in these simulations. Moreover, the size of these vortices appears to increase with increasing bump amplitude $A$. They appear after $\sim 200-400$ orbits, where earlier times correspond to larger values of $A$. 
Numerous small-scale vortices are present as well. These are short-lived and cannot be tracked in our simulations since our time resolution is $\sim 50$ orbits. In previous works \citep{richard2016,manger2018} similarly small vortices were found and their survival times were estimated to be several orbits. 
 Although it is not entirely clear if the correspondence between vortex size and bump amplitude is physical or a coincidental outcome of our simulations, a possible explanation is that large vortices grow by absorption of small-scale vortices and that this can happen more efficiently in the vicinity of a pressure bump. \citet{paardekooper2010} have shown that in razor-thin disc models, vortices undergo migration by emitting sound waves inwards and outwards in an asymmetrical fashion, similar (but not identical) to planets embedded in a disc. The direction of migration is related to the background surface density gradient, such that vortices migrate toward regions of larger surface density. This should also imply migration towards the mild pressure maximum corresponding to a large vortex. However, since the small-scale vortices have a small vertical extent of considerably less than one scaleheight (\citet{richard2016}) the vertically-averaged results of \citet{paardekooper2010} potentially do not apply to these vortices. It is therefore not clear whether or not they can migrate over any significant distance and be absorbed by a large vortex before they dissipate. On the other hand, migration and also merging of larger vortices is directly observed in our simulations, similar to what was reported by \citet{manger2020}.

\begin{figure}[h!]
\centering
\includegraphics[width = 0.35 \textwidth]{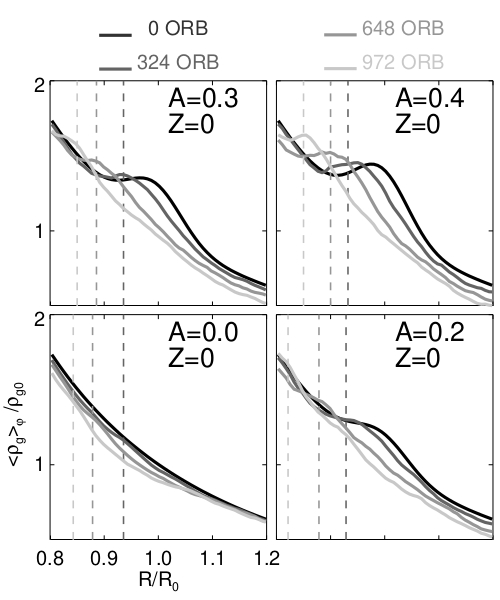}
\caption{Evolution of the mid-plane gas density for dust-free simulations with an initial pressure bump of varying amplitude $A$. The dashed lines indicate the location of the large vortex in these simulations at the three latest times.}
\label{fig:pres_evol}
\end{figure}

 Figure \ref{fig:pres_evol} shows the time evolution of the radial gas density profile (azimuthally averaged) at the mid-plane for the same simulations as in Figure \ref{fig:vort_dustfree}. The dashed vertical lines indicate the radial location of the large vortex for the three latest times. In all simulations the vortex survives for hundreds of orbits and migrates inwards. This is in agreement with the results reported by \citet{manger2018}, \citet{manger2020} and also \citet{pfeil2020}. In addition, the density bump smears out through turbulent diffusion and redistribution of mass within the disc in response to the (turbulent) angular momentum transport generated by the VSI, an aspect recently considered by \citet{manger2021}.
 Indeed, we find a pileup of mass around $R\sim 0.6R_0$ (outside the diagnostic domain) in a similar manner as shown in Figure 9 (bottom left panel) of \citet{manger2021}.

 Apart from the size of the vortices, the results in Figure \ref{fig:pres_evol} do not provide evidence that the pressure bump has an effect on migration of large vortices, or that it is a preferred location for the formation of the latter, since also the vortex in the simulation with $A=0$ forms at a very similar radius as in all other simulations with $A>0$. Nevertheless, we hypothesize that large vortices result from merging of smaller vortices and that this is more likely to occur if vortex radial migration is mitigated. The latter is expected to depend on the surface density gradient (\citep{paardekooper2010}) and we therefore expect a more efficient growth of vortices at a pressure bump.
 To further test this hypothesis we ran additional simulations with different surface density slopes $s$ (and hence different mid-plane gas density slopes $p$). The results of these simulations, which are presented in Appendix \ref{sec:pvar}, show that a flatter mid-plane density profile leads to slower migration speeds and larger vortices, supporting our hypothesis. 


\subsection{Dusty Simulations without a Pressure Bump}\label{sec:nopbdust}

\begin{figure*}[h!]
\centering
\includegraphics[width = 0.99 \textwidth]{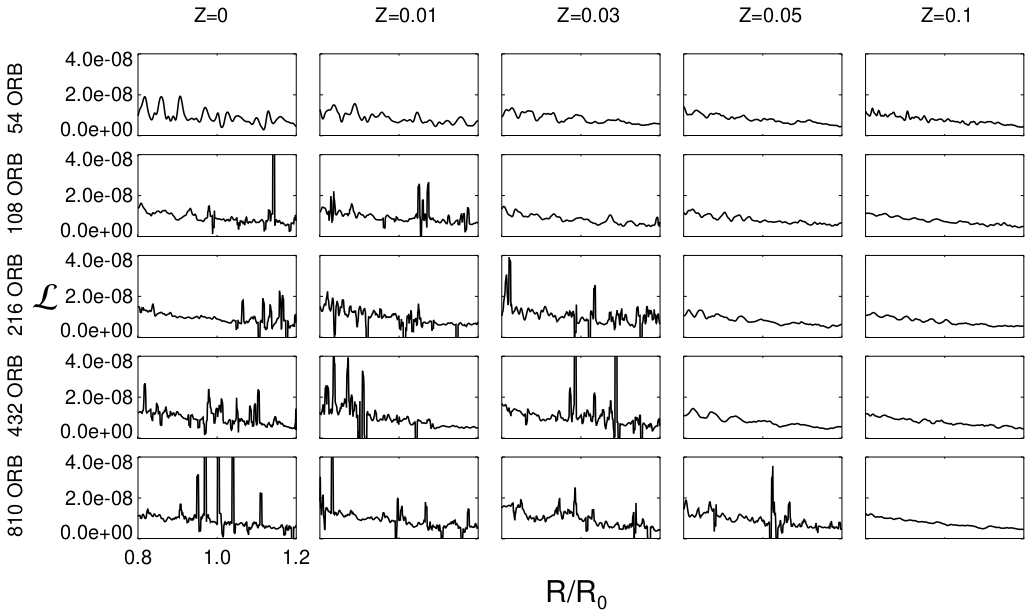}
\caption{Radial profiles (azimuthally averaged) of the vortensity (\ref{eq:vorten}) at different times in simulations with Stokes number $\tau_{0}=10^{-3}$ and different metallicities $Z$. The occurrence of sharp peaks in this quantity suggests that the smallest scale structures appearing in our simulations are most likely not sufficiently resolved in order to properly compute numerical derivatives contained in the quantity $\mathcal{L}$ for all grid points. This should however not affect the qualitative interpretation of the plots.}
\label{fig:vortensity}
\end{figure*}

 Our dusty 3D simulations reveal that large ($\Delta R \sim H_{g0}$) vortices collect dust and survive for typically hundreds of orbital periods, which is similar to the survival time of pure gas vortices discussed above. 
Within a simulation time of 1000 reference orbits, we find large, long-lived vortices only for metallicities $Z\lesssim 0.03$ and if $Z=0.03$ only for small Stokes numbers $\tau_{0} \sim 10^{-3}$. We generally do not observe larger, long-lived vortices in simulations with $Z \geq 0.05$. The exception is in a simulation with $Z=0.05$ and $\tau_{0}=10^{-3}$, where at late simulation times past 1000 orbits we find the formation of a large vortex, but its dust content is negligible compared to ambient turbulent dust lanes that are present in the simulation region as well. 
We attribute the absence of long-lived vortices at high metallicities and Stokes numbers to the corresponding weakening of the VSI by dust feedback and its ability to form vortices. Therefore we also expect weaker vortensity perturbations in simulations with larger $Z$, which is illustrated in Figure \ref{fig:vortensity}, where the vortensity $\mathcal{L}$ is given by (\ref{eq:vort}).

None of the vortices that formed in our simulations without an initial pressure bump attained dust-to-gas ratios of unity or larger. The largest value we find is $\epsilon \sim 0.7$ in a vortex by the end of the simulation with $Z=0.01$ and $\tau_{0} = 10^{-2}$.
This is in strong contrast to the values of $\epsilon$ attained in vortices that formed in the 3D shearing box simulations of the COS by \citet{raettig2021}, where even for an initial value $Z\sim 10^{-4}$ vortices formed rapidly and after tens of orbits values of $\epsilon\sim 10$ were reached within the vortices. The main reason for these differences is certainly the much weaker vertical turbulent velocities generated by the COS ($\sim 10-100$ times smaller than for the VSI) and perhaps to some extent the larger Stokes numbers $\tau_{0} \geq 5\cdot 10^{-2}$ considered in that work. 
That being said, we cannot rule out that larger dust-to-gas ratios could be obtained with Stokes numbers $>10^{-2}$, a larger radial domain, or both; the latter of which would supply more dust for trapping. (An inflow outer boundary condition would have the same effect.) Because of dust drift, we already find significant dust depletion in the diagnostic domain $0.8 \leq R \leq 1.2$ by the end of the aforementioned simulation with $\tau_{0} = 10^{-2}$, such that the collection of dust in the vortex is already hampered. This effect will be even more severe in simulations with larger Stokes numbers such that these would either require a substantially larger radial domain size or inflow of dust at the outer boundary.

\subsection{Dusty Simulations including a Pressure Bump}\label{sec:pbdust}

\begin{figure}[h!]
\centering
\includegraphics[width = 0.35 \textwidth]{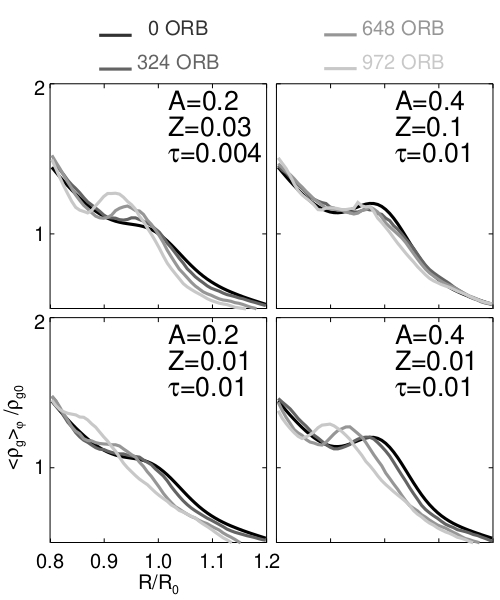}
\caption{Evolution of the mid-plane gas density in dusty simulations with different metallicity $Z$, Stokes number $\tau_{0}$ and pressure bump amplitude $A$.}
\label{fig:pres_evol_dust}
\end{figure}

 In the presence of dust the evolution of a pressure bump can differ significantly from the dust-free evolution discussed in Section \ref{sec:vort_dustfree}, which is illustrated in Figure \ref{fig:pres_evol_dust}. In Section \ref{sec:dust rings}, we had seen that pressure bumps give rise to the formation of dust rings, vortices, or a combination of both.  
 While pressure bumps are expected to either trap dust or cause `traffic jams' to produce dust rings (see \S\ref{sec:pbump}), the frequent formation of vortices at the pressure bump is less obvious based on above dust-free results in Section \ref{sec:vort_dustfree}. (Recall that our initial pressure bumps are stable against the vortex-forming RWI.) However, in Section \ref{sec:instab} we found that dust rings at low and intermediate metallicities are locations of increased hydrodynamic activity due to violations of one of the Solberg-H\o iland criteria [Eqs. (\ref{eq:sh1})-(\ref{eq:sh2})], which are associated with corrugations of the dust-layer by the VSI. In 3D, we can therefore expect such dust rings to be preferred locations for the formation of vortices, for a given range of dust-parameters. Similar to 2D simulations shown in Figure \ref{fig:sptd_zcomp}, such unstable dust rings also go along with an increased gas density such that the pressure bump is effectively being amplified, which is clearly seen in the lower right and upper left panel of Figure \ref{fig:pres_evol_dust}. 
 Within this scenario, it is also consistent that no vortices form at dust rings in simulations with sufficiently large $Z\gtrsim 0.05$, due to effective weakening of the VSI by dusty buoyancy. This can clearly be seen by comparing Figure \ref{fig:instab} with Figure \ref{fig:dust ring_1dm3_amp04}.

\begin{figure}[h!]
\centering
\includegraphics[width = 0.49 \textwidth]{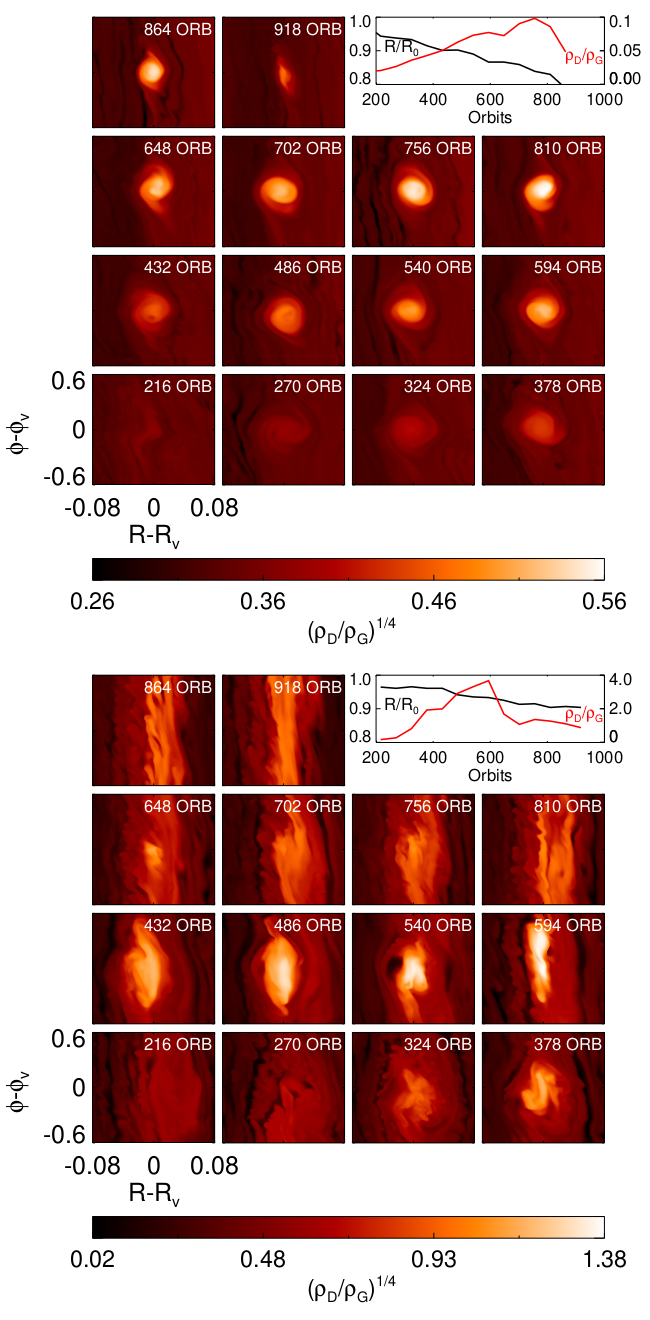}
\caption{Example evolution of large dusty vortices in 3D simulations with metallicity $Z=0.01$ and Stokes numbers $\tau_{0}=10^{-3}$ (upper panels) and $\tau_{0}=10^{-2}$ (lower panels). The contour plots show the dust-to-gas ratio. The insert plot displays the vortex location (black curve) and its maximum dust-to-gas ratio (red curve) as functions of time.}
\label{fig:vort_evol}
\end{figure}

\begin{figure}[h!]
\centering
\includegraphics[width = 0.49 \textwidth]{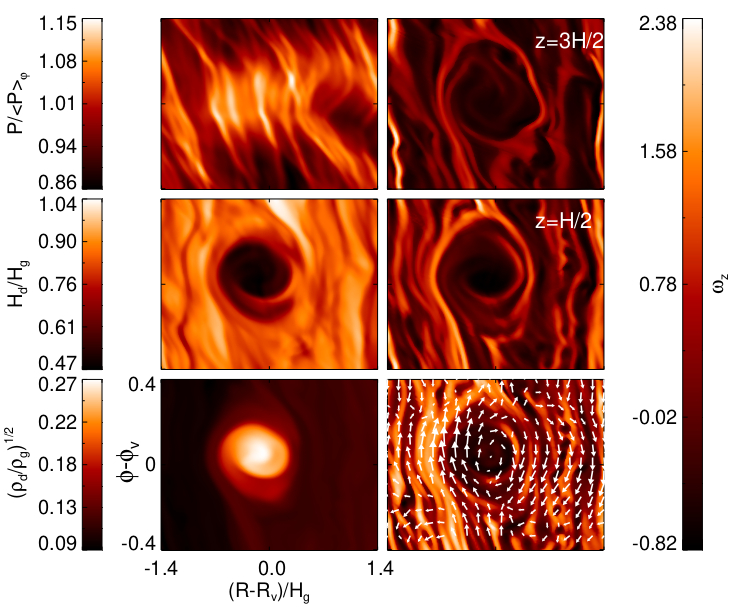}
\caption{Detailed view of the large dusty vortex corresponding to the upper panels of Figure \ref{fig:vort_evol}. Shown are snapshots of the dust-to-gas ratio (lower left panel), the dust-to-gas scaleheight ratio (middle left panel) and the mid-plane gas pressure scaled with its azimuthally averaged value at each radius. The right panels show the dust vorticity (\ref{eq:vort}) at different heights ($z=0$ in the lower right panel). The arrows in the lower right panel illustrate the dust velocity in the frame co-moving with the Keplerian shear and where in addition the spatial average of the azimuthal velocity over the displayed region has been subtracted to suppress the contribution of the vortex bulk motion.}
\label{fig:vort_details}
\end{figure}

 Examples of dusty vortices that form at a pressure bump are found in Figures \ref{fig:dust ring_1dm3_amp04}, \ref{fig:dustring_us},  and \ref{fig:vort_details}. The latter figure also demonstrates that dust is more settled in the vortex core, which is clearly seen in the contour plot of scale height ratio $H_{d}/H_{g}$.
 Unlike bump-free simulations, for moderate initial pressure bumps ($A=0.2-0.4$) we do find vortices that collect sufficient amounts of dust such that the SI should be triggered within these, would it be resolved. 
Figure \ref{fig:vort_evol} illustrates the evolution of large, long-lived vortices that form at a pressure bump with an amplitude $A=0.4$\footnote{A similar result is obtained with $A=0.2$.} in simulations with metallicity $Z=0.01$ and Stokes numbers $\tau_{0}=10^{-3}$ and $\tau_{0}=10^{-2}$, respectively. These vortices survive a considerable time span of about 500 orbits. For $\tau_{0}=10^{-3}$ the vortex core does not attain unity dust-to-gas ratios within the simulation timescale and steadily migrates inward with a rate $\sim 0.4 H_{g0}$/(100 orbits), comparable to that of dust-free vortices (cf. Appendix \ref{sec:pvar}), until it eventually dissolves in the background flow. 

On the other hand, the vortex formed in the simulation with $\tau_{0}=10^{-2}$ reaches unity dust-to-gas ratio at about 200 orbits following its formation and a maximum value of almost 4 after 400 orbits. Therefore it can be expected that the SI would be triggered with higher grid resolutions. Due to the larger dust-to-gas ratio the inward drift of the vortex is substantially slower in this case. Furthermore, the shape of the dust distribution within the vortex is more elongated in the case of larger Stokes number (or equivalently, larger particle size). By the end of the simulation the vortex dissolves into a turbulent, non-axisymmetric circumferential dust ring. A similar trend in the distribution of dust particles of different sizes in vortices was reported by \citet{rubsamen2015} and \citet{surville2019} for vortices in 2D shearing sheet simulations.

\subsection{Discussion}

The small-scale, compact vortices found in our simulations and those of \citet{richard2016} and \citet{manger2020} are assumed to be the result of parasitic KHI that inflict the VSI modes, thereby controlling their nonlinear saturation amplitudes \citep{latter2018}. The rapid dissipation of these vortices is likely caused by the elliptic instability \citep{lesur2009,railton14}, which is more vigorous for small vortex aspect-ratios. Larger vortices with aspect ratios of around 5 or larger do not fit in the scenario just described. First of all, their properties are different from the vortices that were described in \citet{latter2018}. For instance, the large scale vortices found here are vertically extended structures where the vorticity perturbation extends over more than one gas scale height, which is illustrated in Figure \ref{fig:vort_details}.  Moreover, these elongated vortices are likely less affected by the elliptic instability in our simulations since for such vortices this instability grows slowly and requires high radial resolution.

\citet{richard2016} found that long-lived vortices with larger aspect ratios can form in their simulations if the disc possesses relatively long cooling times $\gtrsim 0.5 \Omega^{-1}$ and steep temperature profiles with $q = 2$. %
They argued that for such disc-parameters the SBI \citep{peterson07a,peterson07b,lesur2010} can amplify the vortices, whereas the SBI is inactive for shorter cooling times (such as our locally isothermal discs), shallower temperature profiles, or both. However, since the VSI most likely appears in disc regions that are not expected to fulfill the criteria for the SBI (although the simulations of \citet{pfeil2020} suggest that the VSI can exist in the inner, optically thick part of a PPD, where also the SBI is expected to be active) \citet{richard2016} concluded that the vortices resulting from the VSI are unlikely to promote planetesimal formation on global scales. It was later shown by \citet{manger2018} that the VSI can generate large, long-lived vortices in an isothermal disc and that the reason for the absence of such vortices in the simulations of \citet{richard2016} was a too small azimuthal simulation domain.

Furthermore, it is unlikely that the mechanism which eventually destroys the large-scale vortices in our simulations is the same as that in previous 2D simulations of dusty vortices  \citep{fu2014,rubsamen2015,raettig2015,surville2019} , which might be the heavy core instability \citep{chang2010}. The results presented in Figure \ref{fig:vort_evol} rather suggest that the lifetime of the large scale vortices in our simulations is independent of the amount of dust concentrated within them.
It can be expected that whatever instability attacked vortices in the aforementioned high resolution shearing sheet simulations is not resolved here. Since the vortices do disappear on time scales that are in agreement with the lifetimes of dust-free vortices as discussed in Section \ref{sec:vort_dustfree}, this suggests that the mechanism of destruction is also the same. Most likely this is a combination of turbulent diffusion and Keplerian shear that disrupts large vortices in our simulations. The influence of the elliptical instability \citep{lesur2009}, the parasitic KHI \citep{latter2018}, or the VSI itself in the vortex core must be clarified in future high resolution 3D simulations.

\section{Summary and Conclusion}\label{sec:summary}

We studied the effect of a pressure bump on the evolution of dust in a PPD with fully developed turbulence generated by the VSI. We conducted global 2D axisymmetric and 3D, locally isothermal simulations of gas and dust, modelling the outer parts ($R\gtrsim 10$ au) of PPDs that are heated by stellar irradiation. Dust particles were treated as a pressureless fluid and its back reaction onto gas through drag was taken into account.
In particular, we investigated the influence of the particle's Stokes number $\tau_{0}$, the disc's initial dust abundance $Z=\Sigma_{d}/\Sigma_{g}$, and the amplitude $A$ of the pressure bump on dust accumulation at the bump site. In our models $A$ corresponds to the fractional increase in the gas surface density over a bump width of two gas scale-heights compared to the ambient background. The range of considered Stokes numbers $\tau_{0}=10^{-3}-10^{-2}$ corresponds to particle sizes from $\sim 280 \, \mathrm{\mu m}$ at 10 au to $\sim 9\, \mathrm{\mu m}$ at 100 au for $\tau_{0}=10^{-3}$, or between $\sim 3\,\mathrm{mm}$ and $90 \,\mathrm{\mu m}$ for $\tau_{0}=10^{-2}$.

The VSI is a robust, purely hydrodynamical instability in PPDs that drives weak to moderate turbulence and transport in PPD dead zones \citep{nelson2013,stoll2014,stoll2016,manger2021}. We find $\alpha$-viscosity parameter values $\sim 10^{-4}-10^{-3}$, where lower values correspond to larger $Z$. These $\alpha$ values are consistent with those reported in the aforementioned studies. VSI activity weakens with dust-loading because of stabilization provided by dust-induced buoyancy \citep{lin2019}. 

When active, the VSI results in strong vertical mixing of dust particles and is therefore a suitable source of turbulence to test the robustness of planetesimal formation in the vicinity of a pressure bump \citep{flock2017,lin2019,flock2020}. We find that a moderate pressure bump $A\gtrsim 0.2$ collects sufficient amounts of dust to reach local dust-to-gas density ratios $\epsilon >1$ at solar metallicity $Z=0.01$ and Stokes numbers $\tau_{0} \sim 10^{-2}$.

We find morphology of dust concentrations at the pressure bump transition from non-axisymmetric to axisymmetric  with increasing background metallicity $Z$ or Stokes number $\tau_{0}$. In particular, at lower $Z \lesssim 0.01-0.03$ and $\tau_{0}\sim 10^{-3}$ a weak dust ring forms that gives rise to the formation of one or more vortices with enhanced (but still much lower than unity) dust-to-gas ratio. These vortices subsequently migrate inwards until they leave the computational domain or dissolve in the background flow. In these simulations the pressure bump decays due to turbulent diffusion. 
At low metallicities and larger Stokes numbers $\tau_{0}\gtrsim 5\cdot 10^{-3}$ dust rings of significant  dust-to-gas ratios (order unity) form that are generally highly turbulent, non-axisymmetric in appearance, and undergo inward drift. In some cases, such as the one in the lower panel of Figure \ref{fig:vort_evol}, these rings spawn a vortex which concentrates most of the dust within the ring before it shears out into a new dust ring after some hundreds of orbits. 
Moreover, in certain parameter regimes (intermediate values of $\tau_{0}$ and larger $A=0.4$) such dust rings can spawn multiple vortices, which either split off and migrate inward, or, as described above concentrate most of the dust within the ring throughout the entire simulation time (cf. Figure \ref{fig:dustring_us}). In such simulations the pressure bump remains intact or even gets significantly amplified. The behavior of dust rings just described can be explained by the presence of an instability of dusty rings that is fed by gradients in the dust-to-gas ratio that arise from the corrugation motion of the dust-layer adjacent to the ring due to the VSI.

At large metallicities $Z\gtrsim 0.05$, we find no notable long-lived vortices form. Instead, high density dust rings with weak or negligible non-axisymmetries develop, with dust-to-gas ratios that surpass unity by a large margin for Stokes numbers $\tau_{0} \gtrsim 5 \cdot 10^{-3}$. 

Our 2D axisymmetric simulations capture large parts of the findings described above. That is, they adequately describe the settling of dust in the disc mid-plane as well as the formation of dust rings at pressure bumps. These dust rings can undergo inward drift and disruption due to a dusty gas instability powered by the VSI. Moreover, critical combinations of $Z$ and $\tau_{0}$ required to surpass unity dust-to-gas ratios agree fairly well with those inferred from 3D simulations. For example, in the absence of a pressure bump $Z \gtrsim 0.05$ is required to achieve this for $\tau_{0} \gtrsim 10^{-2}$, while in the presence of a mild pressure bump ($A\gtrsim 0.2$) a solar metallicity $Z \gtrsim 0.01$ is sufficient.
On the other hand, the formation of dust-collecting vortices and other possible non-axisymmetric structures are obviously not captured in axisymmetric simulations. Also, due to the lack of turbulent planar ($\alpha$) diffusion in the latter simulations dust rings can become very thin at large $Z$ values with dust-to-gas ratios much higher than in 3D simulations.

Our results are potentially relevant to the  small -- though non-negligible-- fraction of asymmetric dust distributions compared to axisymmetric ones, as seen in ALMA images of PPDs \citep{vdmarel2021}. From our results it may be possible to infer the background metallicities or particle sizes in observed PPDs that contain dust rings based on the degree of asymmetry. As we have shown, asymmetries in dust rings `naturally' occur in a VSI-turbulent disc when the metallicity in the region where a pressure bump starts to accumulate dust is sufficiently low ($Z \lesssim 0.03$). A direct triggering of the RWI is not required in this scenario. 

Our findings do support the idea that vortices are a possible explanation for observed asymmetries of dust rings in PPDs as has recently been debated in \citet{vdmarel2021}, since it is possible for reasonable dust parameters, that new vortices are continuously being formed in unstable dust rings, such as the one presented in Figure \ref{fig:dustring_us}.
Moreover, our results indicate that vortices formed by the VSI at a pressure bump of moderate amplitude $A=0.2-0.4$ could under the right circumstances play a role in planetesimal formation, since order unity dust-to-gas ratios can be achieved within them for small particle sizes $\tau \sim 10^{-2}$ and solar metallicity $Z=0.01$.

In this study we did not consider the possibility of a reforcing mechanism for the pressure bump, such as in \citet{carrera2021a} and \citet{carrera2021b}. 
Based on our finding that dust rings, which form at a pressure bump, drift inwards for lower metallicities $Z\lesssim 0.03$ (Section \ref{sec:dust rings}), we speculate that a single pressure bump, were it reinforced, could in principle spawn multiple dust rings that would end up at smaller disc radii. As long as dust keeps flowing in from larger radii, we would expect a continuous cycle in which the pressure bump collects dust into a ring and, as soon as the dust-to-gas ratio reaches a certain (low) level, instabilities as discussed in Section \ref{sec:pbumpres} set in and the dust ring will drift inward. Verifying this scenario would require either a considerably larger domain size or a continuous inflow of dust at the outer domain boundary. Furthermore, the inward drift of dust rings (cf. Figure \ref{fig:drift_rings}) may be important when inferring the locations of bump-generating structures such as planet gaps or snow lines. That is, the observed dust rings may not be formed in situ.

The spatial resolution of our simulations is insufficient to resolve several small-scale instabilities that can affect the evolution of dusty rings and vortices. In particular, we did not resolve the SI \citep{youdin2005}, which will start playing a role when unity dust-to-gas density ratios are attained. Previous 2D simulations \citep{fu2014,raettig2015,rubsamen2015,surville2019} suggest that vortices are destroyed by small-scale dust-gas instabilities once dust-to-gas ratios (and hence dust feedback) become sufficiently high. On the other hand, the 3D shearing box simulations by \citet{lyra2018} and \citet{raettig2021} show that vertically extended vortices can remain intact even when high values of $\epsilon$ are reached 
, as feedback is only important at the mid-plane. In 3D, however, the elliptical instability \citep{lesur2009} should operate but is most likely unresolved in our models, which may shorten the lifetime of the large vortices seen in our simulations. Other effects, such as the dust settling instability \citep{squire2018,krapp2020}, the KHI of the mid-plane dust-layer \citep{chiang08,barranco2009}, and vertically shearing streaming instabilities \citep{ishitsu09,lin21} are also not captured. Conducting global 3D simulations of the VSI that also resolve the SI and these small-scale instabilities are computationally too expensive at this time. \citet{schaefer2020} performed global 2D axisymmetric simulations of the VSI and the SI and this setup could be used as a first step to investigate some aspects of the common effect of the SI and VSI on dust rings. 


For a realistic modelling of dust concentration in rings and vortices one will ultimately need to consider simulations which employ a distribution of particle sizes that are in addition allowed to change due coagulation and fragmentation, since particle growth is expected to be enhanced in these structures. Even in the absence of the latter the presence of multiple dust species can have a significant impact on the evolution of the SI \citep{bai2010,krapp2019,paardekooper2020,zhu2021,schaffer2021}, which in turn can affect the evolution of dust rings and dust-laden vortices. 

Throughout our paper, we assumed a fixed temperature profile with power-law index $q=1$. Decreasing this value will result in a weakening of the VSI which is then much more sensitive to dusty buoyancy \citep{lin2019}, shifting critical values of the metallicity $Z$ encountered in our study to lower values. The opposite will be true if $q$ is increased. Eventually, a more realistic equation of state including a finite cooling time should be considered. For instance, \citet{fung2021} showed in 2D shearing sheet simulations that a finite cooling time can have adverse effects on vortex life times at disc radii that are relevant to our work. On the other hand, the dust-free simulations of \citet{richard2016} and \citet{pfeil2020} suggest that large long-lived vortices can be supported by the SBI \citep{lesur2010} when the cooling time is increased. 

In a follow up work we will explore the impact of finite cooling times and multiple grain sizes on the evolution of pressure bumps in turbulent PPDs. 

\begin{acknowledgements}
We thank the anonymous reviewer for their insightful report and C. Cui for useful discussions. This research is supported by the Ministry of Science and Technology of Taiwan (grants 107-2112-M-001-043-MY3, 110-2112-M-001-034-, 110-2124-M-002-012-) and an Academia Sinica Career Development Award (AS-CDA-110-M06). 
Simulations were conducted at the TIARA High-
Performance Computing cluster and the TAIWANIA cluster hosted
by the National Center for High-performance Computing (NCHC).

\end{acknowledgements}

%
%

\bibliographystyle{apalike}

\bibliography{lit}{}

\begin{thebibliography}{}

\bibitem[{ALMA Partnership} et~al., 2015]{alma2015}
{ALMA Partnership}, {Brogan}, C.~L., {P{\'e}rez}, L.~M., {Hunter}, T.~R.,
  {Dent}, W.~R.~F., {Hales}, A.~S., {Hills}, R.~E., {Corder}, S., {Fomalont},
  E.~B., {Vlahakis}, C., {Asaki}, Y., {Barkats}, D., {Hirota}, A., {Hodge},
  J.~A., {Impellizzeri}, C.~M.~V., {Kneissl}, R., {Liuzzo}, E., {Lucas}, R.,
  {Marcelino}, N., {Matsushita}, S., {Nakanishi}, K., {Phillips}, N.,
  {Richards}, A.~M.~S., {Toledo}, I., {Aladro}, R., {Broguiere}, D., {Cortes},
  J.~R., {Cortes}, P.~C., {Espada}, D., {Galarza}, F., {Garcia-Appadoo}, D.,
  {Guzman-Ramirez}, L., {Humphreys}, E.~M., {Jung}, T., {Kameno}, S., {Laing},
  R.~A., {Leon}, S., {Marconi}, G., {Mignano}, A., {Nikolic}, B., {Nyman},
  L.~A., {Radiszcz}, M., {Remijan}, A., {Rod{\'o}n}, J.~A., {Sawada}, T.,
  {Takahashi}, S., {Tilanus}, R.~P.~J., {Vila Vilaro}, B., {Watson}, L.~C.,
  {Wiklind}, T., {Akiyama}, E., {Chapillon}, E., {de Gregorio-Monsalvo}, I.,
  {Di Francesco}, J., {Gueth}, F., {Kawamura}, A., {Lee}, C.~F., {Nguyen
  Luong}, Q., {Mangum}, J., {Pietu}, V., {Sanhueza}, P., {Saigo}, K.,
  {Takakuwa}, S., {Ubach}, C., {van Kempen}, T., {Wootten}, A.,
  {Castro-Carrizo}, A., {Francke}, H., {Gallardo}, J., {Garcia}, J.,
  {Gonzalez}, S., {Hill}, T., {Kaminski}, T., {Kurono}, Y., {Liu}, H.~Y.,
  {Lopez}, C., {Morales}, F., {Plarre}, K., {Schieven}, G., {Testi}, L.,
  {Videla}, L., {Villard}, E., {Andreani}, P., {Hibbard}, J.~E., and
  {Tatematsu}, K. (2015).
\newblock {The 2014 ALMA Long Baseline Campaign: First Results from High
  Angular Resolution Observations toward the HL Tau Region}.
\newblock {\em \apjl}, 808(1):L3.

\bibitem[{Andrews}, 2020]{andrews2020}
{Andrews}, S.~M. (2020).
\newblock {Observations of Protoplanetary Disk Structures}.
\newblock {\em \araa}, 58:483--528.

\bibitem[{Andrews} et~al., 2018]{andrews18}
{Andrews}, S.~M., {Huang}, J., {P{\'e}rez}, L.~M., {Isella}, A., {Dullemond},
  C.~P., {Kurtovic}, N.~T., {Guzm{\'a}n}, V.~V., {Carpenter}, J.~M., {Wilner},
  D.~J., {Zhang}, S., {Zhu}, Z., {Birnstiel}, T., {Bai}, X.-N., {Benisty}, M.,
  {Hughes}, A.~M., {{\"O}berg}, K.~I., and {Ricci}, L. (2018).
\newblock {The Disk Substructures at High Angular Resolution Project (DSHARP).
  I. Motivation, Sample, Calibration, and Overview}.
\newblock {\em \apjl}, 869(2):L41.

\bibitem[{Andrews} et~al., 2009]{andrews2009}
{Andrews}, S.~M., {Wilner}, D.~J., {Hughes}, A.~M., {Qi}, C., and {Dullemond},
  C.~P. (2009).
\newblock {Protoplanetary Disk Structures in Ophiuchus}.
\newblock {\em \apj}, 700(2):1502--1523.

\bibitem[{Armitage}, 2011]{armitage2011}
{Armitage}, P.~J. (2011).
\newblock {Dynamics of Protoplanetary Disks}.
\newblock {\em \araa}, 49(1):195--236.

\bibitem[{Bai}, 2015]{bai2015}
{Bai}, X.-N. (2015).
\newblock {Hall Effect Controlled Gas Dynamics in Protoplanetary Disks. II.
  Full 3D Simulations toward the Outer Disk}.
\newblock {\em \apj}, 798(2):84.

\bibitem[{Bai}, 2017]{bai2017}
{Bai}, X.-N. (2017).
\newblock {Global Simulations of the Inner Regions of Protoplanetary Disks with
  Comprehensive Disk Microphysics}.
\newblock {\em \apj}, 845(1):75.

\bibitem[{Bai} and {Stone}, 2010]{bai2010}
{Bai}, X.-N. and {Stone}, J.~M. (2010).
\newblock {Dynamics of Solids in the Midplane of Protoplanetary Disks:
  Implications for Planetesimal Formation}.
\newblock {\em \apj}, 722(2):1437--1459.

\bibitem[{Balbus} and {Hawley}, 1991]{balbus1991}
{Balbus}, S.~A. and {Hawley}, J.~F. (1991).
\newblock {A Powerful Local Shear Instability in Weakly Magnetized Disks. I.
  Linear Analysis}.
\newblock {\em \apj}, 376:214.

\bibitem[{Barge} and {Sommeria}, 1995]{barge1995}
{Barge}, P. and {Sommeria}, J. (1995).
\newblock {Did planet formation begin inside persistent gaseous vortices?}
\newblock {\em \aap}, 295:L1--L4.

\bibitem[{Barker} and {Latter}, 2015]{barker2015}
{Barker}, A.~J. and {Latter}, H.~N. (2015).
\newblock {On the vertical-shear instability in astrophysical discs}.
\newblock {\em \mnras}, 450(1):21--37.

\bibitem[{Barranco}, 2009]{barranco2009}
{Barranco}, J.~A. (2009).
\newblock {Three-Dimensional Simulations of Kelvin-Helmholtz Instability in
  Settled Dust Layers in Protoplanetary Disks}.
\newblock {\em \apj}, 691(2):907--921.

\bibitem[{Ben{\'\i}tez-Llambay} et~al., 2019]{llambay2019}
{Ben{\'\i}tez-Llambay}, P., {Krapp}, L., and {Pessah}, M.~E. (2019).
\newblock {Asymptotically Stable Numerical Method for Multispecies Momentum
  Transfer: Gas and Multifluid Dust Test Suite and Implementation in FARGO3D}.
\newblock {\em \apjs}, 241(2):25.

\bibitem[Benítez-Llambay and Masset, 2016]{fargo3d}
Benítez-Llambay, P. and Masset, F.~S. (2016).
\newblock Fargo3d: A new gpu-oriented mhd code.
\newblock {\em The Astrophysical Journal Supplement Series}, 223(1):11.

\bibitem[{Blum}, 2018]{blum2018}
{Blum}, J. (2018).
\newblock {Dust Evolution in Protoplanetary Discs and the Formation of
  Planetesimals. What Have We Learned from Laboratory Experiments?}
\newblock {\em \ssr}, 214(2):52.

\bibitem[{Carrera} et~al., 2015]{carrera15}
{Carrera}, D., {Johansen}, A., and {Davies}, M.~B. (2015).
\newblock {How to form planetesimals from mm-sized chondrules and chondrule
  aggregates}.
\newblock {\em \aap}, 579:A43.

\bibitem[{Carrera} et~al., 021a]{carrera2021a}
{Carrera}, D., {Simon}, J.~B., {Li}, R., {Kretke}, K.~A., and {Klahr}, H.
  (2021a).
\newblock {Protoplanetary Disk Rings as Sites for Planetesimal Formation}.
\newblock {\em \aj}, 161(2):96.

\bibitem[{Carrera} et~al., 021b]{carrera2021b}
{Carrera}, D., {Thomas}, A., {Simon}, J.~B., {Small}, M.~A., {Kretke}, K.~A.,
  and {Klahr}, H. (2021b).
\newblock {Resilience of Planetesimal Formation in Weakly-Reinforced Pressure
  Bumps}.
\newblock {\em arXiv e-prints}, page arXiv:2108.08315.

\bibitem[{Chang} and {Oishi}, 2010]{chang2010}
{Chang}, P. and {Oishi}, J.~S. (2010).
\newblock {On the Stability of Dust-laden Protoplanetary Vortices}.
\newblock {\em \apj}, 721(2):1593--1602.

\bibitem[{Chen} and {Lin}, 2018]{chen2018}
{Chen}, J.-W. and {Lin}, M.-K. (2018).
\newblock {Dusty disc-planet interaction with dust-free simulations}.
\newblock {\em \mnras}, 478(2):2737--2752.

\bibitem[{Chiang}, 2008]{chiang08}
{Chiang}, E. (2008).
\newblock {Vertical Shearing Instabilities in Radially Shearing Disks: The
  Dustiest Layers of the Protoplanetary Nebula}.
\newblock {\em \apj}, 675:1549--1558.

\bibitem[{Crnkovic-Rubsamen} et~al., 2015]{rubsamen2015}
{Crnkovic-Rubsamen}, I., {Zhu}, Z., and {Stone}, J.~M. (2015).
\newblock {Survival and structure of dusty vortices in protoplanetary discs}.
\newblock {\em \mnras}, 450(4):4285--4291.

\bibitem[{Flock} et~al., 2017]{flock2017}
{Flock}, M., {Nelson}, R.~P., {Turner}, N.~J., {Bertrang}, G. H.~M.,
  {Carrasco-Gonz{\'a}lez}, C., {Henning}, T., {Lyra}, W., and {Teague}, R.
  (2017).
\newblock {Radiation Hydrodynamical Turbulence in Protoplanetary Disks:
  Numerical Models and Observational Constraints}.
\newblock {\em \apj}, 850(2):131.

\bibitem[{Flock} et~al., 2020]{flock2020}
{Flock}, M., {Turner}, N.~J., {Nelson}, R.~P., {Lyra}, W., {Manger}, N., and
  {Klahr}, H. (2020).
\newblock {Gas and Dust Dynamics in Starlight-heated Protoplanetary Disks}.
\newblock {\em \apj}, 897(2):155.

\bibitem[{Fu} et~al., 2014]{fu2014}
{Fu}, W., {Li}, H., {Lubow}, S., {Li}, S., and {Liang}, E. (2014).
\newblock {Effects of Dust Feedback on Vortices in Protoplanetary Disks}.
\newblock {\em \apjl}, 795(2):L39.

\bibitem[{Fung} and {Ono}, 2021]{fung2021}
{Fung}, J. and {Ono}, T. (2021).
\newblock {Cooling-induced Vortex Decay in Keplerian Disks}.
\newblock {\em \apj}, 922(1):13.

\bibitem[{Gammie}, 1996]{gammie1996}
{Gammie}, C.~F. (1996).
\newblock {Layered Accretion in T Tauri Disks}.
\newblock {\em \apj}, 457:355.

\bibitem[{Goldreich} and {Ward}, 1973]{goldreich1973}
{Goldreich}, P. and {Ward}, W.~R. (1973).
\newblock {The Formation of Planetesimals}.
\newblock {\em \apj}, 183:1051--1062.

\bibitem[{Gressel} et~al., 2015]{gressel2015}
{Gressel}, O., {Turner}, N.~J., {Nelson}, R.~P., and {McNally}, C.~P. (2015).
\newblock {Global Simulations of Protoplanetary Disks With Ohmic Resistivity
  and Ambipolar Diffusion}.
\newblock {\em \apj}, 801(2):84.

\bibitem[{Haghighipour} and {Boss}, 003a]{haghighipour2003a}
{Haghighipour}, N. and {Boss}, A.~P. (2003a).
\newblock {On Gas Drag-Induced Rapid Migration of Solids in a Nonuniform Solar
  Nebula}.
\newblock {\em \apj}, 598(2):1301--1311.

\bibitem[{Haghighipour} and {Boss}, 003b]{haghighipour2003b}
{Haghighipour}, N. and {Boss}, A.~P. (2003b).
\newblock {On Pressure Gradients and Rapid Migration of Solids in a Nonuniform
  Solar Nebula}.
\newblock {\em \apj}, 583(2):996--1003.

\bibitem[{Huang} et~al., 2020]{huang2020}
{Huang}, P., {Li}, H., {Isella}, A., {Miranda}, R., {Li}, S., and {Ji}, J.
  (2020).
\newblock {Meso-scale Instability Triggered by Dust Feedback in Dusty Rings:
  Origin and Observational Implications}.
\newblock {\em \apj}, 893(2):89.

\bibitem[{Ishitsu} et~al., 2009]{ishitsu09}
{Ishitsu}, N., {Inutsuka}, S.-i., and {Sekiya}, M. (2009).
\newblock {Two-fluid Instability of Dust and Gas in the Dust Layer of a
  Protoplanetary Disk}.
\newblock {\em arXiv e-prints}, page arXiv:0905.4404.

\bibitem[{Jacquet} et~al., 2011]{jacquet2011}
{Jacquet}, E., {Balbus}, S., and {Latter}, H. (2011).
\newblock {On linear dust-gas streaming instabilities in protoplanetary discs}.
\newblock {\em \mnras}, 415:3591--3598.

\bibitem[{Johansen} et~al., 2004]{johansen2004}
{Johansen}, A., {Andersen}, A.~C., and {Brandenburg}, A. (2004).
\newblock {Simulations of dust-trapping vortices in protoplanetary discs}.
\newblock {\em \aap}, 417:361--374.

\bibitem[{Johansen} et~al., 2014]{johansen2014}
{Johansen}, A., {Blum}, J., {Tanaka}, H., {Ormel}, C., {Bizzarro}, M., and
  {Rickman}, H. (2014).
\newblock {The Multifaceted Planetesimal Formation Process}.
\newblock In {Beuther}, H., {Klessen}, R.~S., {Dullemond}, C.~P., and
  {Henning}, T., editors, {\em Protostars and Planets VI}, page 547.

\bibitem[{Johansen} et~al., 2006]{johansen2006}
{Johansen}, A., {Henning}, T., and {Klahr}, H. (2006).
\newblock {Dust Sedimentation and Self-sustained Kelvin-Helmholtz Turbulence in
  Protoplanetary Disk Midplanes}.
\newblock {\em \apj}, 643(2):1219--1232.

\bibitem[{Johansen} and {Youdin}, 2007]{johansen2007}
{Johansen}, A. and {Youdin}, A. (2007).
\newblock {Protoplanetary Disk Turbulence Driven by the Streaming Instability:
  Nonlinear Saturation and Particle Concentration}.
\newblock {\em \apj}, 662(1):627--641.

\bibitem[{Johansen} et~al., 009b]{johansen2009b}
{Johansen}, A., {Youdin}, A., and {Klahr}, H. (2009b).
\newblock {Zonal Flows and Long-lived Axisymmetric Pressure Bumps in
  Magnetorotational Turbulence}.
\newblock {\em \apj}, 697(2):1269--1289.

\bibitem[{Johansen} et~al., 009a]{johansen2009a}
{Johansen}, A., {Youdin}, A., and {Mac Low}, M.-M. (2009a).
\newblock {Particle Clumping and Planetesimal Formation Depend Strongly on
  Metallicity}.
\newblock {\em \apjl}, 704(2):L75--L79.

\bibitem[{Kanagawa} et~al., 2017]{kanagawa2017}
{Kanagawa}, K.~D., {Ueda}, T., {Muto}, T., and {Okuzumi}, S. (2017).
\newblock {Effect of Dust Radial Drift on Viscous Evolution of Gaseous Disk}.
\newblock {\em \apj}, 844(2):142.

\bibitem[{Kenyon} and {Luu}, 1999]{kenyon1999}
{Kenyon}, S.~J. and {Luu}, J.~X. (1999).
\newblock {Accretion in the Early Kuiper Belt. II. Fragmentation}.
\newblock {\em \aj}, 118(2):1101--1119.

\bibitem[{Klahr} and {Bodenheimer}, 2006]{klahr2006}
{Klahr}, H. and {Bodenheimer}, P. (2006).
\newblock {Formation of Giant Planets by Concurrent Accretion of Solids and Gas
  inside an Anticyclonic Vortex}.
\newblock {\em \apj}, 639(1):432--440.

\bibitem[{Klahr} and {Hubbard}, 2014]{klahr2014}
{Klahr}, H. and {Hubbard}, A. (2014).
\newblock {Convective Overstability in Radially Stratified Accretion Disks
  under Thermal Relaxation}.
\newblock {\em \apj}, 788(1):21.

\bibitem[{Klahr} and {Bodenheimer}, 2003]{klahr2003}
{Klahr}, H.~H. and {Bodenheimer}, P. (2003).
\newblock {Turbulence in Accretion Disks: Vorticity Generation and Angular
  Momentum Transport via the Global Baroclinic Instability}.
\newblock {\em \apj}, 582(2):869--892.

\bibitem[{Krapp} et~al., 2019]{krapp2019}
{Krapp}, L., {Ben{\'\i}tez-Llambay}, P., {Gressel}, O., and {Pessah}, M.~E.
  (2019).
\newblock {Streaming Instability for Particle-size Distributions}.
\newblock {\em \apjl}, 878(2):L30.

\bibitem[{Krapp} et~al., 2020]{krapp2020}
{Krapp}, L., {Youdin}, A.~N., {Kratter}, K.~M., and {Ben{\'\i}tez-Llambay}, P.
  (2020).
\newblock {Dust settling instability in protoplanetary discs}.
\newblock {\em \mnras}, 497(3):2715--2729.

\bibitem[{Laibe} and {Price}, 2014]{laibe2014}
{Laibe}, G. and {Price}, D.~J. (2014).
\newblock {Dusty gas with one fluid}.
\newblock {\em \mnras}, 440(3):2136--2146.

\bibitem[{Latter}, 2016]{latter2016}
{Latter}, H.~N. (2016).
\newblock {On the convective overstability in protoplanetary discs}.
\newblock {\em \mnras}, 455(3):2608--2618.

\bibitem[{Latter} and {Papaloizou}, 2018]{latter2018}
{Latter}, H.~N. and {Papaloizou}, J. (2018).
\newblock {Vortices and the saturation of the vertical shear instability in
  protoplanetary discs}.
\newblock {\em \mnras}, 474(3):3110--3124.

\bibitem[{Lee} et~al., 2010]{lee2010}
{Lee}, A.~T., {Chiang}, E., {Asay-Davis}, X., and {Barranco}, J. (2010).
\newblock {Forming Planetesimals by Gravitational Instability. I. The Role of
  the Richardson Number in Triggering the Kelvin-Helmholtz Instability}.
\newblock {\em \apj}, 718(2):1367--1377.

\bibitem[{Lesur} and {Papaloizou}, 2009]{lesur2009}
{Lesur}, G. and {Papaloizou}, J.~C.~B. (2009).
\newblock {On the stability of elliptical vortices in accretion discs}.
\newblock {\em \aap}, 498(1):1--12.

\bibitem[{Lesur} and {Papaloizou}, 2010]{lesur2010}
{Lesur}, G. and {Papaloizou}, J.~C.~B. (2010).
\newblock {The subcritical baroclinic instability in local accretion disc
  models}.
\newblock {\em \aap}, 513:A60.

\bibitem[{Lesur} and {Latter}, 2016]{lesur2016}
{Lesur}, G. R.~J. and {Latter}, H. (2016).
\newblock {On the survival of zombie vortices in protoplanetary discs}.
\newblock {\em \mnras}, 462(4):4549--4554.

\bibitem[{Li} et~al., 2001]{li01}
{Li}, H., {Colgate}, S.~A., {Wendroff}, B., and {Liska}, R. (2001).
\newblock {Rossby Wave Instability of Thin Accretion Disks. III. Nonlinear
  Simulations}.
\newblock {\em \apj}, 551:874--896.

\bibitem[{Li} and {Youdin}, 2021]{li21}
{Li}, R. and {Youdin}, A. (2021).
\newblock {Thresholds for Particle Clumping by the Streaming Instability}.
\newblock {\em arXiv e-prints}, page arXiv:2105.06042.

\bibitem[{Lin}, 2019]{lin2019}
{Lin}, M.-K. (2019).
\newblock {Dust settling against hydrodynamic turbulence in protoplanetary
  discs}.
\newblock {\em \mnras}, 485(4):5221--5234.

\bibitem[{Lin}, 2021]{lin21}
{Lin}, M.-K. (2021).
\newblock {Stratified and Vertically Shearing Streaming Instabilities in
  Protoplanetary Disks}.
\newblock {\em \apj}, 907(2):64.

\bibitem[{Lin} and {Youdin}, 2015]{lin2015}
{Lin}, M.-K. and {Youdin}, A.~N. (2015).
\newblock {Cooling Requirements for the Vertical Shear Instability in
  Protoplanetary Disks}.
\newblock {\em \apj}, 811(1):17.

\bibitem[{Lin} and {Youdin}, 2017]{lin2017}
{Lin}, M.-K. and {Youdin}, A.~N. (2017).
\newblock {A Thermodynamic View of Dusty Protoplanetary Disks}.
\newblock {\em \apj}, 849(2):129.

\bibitem[{Long} et~al., 2018]{long18}
{Long}, F., {Pinilla}, P., {Herczeg}, G.~J., {Harsono}, D., {Dipierro}, G.,
  {Pascucci}, I., {Hendler}, N., {Tazzari}, M., {Ragusa}, E., {Salyk}, C.,
  {Edwards}, S., {Lodato}, G., {van de Plas}, G., {Johnstone}, D., {Liu}, Y.,
  {Boehler}, Y., {Cabrit}, S., {Manara}, C.~F., {Menard}, F., {Mulders}, G.~D.,
  {Nisini}, B., {Fischer}, W.~J., {Rigliaco}, E., {Banzatti}, A., {Avenhaus},
  H., and {Gully-Santiago}, M. (2018).
\newblock {Gaps and Rings in an ALMA Survey of Disks in the Taurus Star-forming
  Region}.
\newblock {\em \apj}, 869(1):17.

\bibitem[{Lor{\'e}n-Aguilar} and {Bate}, 2015]{aguilar2015}
{Lor{\'e}n-Aguilar}, P. and {Bate}, M.~R. (2015).
\newblock {Toroidal vortices and the conglomeration of dust into rings in
  protoplanetary discs}.
\newblock {\em \mnras}, 453(1):L78--L82.

\bibitem[{Lovascio} and {Paardekooper}, 2019]{lovascio2019}
{Lovascio}, F. and {Paardekooper}, S.-J. (2019).
\newblock {Dynamics of dusty vortices - I. Extensions and limitations of the
  terminal velocity approximation}.
\newblock {\em \mnras}, 488(4):5290--5299.

\bibitem[{Lovelace} et~al., 1999]{lovelace1999}
{Lovelace}, R.~V.~E., {Li}, H., {Colgate}, S.~A., and {Nelson}, A.~F. (1999).
\newblock {Rossby Wave Instability of Keplerian Accretion Disks}.
\newblock {\em \apj}, 513(2):805--810.

\bibitem[{Lyra}, 2014]{lyra2014}
{Lyra}, W. (2014).
\newblock {Convective Overstability in Accretion Disks: Three-dimensional
  Linear Analysis and Nonlinear Saturation}.
\newblock {\em \apj}, 789(1):77.

\bibitem[{Lyra} and {Klahr}, 2011]{lyra2011}
{Lyra}, W. and {Klahr}, H. (2011).
\newblock {The baroclinic instability in the context of layered accretion.
  Self-sustained vortices and their magnetic stability in local compressible
  unstratified models of protoplanetary disks}.
\newblock {\em \aap}, 527:A138.

\bibitem[{Lyra} et~al., 2018]{lyra2018}
{Lyra}, W., {Raettig}, N., and {Klahr}, H. (2018).
\newblock {Pebble-trapping Backreaction Does Not Destroy Vortices}.
\newblock {\em Research Notes of the American Astronomical Society}, 2(4):195.

\bibitem[{Manger} and {Klahr}, 2018]{manger2018}
{Manger}, N. and {Klahr}, H. (2018).
\newblock {Vortex formation and survival in protoplanetary discs subject to
  vertical shear instability}.
\newblock {\em \mnras}, 480(2):2125--2136.

\bibitem[{Manger} et~al., 2020]{manger2020}
{Manger}, N., {Klahr}, H., {Kley}, W., and {Flock}, M. (2020).
\newblock {High resolution parameter study of the vertical shear instability}.
\newblock {\em \mnras}, 499(2):1841--1853.

\bibitem[{Manger} et~al., 2021]{manger2021}
{Manger}, N., {Pfeil}, T., and {Klahr}, H. (2021).
\newblock {High-resolution parameter study of the vertical shear instability -
  II: dependence on temperature gradient and cooling time}.
\newblock {\em \mnras}, 508(4):5402--5409.

\bibitem[{Marcus} et~al., 2015]{marcus2015}
{Marcus}, P.~S., {Pei}, S., {Jiang}, C.-H., {Barranco}, J.~A., {Hassanzadeh},
  P., and {Lecoanet}, D. (2015).
\newblock {Zombie Vortex Instability. I. A Purely Hydrodynamic Instability to
  Resurrect the Dead Zones of Protoplanetary Disks}.
\newblock {\em \apj}, 808(1):87.

\bibitem[{Meheut} et~al., 2012]{meheut2012}
{Meheut}, H., {Meliani}, Z., {Varniere}, P., and {Benz}, W. (2012).
\newblock {Dust-trapping Rossby vortices in protoplanetary disks}.
\newblock {\em \aap}, 545:A134.

\bibitem[{Miranda} et~al., 2017]{miranda2017}
{Miranda}, R., {Li}, H., {Li}, S., and {Jin}, S. (2017).
\newblock {Long-lived Dust Asymmetries at Dead Zone Edges in Protoplanetary
  Disks}.
\newblock {\em \apj}, 835(2):118.

\bibitem[{Mizuno}, 1980]{mizuno1980}
{Mizuno}, H. (1980).
\newblock {Formation of the Giant Planets}.
\newblock {\em Progress of Theoretical Physics}, 64(2):544--557.

\bibitem[{Nakagawa} et~al., 1986]{nakagawa1986}
{Nakagawa}, Y., {Sekiya}, M., and {Hayashi}, C. (1986).
\newblock {Settling and growth of dust particles in a laminar phase of a
  low-mass solar nebula}.
\newblock {\em \icarus}, 67(3):375--390.

\bibitem[{Nelson} et~al., 2013]{nelson2013}
{Nelson}, R.~P., {Gressel}, O., and {Umurhan}, O.~M. (2013).
\newblock {Linear and non-linear evolution of the vertical shear instability in
  accretion discs}.
\newblock {\em \mnras}, 435(3):2610--2632.

\bibitem[{Onishi} and {Sekiya}, 2017]{onishi2017}
{Onishi}, I.~K. and {Sekiya}, M. (2017).
\newblock {Planetesimal formation by an axisymmetric radial bump of the column
  density of the gas in a protoplanetary disk}.
\newblock {\em Earth, Planets, and Space}, 69(1):50.

\bibitem[{Ono} et~al., 2016]{ono2016}
{Ono}, T., {Muto}, T., {Takeuchi}, T., and {Nomura}, H. (2016).
\newblock {Parametric Study of the Rossby Wave Instability in a Two-dimensional
  Barotropic Disk}.
\newblock {\em \apj}, 823(2):84.

\bibitem[{Paardekooper} et~al., 2010]{paardekooper2010}
{Paardekooper}, S.-J., {Lesur}, G., and {Papaloizou}, J. C.~B. (2010).
\newblock {Vortex Migration in Protoplanetary Disks}.
\newblock {\em \apj}, 725(1):146--158.

\bibitem[{Paardekooper} et~al., 2020]{paardekooper2020}
{Paardekooper}, S.-J., {McNally}, C.~P., and {Lovascio}, F. (2020).
\newblock {Polydisperse streaming instability - I. Tightly coupled particles
  and the terminal velocity approximation}.
\newblock {\em \mnras}, 499(3):4223--4238.

\bibitem[{Petersen} et~al., 2007a]{peterson07a}
{Petersen}, M.~R., {Julien}, K., and {Stewart}, G.~R. (2007a).
\newblock {Baroclinic Vorticity Production in Protoplanetary Disks. I. Vortex
  Formation}.
\newblock {\em \apj}, 658:1236--1251.

\bibitem[{Petersen} et~al., 2007b]{peterson07b}
{Petersen}, M.~R., {Stewart}, G.~R., and {Julien}, K. (2007b).
\newblock {Baroclinic Vorticity Production in Protoplanetary Disks. II. Vortex
  Growth and Longevity}.
\newblock {\em \apj}, 658:1252--1263.

\bibitem[{Pfeil} and {Klahr}, 2021]{pfeil2020}
{Pfeil}, T. and {Klahr}, H. (2021).
\newblock {The Sandwich Mode for Vertical Shear Instability in Protoplanetary
  Disks}.
\newblock {\em \apj}, 915(2):130.

\bibitem[{Picogna} et~al., 2018]{picogna2018}
{Picogna}, G., {Stoll}, M. H.~R., and {Kley}, W. (2018).
\newblock {Particle accretion onto planets in discs with hydrodynamic
  turbulence}.
\newblock {\em \aap}, 616:A116.

\bibitem[{Raettig} et~al., 2015]{raettig2015}
{Raettig}, N., {Klahr}, H., and {Lyra}, W. (2015).
\newblock {Particle Trapping and Streaming Instability in Vortices in
  Protoplanetary Disks}.
\newblock {\em \apj}, 804(1):35.

\bibitem[{Raettig} et~al., 2021]{raettig2021}
{Raettig}, N., {Lyra}, W., and {Klahr}, H. (2021).
\newblock {Pebble Trapping in Vortices: Three-dimensional Simulations}.
\newblock {\em \apj}, 913(2):92.

\bibitem[{Ragusa} et~al., 2017]{ragusa2017}
{Ragusa}, E., {Dipierro}, G., {Lodato}, G., {Laibe}, G., and {Price}, D.~J.
  (2017).
\newblock {On the origin of horseshoes in transitional discs}.
\newblock {\em \mnras}, 464(2):1449--1455.

\bibitem[{Railton} and {Papaloizou}, 2014]{railton14}
{Railton}, A.~D. and {Papaloizou}, J.~C.~B. (2014).
\newblock {On the local stability of vortices in differentially rotating
  discs}.
\newblock {\em \mnras}, 445(4):4409--4426.

\bibitem[{Richard} et~al., 2016]{richard2016}
{Richard}, S., {Nelson}, R.~P., and {Umurhan}, O.~M. (2016).
\newblock {Vortex formation in protoplanetary discs induced by the vertical
  shear instability}.
\newblock {\em \mnras}, 456(4):3571--3584.

\bibitem[{Safronov}, 1972]{safronov1972}
{Safronov}, V.~S. (1972).
\newblock {\em {Evolution of the protoplanetary cloud and formation of the
  earth and planets.}}

\bibitem[{Sch{\"a}fer} et~al., 2020]{schaefer2020}
{Sch{\"a}fer}, U., {Johansen}, A., and {Banerjee}, R. (2020).
\newblock {The coexistence of the streaming instability and the vertical shear
  instability in protoplanetary disks}.
\newblock {\em \aap}, 635:A190.

\bibitem[{Schaffer} et~al., 2021]{schaffer2021}
{Schaffer}, N., {Johansen}, A., and {Lambrechts}, M. (2021).
\newblock {Streaming instability of multiple particle species. II. Numerical
  convergence with increasing particle number}.
\newblock {\em \aap}, 653:A14.

\bibitem[{Shakura} and {Sunyaev}, 1973]{shakura1973}
{Shakura}, N.~I. and {Sunyaev}, R.~A. (1973).
\newblock {Reprint of 1973A\&A....24..337S. Black holes in binary systems.
  Observational appearance.}
\newblock {\em \aap}, 500:33--51.

\bibitem[{Shi} and {Chiang}, 2013]{shi2013}
{Shi}, J.-M. and {Chiang}, E. (2013).
\newblock {From Dust to Planetesimals: Criteria for Gravitational Instability
  of Small Particles in Gas}.
\newblock {\em \apj}, 764(1):20.

\bibitem[{Simon} et~al., 2016]{simon2016}
{Simon}, J.~B., {Armitage}, P.~J., {Li}, R., and {Youdin}, A.~N. (2016).
\newblock {The Mass and Size Distribution of Planetesimals Formed by the
  Streaming Instability. I. The Role of Self-gravity}.
\newblock {\em \apj}, 822(1):55.

\bibitem[{Squire} and {Hopkins}, 2018]{squire2018}
{Squire}, J. and {Hopkins}, P.~F. (2018).
\newblock {Resonant drag instabilities in protoplanetary discs: the streaming
  instability and new, faster growing instabilities}.
\newblock {\em \mnras}, 477(4):5011--5040.

\bibitem[{Stoll} and {Kley}, 2014]{stoll2014}
{Stoll}, M. H.~R. and {Kley}, W. (2014).
\newblock {Vertical shear instability in accretion disc models with radiation
  transport}.
\newblock {\em \aap}, 572:A77.

\bibitem[{Stoll} and {Kley}, 2016]{stoll2016}
{Stoll}, M. H.~R. and {Kley}, W. (2016).
\newblock {Particle dynamics in discs with turbulence generated by the vertical
  shear instability}.
\newblock {\em \aap}, 594:A57.

\bibitem[{Stoll} et~al., 2017]{stoll2017}
{Stoll}, M. H.~R., {Kley}, W., and {Picogna}, G. (2017).
\newblock {Anisotropic hydrodynamic turbulence in accretion disks}.
\newblock {\em \aap}, 599:L6.

\bibitem[{Surville} and {Mayer}, 2019]{surville2019}
{Surville}, C. and {Mayer}, L. (2019).
\newblock {Dust-vortex Instability in the Regime of Well-coupled Grains}.
\newblock {\em \apj}, 883(2):176.

\bibitem[{Takeuchi} and {Lin}, 2002]{takeuchi2002}
{Takeuchi}, T. and {Lin}, D.~N.~C. (2002).
\newblock {Radial Flow of Dust Particles in Accretion Disks}.
\newblock {\em \apj}, 581(2):1344--1355.

\bibitem[{Taki} et~al., 2016]{taki2016}
{Taki}, T., {Fujimoto}, M., and {Ida}, S. (2016).
\newblock {Dust and gas density evolution at a radial pressure bump in
  protoplanetary disks}.
\newblock {\em \aap}, 591:A86.

\bibitem[{Tassoul}, 1978]{tassoul1978}
{Tassoul}, J. (1978).
\newblock {\em {Theory of rotating stars}}.

\bibitem[{Turner} and {Drake}, 2009]{turner2009}
{Turner}, N.~J. and {Drake}, J.~F. (2009).
\newblock {Energetic Protons, Radionuclides, and Magnetic Activity in
  Protostellar Disks}.
\newblock {\em \apj}, 703(2):2152--2159.

\bibitem[{Turner} et~al., 2014]{turner2014}
{Turner}, N.~J., {Fromang}, S., {Gammie}, C., {Klahr}, H., {Lesur}, G.,
  {Wardle}, M., and {Bai}, X.~N. (2014).
\newblock {Transport and Accretion in Planet-Forming Disks}.
\newblock In {Beuther}, H., {Klessen}, R.~S., {Dullemond}, C.~P., and
  {Henning}, T., editors, {\em Protostars and Planets VI}, page 411.

\bibitem[{Urpin}, 2003]{urpin2003}
{Urpin}, V. (2003).
\newblock {A comparison study of the vertical and magnetic shear instabilities
  in accretion discs}.
\newblock {\em \aap}, 404:397--403.

\bibitem[{Urpin} and {Brandenburg}, 1998]{urpin1998}
{Urpin}, V. and {Brandenburg}, A. (1998).
\newblock {Magnetic and vertical shear instabilities in accretion discs}.
\newblock {\em \mnras}, 294(3):399--406.

\bibitem[{van der Marel} et~al., 2021]{vdmarel2021}
{van der Marel}, N., {Birnstiel}, T., {Garufi}, A., {Ragusa}, E.,
  {Christiaens}, V., {Price}, D.~J., {Sallum}, S., {Muley}, D., {Francis}, L.,
  and {Dong}, R. (2021).
\newblock {On the Diversity of Asymmetries in Gapped Protoplanetary Disks}.
\newblock {\em \aj}, 161(1):33.

\bibitem[{van der Marel} et~al., 2013]{vdmarel2013}
{van der Marel}, N., {van Dishoeck}, E.~F., {Bruderer}, S., {Birnstiel}, T.,
  {Pinilla}, P., {Dullemond}, C.~P., {van Kempen}, T.~A., {Schmalzl}, M.,
  {Brown}, J.~M., {Herczeg}, G.~J., {Mathews}, G.~S., and {Geers}, V. (2013).
\newblock {A Major Asymmetric Dust Trap in a Transition Disk}.
\newblock {\em Science}, 340(6137):1199--1202.

\bibitem[{Weidenschilling}, 1977]{weidenschilling1977}
{Weidenschilling}, S.~J. (1977).
\newblock {Aerodynamics of solid bodies in the solar nebula.}
\newblock {\em \mnras}, 180:57--70.

\bibitem[{Weidenschilling} and {Cuzzi}, 1993]{weidenschilling1993}
{Weidenschilling}, S.~J. and {Cuzzi}, J.~N. (1993).
\newblock {Formation of Planetesimals in the Solar Nebula}.
\newblock In {Levy}, E.~H. and {Lunine}, J.~I., editors, {\em Protostars and
  Planets III}, page 1031.

\bibitem[{Wetherill}, 1990]{wetherill1990}
{Wetherill}, G.~W. (1990).
\newblock {Formation of the Earth and Terrestrial Planets}.
\newblock In {\em Bulletin of the American Astronomical Society}, volume~22,
  page 1335.

\bibitem[{Yang} et~al., 2017]{yang17}
{Yang}, C.~C., {Johansen}, A., and {Carrera}, D. (2017).
\newblock {Concentrating small particles in protoplanetary disks through the
  streaming instability}.
\newblock {\em \aap}, 606:A80.

\bibitem[{Youdin} and {Johansen}, 2007]{youdin2007}
{Youdin}, A. and {Johansen}, A. (2007).
\newblock {Protoplanetary Disk Turbulence Driven by the Streaming Instability:
  Linear Evolution and Numerical Methods}.
\newblock {\em \apj}, 662(1):613--626.

\bibitem[{Youdin} and {Goodman}, 2005]{youdin2005}
{Youdin}, A.~N. and {Goodman}, J. (2005).
\newblock {Streaming Instabilities in Protoplanetary Disks}.
\newblock {\em \apj}, 620(1):459--469.

\bibitem[{Zhu} and {Stone}, 2014]{zhu2014}
{Zhu}, Z. and {Stone}, J.~M. (2014).
\newblock {Dust Trapping by Vortices in Transitional Disks: Evidence for
  Non-ideal Magnetohydrodynamic Effects in Protoplanetary Disks}.
\newblock {\em \apj}, 795(1):53.

\bibitem[{Zhu} and {Yang}, 2021]{zhu2021}
{Zhu}, Z. and {Yang}, C.-C. (2021).
\newblock {Streaming instability with multiple dust species - I. Favourable
  conditions for the linear growth}.
\newblock {\em \mnras}, 501(1):467--482.

\end{thebibliography}

\begin{appendix}

 \section{Influence of Temperature Slope $q$ on Dusty Gas Instability}\label{sec:qvar}
\begin{figure*}[h!]
\centering
\includegraphics[width = 0.99 \textwidth]{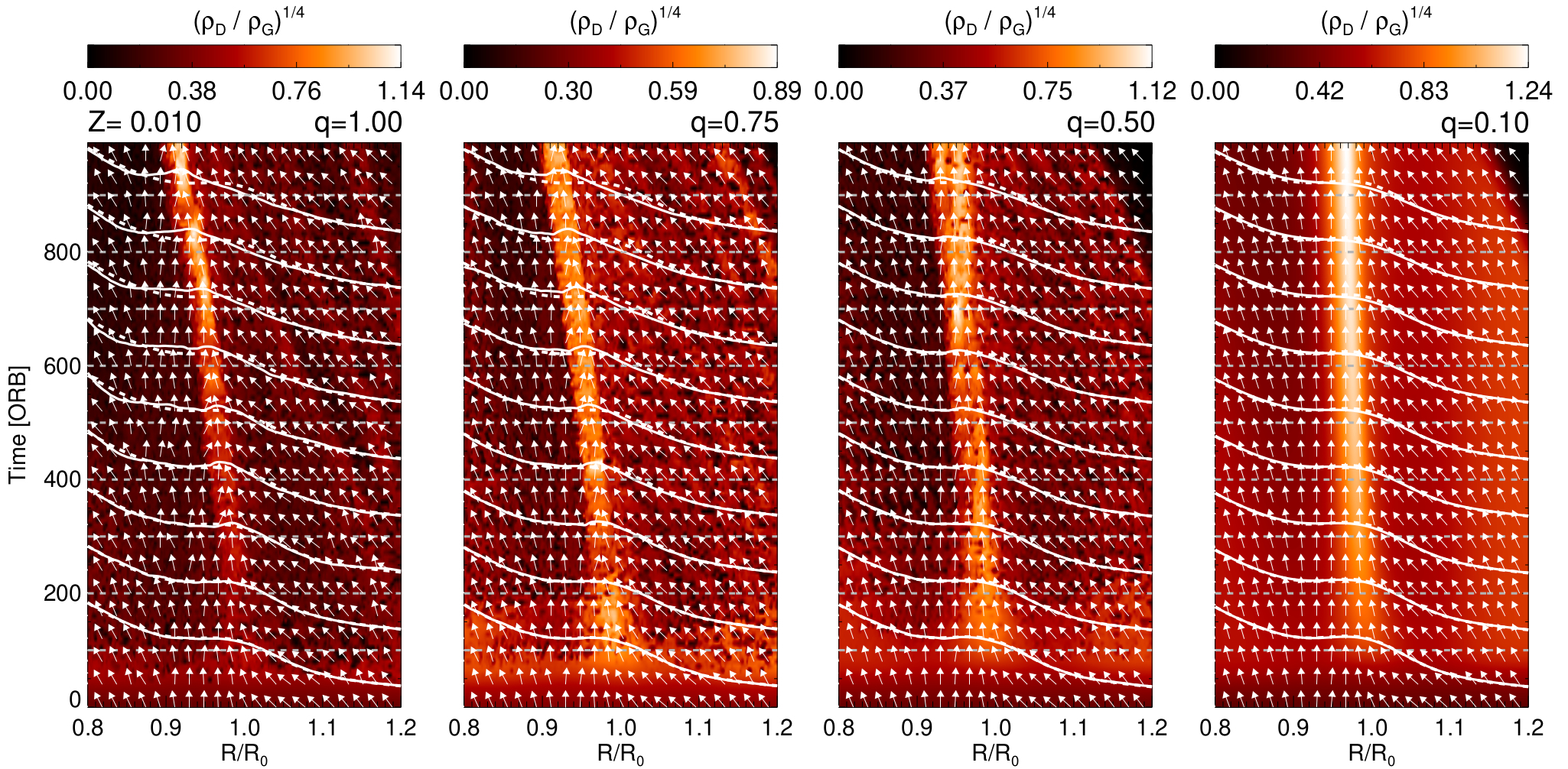}
\caption{Same as Figure \ref{fig:sptd_zcomp} but now the temperature slope $q$ is varied from left to right as indicated. The runs adopted a metallicity $Z=0.01$, Stokes number $\tau_{0}=6\cdot 10^{-3}$ and an initial pressure bump with $A=0.4$.}
\label{fig:qcomp_z001}
\end{figure*}

\begin{figure*}[h!]
\centering
\includegraphics[width = 0.99 \textwidth]{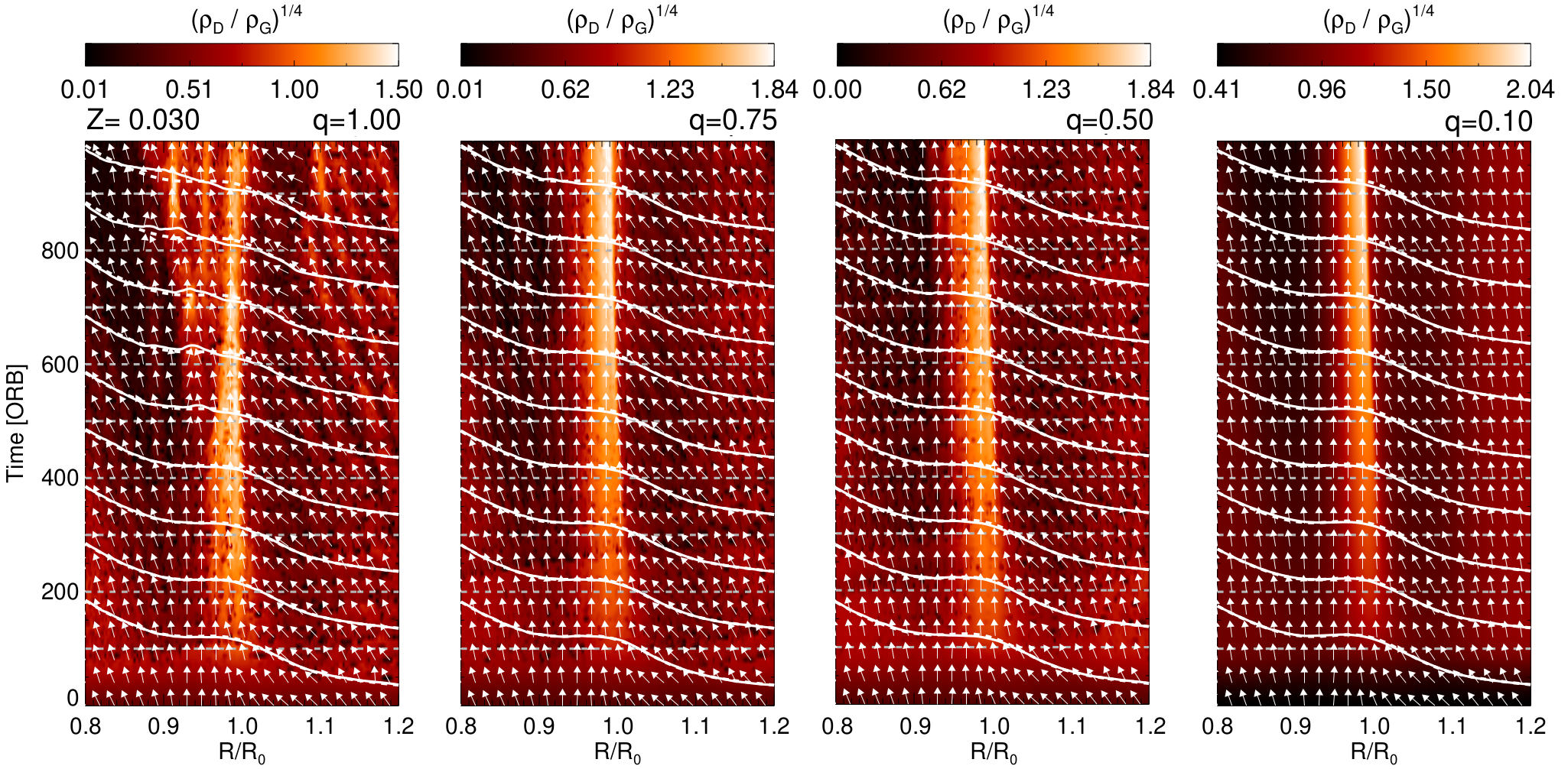}
\caption{Same as Figure \ref{fig:qcomp_z001} except that $Z=0.03$.}
\label{fig:qcomp_z003}
\end{figure*}

 In Section \ref{sec:instab} we described how the VSI leads to the emergence of a dusty gas instability by inducing a corrugation motion of the dusty mid-plane layer. To further substantiate this hypothesis we conducted additional 2D axisymmetric simulations including an initial pressure bump with $A=0.4$ and with reduced values of the temperature slope $q$ as this should result in a weaker VSI and hence a weaker dusty gas instability. The results of these simulations are presented in Figures \ref{fig:qcomp_z001} and \ref{fig:qcomp_z003} for the cases $\tau_{0}=6\cdot 10^{-3}$ with $z=0.01$ and $z=0.03$, respectively. From left to right the simulations adopt a temperature slope $q=1$ (the fiducial value), $q=0.75$, $q=0.5$ and $q=0.1$. According to observational studies \citep{andrews2009} the two intermediate values $q=0.75$ and $q=0.5$ should be the most realistic ones to model the outer parts of PPDs. With decreasing value of $q$ we find that the inward drifting motion as well as the splitting of dust rings is increasingly mitigated. However, the inward drift remains significant for all but the smallest value of $q$ considered here. The simulations with the smallest value $q=0.1$ are almost entirely laminar with only very weak VSI activity. Additional unstratified 2D simulations as well as 1D simulations with various values of $q\leq 1$ that we carried out are very similar to the cases with $q=0.1$ shown here and will therefore not be presented.


 \section{Influence of Density Slope $p$ on Vortex Evolution}\label{sec:pvar}

\begin{figure*}[h!]
\centering
\includegraphics[width = 0.99 \textwidth]{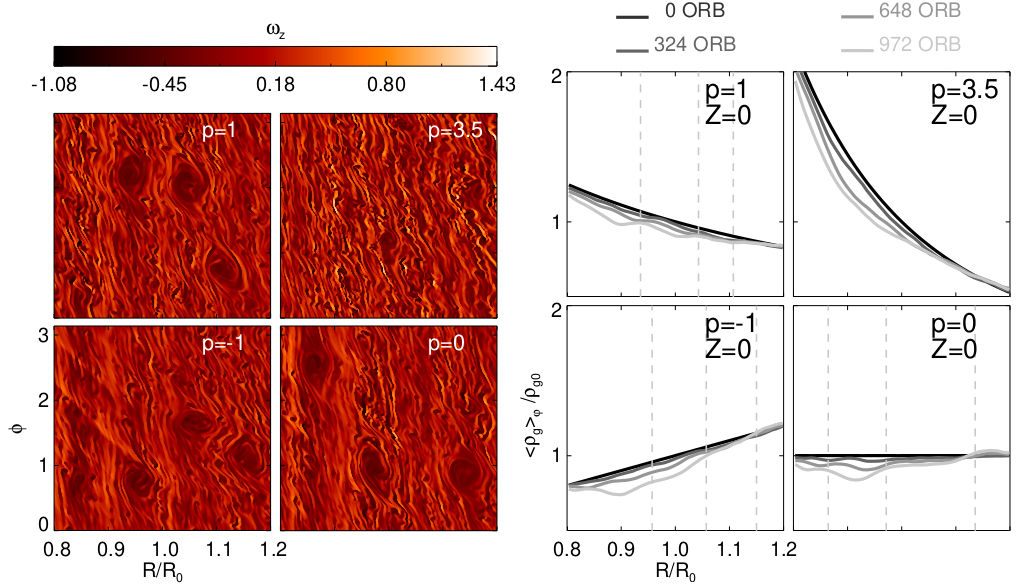}
\caption{\emph{Left}: Same as Figure \ref{fig:vort_dustfree} but now $A=0$ and the mid-plane gas density powerlaw index $p$ is varied through corresponding variation of the powerlaw index $s$ of the surface mass density. The snapshots correspond to $\sim 1000$ orbits. \emph{Right}: Same as Figure \ref{fig:pres_evol} but now the gas density slope $p$ is varied through corresponding variation of the slope $s$ of the surface mass density. Moreover, the dashed lines indicate the locations of three large vortices at  $\sim 1000$ orbits. }
\label{fig:pvar}
\end{figure*}

 As described in Section \ref{sec:vort_dustfree}, we hypothesize that the size of large vortices depends on their capability to absorb small vortices within their vicinity. Based on the study of \citet{paardekooper2010} we expect this capability to be increased for flatter surface density profiles where small vortices should be more efficiently attracted by the mild pressure maximum corresponding to a large vortex (cf. Figure \ref{fig:vort_details}). To test this hypothesis we conducted 3D dust-free simulations with varying gas surface mass density slope $s$ (and hence varying mid-plane gas density slope $p$). These simulations are presented in Figure \ref{fig:pvar}. The results in the left panels correspond to $\sim 1000$ orbits and illustrate that for $|p| \leq 1$ vortices are substantially stronger and more numerous than in the simulation with $p=3.5$. In this regard our fiducial run with $p=2.5$ (lower left panel of Figure \ref{fig:vort_dustfree}) is similar to the case $p=3.5$.  Inspection of the time evolution of the mid-plane gas density in the right panels of Figure \ref{fig:pvar} shows the development of bumps and dips, which again are believed to be a consequence of the VSI turbulent mass transport and which are more prominent in the simulations with smaller $|p|$. We find that large vortices in simulations with $|p|\leq 1$ migrate slowly, in some cases without a detectable preference direction.  On the other hand, the vortices in the simulation with $p=3.5$ undergo inward migration similar to the vortices in Figure \ref{fig:vort_dustfree} where $p=2.5$. This should be a consequence of the overall steeper density profiles in the latter simulations. Typical migration rates in the simulations with $p=2.5$ (Figure \ref{fig:vort_dustfree}) are $\sim 0.25-0.3$ $H_{g0}$/(100 orbits), whereas in the simulation with $p=3.5$ we find values $\sim 0.4$ $H_{g0}$/(100 orbits). Moreover, due to the similar results of the simulations with $|p|\leq 1$ this suggests that it is the mid-plane gas volume density which determines the migration speed of vortices (and therefore their possible growth by absorption of smaller vortices), rather than the surface mass density. That is, for $p={-1,0,1,2.5,3.5}$ the gas surface mass density $\Sigma_{g}$  follows a power-law with indices $s={-2,-1,0,1.5,2.5}$, respectively. Thus, if $\Sigma_{g}$ were the relevant quantity to vortex migration we would expect the strongest vortices to occur for $p=0,1,2.5$ and only little difference between the simulations with $p=0,2.5$ ($p=2.5$ being the fiducial case), which is not what we observe. These results are also compatible with the finding of \citet{manger2020} that vortices in their simulations with $p=0.66$ and $p=1.5$ appeared very similar, although the vortices in the top panels ($p=0.66$) of their Figure 8  do appear slightly larger than those in the bottom panels ($p=1.5$).

\end{appendix}

 \end{document}